\documentclass[11pt,a4paper]{article}
\input amssymb.sty
\usepackage{amssymb}
\usepackage{amscd}
\begin{document}


\newcounter{mo}
\newcommand{\mo}[1]
{\stepcounter{mo}$^{\bf MO\themo}$%
\footnotetext{\hspace{-3.7mm}$^{\blacksquare\!\blacksquare}$
{\bf MO\themo:~}#1}}

\newcounter{bk}
\newcommand{\bk}[1]
{\stepcounter{bk}$^{\bf BK\thebk}$%
\footnotetext{\hspace{-3.7mm}$^{\blacksquare\!\blacksquare}$
{\bf BK\thebk:~}#1}}


\newcommand{\Si}{\Sigma}
\newcommand{\tr}{{\rm tr}}
\newcommand{\ad}{{\rm ad}}
\newcommand{\Ad}{{\rm Ad}}
\newcommand{\ti}[1]{\tilde{#1}}
\newcommand{\om}{\omega}
\newcommand{\Om}{\Omega}
\newcommand{\de}{\delta}
\newcommand{\al}{\alpha}
\newcommand{\te}{\theta}
\newcommand{\vth}{\vartheta}
\newcommand{\be}{\beta}
\newcommand{\la}{\lambda}
\newcommand{\La}{\Lambda}
\newcommand{\D}{\Delta}
\newcommand{\ve}{\varepsilon}
\newcommand{\ep}{\epsilon}
\newcommand{\vf}{\varphi}
\newcommand{\vfh}{\varphi^\hbar}
\newcommand{\vfe}{\varphi^\eta}
\newcommand{\fh}{\phi^\hbar}
\newcommand{\fe}{\phi^\eta}
\newcommand{\G}{\Gamma}
\newcommand{\ka}{\kappa}
\newcommand{\ip}{\hat{\upsilon}}
\newcommand{\Ip}{\hat{\Upsilon}}
\newcommand{\ga}{\gamma}
\newcommand{\ze}{\zeta}
\newcommand{\si}{\sigma}

\def\hS{{\hat{S}}}

\newcommand{\li}{\lim_{n\rightarrow \infty}}
\def\mapright#1{\smash{
\mathop{\longrightarrow}\limits^{#1}}}

\newcommand{\mat}[4]{\left(\begin{array}{cc}{#1}&{#2}\\{#3}&{#4}
\end{array}\right)}
\newcommand{\thmat}[9]{\left(
\begin{array}{ccc}{#1}&{#2}&{#3}\\{#4}&{#5}&{#6}\\
{#7}&{#8}&{#9}
\end{array}\right)}
\newcommand{\beq}[1]{\begin{equation}\label{#1}}
\newcommand{\eq}{\end{equation}}
\newcommand{\beqn}[1]{\begin{small} \begin{eqnarray}\label{#1}}
\newcommand{\eqn}{\end{eqnarray} \end{small}}
\newcommand{\p}{\partial}
\def\sq2{\sqrt{2}}
\newcommand{\di}{{\rm diag}}
\newcommand{\oh}{\frac{1}{2}}
\newcommand{\su}{{\bf su_2}}
\newcommand{\uo}{{\bf u_1}}
\newcommand{\SL}{{\rm SL}(2,{\mathbb C})}
\newcommand{\GLN}{{\rm GL}(N,{\mathbb C})}
\def\sln{{\rm sl}(N, {\mathbb C})}
\def\sl2{{\rm sl}(2, {\mathbb C})}
\def\SLN{{\rm SL}(N, {\mathbb C})}
\def\SLT{{\rm SL}(2, {\mathbb C})}
\def\PSLN{{\rm PSL}(N, {\mathbb C})}
\newcommand{\PGLN}{{\rm PGL}(N,{\mathbb C})}
\newcommand{\gln}{{\rm gl}(N, {\mathbb C})}
\newcommand{\PSL}{{\rm PSL}_2( {\mathbb Z})}
\def\f1#1{\frac{1}{#1}}
\def\lb{\lfloor}
\def\rb{\rfloor}
\def\sn{{\rm sn}}
\def\cn{{\rm cn}}
\def\dn{{\rm dn}}
\newcommand{\rar}{\rightarrow}
\newcommand{\upar}{\uparrow}
\newcommand{\sm}{\setminus}
\newcommand{\ms}{\mapsto}
\newcommand{\bp}{\bar{\partial}}
\newcommand{\bz}{\bar{z}}
\newcommand{\bw}{\bar{w}}
\newcommand{\bA}{\bar{A}}
\newcommand{\bG}{\bar{G}}
\newcommand{\bL}{\bar{L}}
\newcommand{\btau}{\bar{\tau}}

\newcommand{\tie}{\tilde{e}}
\newcommand{\tial}{\tilde{\alpha}}

\newcommand{\Sh}{\hat{S}}
\newcommand{\vtb}{\theta_{2}}
\newcommand{\vtc}{\theta_{3}}
\newcommand{\vtd}{\theta_{4}}

\def\mC{{\mathbb C}}
\def\mZ{{\mathbb Z}}
\def\mR{{\mathbb R}}
\def\mN{{\mathbb N}}

\def\frak{\mathfrak}
\def\gg{{\frak g}}
\def\gJ{{\frak J}}
\def\gS{{\frak S}}
\def\gL{{\frak L}}
\def\gG{{\frak G}}
\def\gE{{\frak E}}
\def\gk{{\frak k}}
\def\gK{{\frak K}}
\def\gl{{\frak l}}
\def\gh{{\frak h}}
\def\gH{{\frak H}}
\def\gt{{\frak t}}
\def\gT{{\frak T}}
\def\gR{{\frak R}}

\def\baal{\bar{\al}}
\def\babe{\bar{\be}}

\def\bfa{{\bf a}}
\def\bfb{{\bf b}}
\def\bfc{{\bf c}}
\def\bfd{{\bf d}}
\def\bfe{{\bf e}}
\def\bff{{\bf f}}
\def\bfg{{\bf g}}
\def\bfm{{\bf m}}
\def\bfn{{\bf n}}
\def\bfp{{\bf p}}
\def\bfu{{\bf u}}
\def\bfv{{\bf v}}
\def\bfr{{\bf r}}
\def\bfs{{\bf s}}
\def\bft{{\bf t}}
\def\bfx{{\bf x}}
\def\bfy{{\bf y}}
\def\bfM{{\bf M}}
\def\bfR{{\bf R}}
\def\bfC{{\bf C}}
\def\bfP{{\bf P}}
\def\bfq{{\bf q}}
\def\bfS{{\bf S}}
\def\bfJ{{\bf J}}
\def\bfz{{\bf z}}
\def\bfnu{{\bf \nu}}
\def\bfsi{{\bf \sigma}}
\def\bfU{{\bf U}}
\def\bfso{{\bf so}}

\def\clA{\mathcal{A}}
\def\clC{\mathcal{C}}
\def\clD{\mathcal{D}}
\def\clE{\mathcal{E}}
\def\clG{\mathcal{G}}
\def\clR{\mathcal{R}}
\def\clU{\mathcal{U}}
\def\clT{\mathcal{T}}
\def\clO{\mathcal{O}}
\def\clH{\mathcal{H}}
\def\clK{\mathcal{K}}
\def\clJ{\mathcal{J}}
\def\clI{\mathcal{I}}
\def\clL{\mathcal{L}}
\def\clM{\mathcal{M}}
\def\clN{\mathcal{N}}
\def\clQ{\mathcal{Q}}
\def\clW{\mathcal{W}}
\def\clZ{\mathcal{Z}}

\def\baf{{\bf f_4}}
\def\bae{{\bf e_6}}
\def\ble{{\bf e_7}}
\def\bag2{{\bf g_2}}
\def\bas8{{\bf so(8)}}
\def\baso{{\bf so(n)}}

\def\sr2{\sqrt{2}}
\newcommand{\ran}{\rangle}
\newcommand{\lan}{\langle}
\def\f1#1{\frac{1}{#1}}
\def\lb{\lfloor}
\def\rb{\rfloor}
\newcommand{\slim}[2]{\sum\limits_{#1}^{#2}}

\newcommand{\sect}[1]{\setcounter{equation}{0}\section{#1}}
\renewcommand{\theequation}{\thesection.\arabic{equation}}
\newtheorem{predl}{Proposition}[section]
\newtheorem{defi}{Definition}[section]
\newtheorem{rem}{Remark}[section]
\newtheorem{cor}{Corollary}[section]
\newtheorem{lem}{Lemma}[section]
\newtheorem{theor}{Theorem}[section]

\begin{flushright}
 ITEP-TH-76/09\\
\end{flushright}
\vspace{10mm}
\begin{center}
{\Large{\bf Characteristic Classes and Integrable Systems. General Construction.}
}\\
\vspace{5mm}

A.Levin \\
{\sf State University - Higher School of Economics, Department of Mathematics, } \\
{\sf 20 Myasnitskaya Ulitsa, Moscow, 101000, Russia } \\
{\em e-mail alevin@hse.ru}\\
M.Olshanetsky\\
{ \sf
Institute of Theoretical and Experimental Physics, Moscow, Russia}
\\
{\em e-mail olshanet@itep.ru}\\
A.Smirnov\\
{\sf Institute of Theoretical and Experimental Physics, Moscow, Russia,}\\
{\em e-mail asmirnov@itep.ru}\\
A.Zotov \\
{\sf Institute of Theoretical and Experimental Physics, Moscow, Russia,}\\
{\em e-mail zotov@itep.ru}\\
\vspace{5mm}
\end{center}

\begin{abstract}

 We consider  topologically non-trivial Higgs bundles over elliptic curves with  marked points
  and construct corresponding integrable systems.
   In the case of one marked point we call them the modified Calogero-Moser systems
(MCM systems). Their phase space has the same dimension as the phase space of
the standard CM systems with spin, but less number of particles and greater number
of spin variables.
 Topology of the holomorphic bundles are defined by their characteristic classes.
Such bundles occur if G has a non-trivial center, i.e. classical simply-connected groups,
$E_6$ and $E_7$. We define the conformal version CG of  G - an analog of
GL(N)  for SL(N), and
relate the  characteristic classes with degrees of CG-bundles.
Starting with these bundles we construct
 Lax operators,
quadratic Hamiltonians, define the  phase spaces and the Poisson structure using dynamical r-matrices.
 To describe the systems we use a special basis in the Lie algebras that generalizes the basis of t'Hooft matrices for sl(N).  We find that
the MCM systems contain the standard CM systems related to some (unbroken)
subalgebras. The configuration space of the CM particles is the moduli space of the  holomorphic bundles with non-trivial  characteristic classes.
\end{abstract}


\tableofcontents


\section{Introduction}
\setcounter{equation}{0}

The paper conventionally speaking contains two types of results. First, we construct
topologically nontrivial holomorphic $G$-bundles over elliptic curves, where $G$ is a complex
Lie group and describe their moduli space.
Second, on the base of these results, we construct a new family of classical integrable systems related
to simple Lie groups. They are analogues of the elliptic Calogero-Moser systems.
We define the corresponding Lax operators, quadratic Hamiltonians and the classical dynamical elliptic $r$-matrices. The latter completes the classification list of classical
elliptic dynamical $r$- matrices \cite{EV}, where the underlying bundles are topologically trivial.

\bigskip

\textbf{1.Non-trivial holomorphic bundles over elliptic curves.}\\
 Let $\clE_G$ be a principle $G$-bundle
 over an elliptic curve $\Si_\tau=\mC/(\tau\mZ+\mZ)$ and $\pi$ is a representation of $G$ in
 $V$.  Following \cite{NS} we define a
  $G$-bundle $E=\clE_G\times_GV$  by the
transition operators $\clQ$ and $\La$ acting on  sections of $E$ as
$$
s(z+1)=\pi(\clQ)s(z)\,,~~~s(z+\tau)=\pi(\La) s(z)\,,~~~\clQ\,,\La\in G\,,
$$
where $\clQ$ and $\La$ take values in $G$.
The compatibility of this system dictates the following equation
for the transition operators
$$
\clQ(z+\tau)\La(z)\clQ(z)^{-1}\La^{-1}(z+1)=Id\,.
$$
Let $\zeta$ be an element of the center $\clZ(G)$ of $G$. Assume that  $\clQ$ and $\La$
 satisfy the equation
$$
\clQ(z+\tau)\La(z)\clQ(z)^{-1}\La^{-1}(z+1)=\zeta\,.
$$
Then  $\clQ$ and $\La$ can serve as transition operators only for a $G^{ad}=G/\clZ(G)$-bundle,
but not for a $G$-bundle and $\zeta$ is an obstruction to lift a $G^{ad}$-bundle to a $G$-bundle.

More generally, consider a $G$-bundle over a Riemann surface $\Si$ and assume that $G$ has a nontrivial center $\clZ(G)$. It means that $G$ is a classical simply-connected group, or some
of its subgroups, or a simply-connected group of type $E_6$ or $E_7$.
 The  topologically non-trivial $G$-bundles are characterized by elements of
  $H^2(\Si,\clZ(G))$. We call them the  \emph{characteristic classes}, since for $G=Spin_n$ they
  coincide with the Shtiefel-Whitney classes.

  It  follows from  \cite{NS} that it is possible to choose the constant transition operators.
   Then we come to the equation on $G$
  \beq{1}
  \clQ\La\clQ^{-1}\La^{-1}=\zeta\,.
  \eq
   We describe the set $\clM(G)=$(solutions of (\ref{1}))$/$(conjugations), when $G=\bG$ is
  a simply-connected group.\footnote{ The case  $G\subset\bG$ and $\clZ(G)$
  is nontrivial is also analyzed.}
   Assume that $\clQ$ is a semisimple element and $\clQ\in\clH_{\bG}$, where $\clH_{\bG}$ is a
   Cartan subgroup. Then $\clM(G)=(\clQ,\La)$ is defined as
   $$
  \clQ=\exp\,\Bigl(2\pi i\frac{\rho^\vee}{h}\Bigr)U\,, ~~~\La=\La^0V\,,
 $$
 $\rho^\vee$ is a half-sum of positive coroots, $h$  is the Coxeter number,
 $\La^0$ is an element of the Weyl group defined by $\zeta$. It is a symmetry of the extended
 Dynkin diagram of $\gg=$Lie$(\bG)$. $V$ and $U$ are arbitrary elements of the
 Cartan subgroup $\ti\clH_0\subset\clH_{\bG}$ commuting with $\La^0$ and
  $\ti\gH_0=$Lie$\,\ti\clH_0$ is a Cartan subalgebra corresponding to a simple Lie subgroup $\ti G_0\subset \bG$.

Since $(\La^0)^l=1$ for some $l$, the adjoint action of  $\La^0$ on $\gg$ is an automorphism of
 order $l$. All such  automorphisms described in \cite{Ka}. Ad$_{(\La^0)}$
induces a $\mu_l=\mZ/l\mZ$ gradation in $\gg$
$$
\gg=\oplus_{k=0}^{l-1}\gg_k\,,
$$
where $\gg_0$ is a reductive subalgebra.
The Lie algebra $\ti\gg_0=$Lie$(\ti G_0)$ in its turn is a subalgebra of $\gg_0$. The concrete forms of invariant
subalgebras are presented in Table 1. They will be calculated in \cite{LOSZII}.

\begin{center}\vspace{5mm}
\begin{tabular}{|c|c|c|c|}

  \hline
  & & & \\
  $G$ & ord $(\La^0)$ & $\ti\gg_0$ & $\gg_0$ \\
  \hline
   $\SLN$ $\,(N=pl)$ & $N/p$ & ${\bf sl_p}$ & ${\bf sl_p}\oplus_{j=1}^{l-1}{\bf gl_p}$\\
  SO$(2n+1)$ & 2 & $ \bfso(2n-1)$&$ \bfso(2n)$  \\
  Sp$(2l)$ & 2 & $\bfso(2l)$ & ${\bf gl}_{2l}$  \\
  Sp$(2l+1)$ & 2 & $\bfso(2l+1)$ & ${\bf gl}_{2l+1}$ \\
  SO$(4l+2)$ &4&  $\bfso(2l-1)$ & $\bfso(2l)\oplus\bfso(2l)\oplus\underline{1}$\\
  SO$(4l+2)$ &2&  $\bfso(4l-1)$ &  $\bfso(4l)\oplus\underline{1}$\\
  SO$(4l)$ & 2 & $\bfso(2l)$& $\bfso(2l)\oplus\bfso(2l)$\\
   SO$(4l)$ & 2 & $\bfso(4l-3)$& $\bfso(4l-2)\oplus\underline{1}$\\
  $E_6$ & 3& ${\bf g}_2$ &$\bfso(8)\oplus 2\cdot\underline{1}$\\
  $E_7$ &2 & ${\bf f}_4$ & ${\bf e}_6\oplus\underline{1}$\\
  \hline
\end{tabular}
\end{center}
\begin{center}
\textbf{Table 1.} $\La^0$-invariant subgroups and subalgebras
\footnote{Sp$(n)$ is a group preserving antisymmetric form in $\mC^{2n}$.}.\\
\small{
Since  $\clZ($SO$(4l))=\mu_2\oplus\mu_2$ we take two different $\La^0_a$, $(a=1,2)$}.
\end{center}

A big cell in the moduli space of trivial holomorphic bundles over an elliptic curve
 is a quotient  of the Cartan subalgebra
 $\gH$ of $G$ under the action of some discrete group.
For $G=GL_N$ the moduli space was described by M.Atiyah \cite{At}.
For trivial $G$-bundles, where $G$ is a complex simple group, it was done in \cite{BS,Lo}.
 Nontrivial $G$-bundles and their moduli space was considered in \cite{FM,Sch}.

 It is important for
 applications to consider holomorphic bundles with quasi-parabolic structures at marked points at
 $\Si_\tau$. It means that the automorphisms of the bundles (the gauge transformations) preserve
 flags $Fl_a$ located at $n$ marked points \cite{Si}.
 The structure of a big cell $\clM^0_{g,n}$ $(g=1)$ in the moduli
 space of these bundles can be extracted from the moduli space $\clM(G)$
of solutions of (\ref{1}).
In the simplest case $n=1$
\beq{2}
\ti\clM^0_{1,1}=(\ti\gH_0/\ti W_{BS})\times (Fl/\ti\clH_0)\,,
\eq
where $\ti W_{BS}$ are the Bernstein-Schwarzman generalizations \cite{BS} of the affine Weyl groups
$W^{aff}(\ti G_0)$, corresponding to different sublattices of the coweight lattice.
Note that for the trivial bundles $\La^0$ can be chosen as $Id$. In this case
\beq{3}
\clM^0_{1,1}=(\gH/ W_{BS})\times (Fl/\clH_{\bG})\,,
\eq
where $\gH=$Lie$(\clH_{\bG})$. Thus,  the big cell $ \ti\clM^0_{1,1}$ for the nontrivial $G$ bundles is the same as the big cell $\clM^0_{1,1}$ for trivial $\ti G_0$-bundles.
 A detail description is given in Section 3.2.

By product, we obtained some additional results related to this subject.
We describe an interrelation between the characteristic classes and degrees of some bundles.
In the $A_{N-1}$ case this relation is simple.
 The center of $G=\SLN$ is the cyclic group $\mu_N=\mZ/N\mZ$.
The cohomology group $H^2(\Si,\clZ(\SLN))$ is isomorphic to $\mu_N$.
Represent elements of $\mu_N$ as $\exp\,\frac{2\pi i}Nj\,,~j=1\ldots,N-1$.
Let $\zeta$ be a generator of $\mu_N$. Consider a principle PGL$(N,\mC)$
bundle with the  characteristic classes $\zeta$. It cannot be lifted to a
$\SLN$-bundle, but  can be lifted to a $\GLN$ bundle. The degree of its determinant bundle
$deg E$ is $-1$ and $\zeta=\exp\,(-\frac{2\pi i}N)=\exp\,(\frac{2\pi i\, deg E}{rank\, E})$.
 We generalize this construction to other simple groups. To this end for a simple group $G$
 we define its conformal version $CG$ (Definition 3.2). In particular,
 for the symplectic and orthogonal groups their conformal versions
 are  groups preserving (anti)symmetric forms up to dilatations.
 It allows us to relate the  characteristic classes of $G$-bundles
  to degrees of the determinant bundles of $CG$ (Theorem 3.1).

We introduce a special basis in  $\gg=Lie\,G$.
In the $A_{N-1}$ case it is the basis of the finite-dimensional sin-algebra \cite{FFZ},
generated by the t'Hooft matrices
$\clQ$, $\La$ $\,(\clQ\La\clQ^{-1}\La^{-1}=\exp\,(\frac{2\pi i}N))$.
We call it the generalized sinus (GS) basis and use it in the context of integrable systems.

\bigskip

\textbf{2.Integrable systems.}\\
Using the above construction we describe a new class of the finite-dimensional classical completely integrable systems
related to simple Lie groups with nontrivial centers. They are  generalizations of the elliptic Calogero-Moser systems, in
general with spin degrees of freedom.
 Calogero-Moser systems (CM) were originally defined in quantum case by Francesco Calogero
\cite{Ca} and in classical case by Jurgen Moser \cite{Mo},
as an integrable model of one-dimensional nuclei.
Now they play an essential role both in mathematics and in theoretical  physics.
  \footnote{The mathematical aspects of the systems are discussed in \cite{Et}.}

Their generalizations as integrable systems related to simple Lie groups has a long history.
It was started more than thirty years ago  \cite{OP},
but the classical integrability was proved there only for the classical groups. It was done
later in \cite{BCS,DF}. They are the so-called spinless CM systems.
The case of the $A_{n-1}$ type (SL(n)) systems
 is very special.
The integrability of these systems for rational and
trigonometric potentials has a natural explanation in terms
of Hamiltonian reduction \cite{KKS,OP1}. Later this approach was generalized for a wide
class of classical integrable systems
- the so-called Hitchin systems \cite{Hi}. It was realized in \cite{ER,GN,LO,Ma} that the
$A_{n-1}$ type  CM systems with elliptic potential are particular examples of the Hitchin systems.

From the point of view of the Hitchin construction it is more natural to consider
CM systems with spin, introduced in the  $A_{n-1}$ case in \cite{GH,Wo}.
\footnote{The spinless CM systems considered in \cite{BCS,DF} was described as Hitchin systems  in \cite{HM}.}
 Their description for all simple Lie algebras can be found in \cite{LX}.
Generically, the Hitchin systems come up as a result of the
Hamiltonian reduction of the Higgs bundles \cite{Hi}.
Upon the reduction we obtain an integrable systems in the Lax form, where the Lax operator
depends on a spectral parameter belonging to the base of the bundle.
The reduced phase is the  cotangent bundle to the moduli space of holomorphic bundles
with the quasi-parabolic structure.
The CM systems with spin appear as a result of the Hamiltonian reduction
of the quasi-parabolic Higgs bundles over an elliptic curve with one marked point.

It turns out that the standard classification of the CM systems is based on topologically trivial bundles.
The primary goal of this paper is a  classification of MCM systems related to topologically non-trivial bundles.
A particular examples related to $\SLN$
are known. If the characteristic class of the bundle $\zeta=\exp\,(-\frac{2\pi i}N)$, instead of interacting CM particles
 we come to the Euler-Arnold (EA) top \cite{Ar} related to  $\SLN$   \cite{KLO,LOZ1,RSTS}.
This top describes the classical degrees of freedom on a vertex in the vertex spin chain. The corresponding classical $r$ matrix is non-dynamical \cite{BD}. But if $N=pl$ there exists an intermediate situation  \cite{LZ} described in column 2:

\bigskip
\begin{tabular}{|c|c|c|c|}
  \hline
  & 1& 2& 3\\
  \hline
  &&&\\
  $\zeta$ & 1 & $\exp\,(-\frac{2\pi ip}N)\,,$\,~~N=pl$ $&$\exp\,(\frac{-2\pi i}N)$ \\
   \hline
   &&&\\
 System & SL$_N$-CM system & SL$_p$-CM-system $+l$ interacting EA-tops & SL$_N$-EA-top\\
  \hline
\end{tabular}
\begin{center}
\textbf{Table 2.} \\
Integrable systems corresponding to different characteristic classes
of SL$(N)$ bundles.
\end{center}
\bigskip

 In this paper we construct  Lax operators, quadratic Hamiltonians and
 corresponding classical dynamical $r$-matrices for any
simple  complex Lie group $G$ with a non-trivial center and arbitrary  characteristic classes
$\zeta\in H^2(\Si_\tau,\clZ(G))$. The obtained elliptic $r$-matrices  are completion
of the list \cite{EV,LX}, because the dynamical parameters belongs
to the Cartan subalgebra  $\ti\gH_0\subset\gH_G$.
This type of  $r$-matrices in the trigonometric case were constructed
in \cite{ES,Sh},  using an algebraic approach.

In fact, $\ti\gH_0$ is the same Cartan subalgebra that participates
 in the definition of the moduli space
 (\ref{2}). Let us explain this phenomena.
The phase space of the  Hitchin systems is the moduli space $\clM^H_{\Si_n}$
of the Higgs bundles over a curve $\Si_n$ with the quasi-parabolic structure at $n$ marked points.
It is a bundle over the moduli space $\clM^H_{\Si}$ of the Higgs bundles over the compact curve
$\Si$. The base  $\clM^H_{\Si}$ can be interpreted as the phase space of interacting
particles. It is the cotangent bundle to the moduli space
$\clM_{\Si}$ holomorphic bundles over $\Si$.
 The fibers $\clM^H_{\Si_n}\to \clM^H_{\Si}$ are   coadjoint $G$-orbits
 located at the marked points.
The coordinates on the orbits are called the spin variables.
\footnote{For elliptic curves the phase space of the spin variables is a result of a
Hamiltonian reduction of the coadjoint orbits with respect to action of the Cartan subgroup.}
If the number of the marked points $n=1$ and the $G$-bundle
over the elliptic curve has a trivial characteristic class, then
the spin variables can be identify with angular velocities of the EA top related to $G$.
The inertia tensor of the top depends on coordinates of CM particles related to the same group  $G$. The configuration spaces of paricles are the quotient of the Cartan algebra as in (\ref{3}).
It the space  of dynamical parameters of $r$.

For non-trivial bundles the configuration space of particles
is quotient of the Cartan subalgebra $\ti\gH_0\subset\gH$ and the dynamical $r$-matrices depends on variables belonging $\ti\gH_0$.
The integrable system looks as interacting EA tops with parameters depending on
coordinates of the CM system related to $\ti G_0$.
For this reason we call $\ti G_0$ the unbroken subgroup (see Table 1).

Solutions of (\ref{1}) allows us to define the Lax operators for
non-trivial bundles. We describe the Poisson brackets for the Lax
operators in terms of classical dynamical r-matrix following the
papers \cite{AT,BAB,BSu,Feh,LX}. This form contains an anomalous term
preventing the integrability of the system upon the Hamiltonian
reduction with respect to action the Cartan subgroup $\ti\clH_0$. We
prove the classical dynamical Yang-Baxter equation for the
$r$-matrix defined in \cite{Fe,Skl}.

 It is worthwhile to emphasize that for  the standard CM systems we deal in fact with
a few different systems.
More exactly, we have as many configuration spaces as a number of non-isomorphic moduli
spaces. It amounts to existence of different sublattices in the coweight lattice containing the coroot lattice.
A naive explanation of this fact is as follows.
The potential of the system has the form $\wp(\lan\bfu,\al\ran)$,
where $\bfu$ is a coordinate vector,  $\al$ is a root and $\wp$ is the Weierstrass function.
\footnote{In what follows we use the second Eisenstein function $E_2(z)$. It differs from $\wp$
on a constant.}
 Adding to $\bfu$ any combination
$\ga_1+\ga_2\tau$, where $\ga_j\in Q^\vee$-coroot lattice,
does not change the potential, because $\wp(\lan\bfu,\al\ran)$ is a double-periodic on the lattice $\tau\mZ\oplus\mZ$
 and $\lan\ga,\al\ran$ is an integer.
Thus, the configuration space is the quotient $\gH/(\tau Q^\vee\oplus Q^\vee)$.
It is the most big configuration space. But we can harmlessly shift as well by the coweight
lattice $\tau P^\vee\oplus P^\vee$. Then we come to a different configuration space (the smallest one).
For $A_{N-1}$ root systems we describe in this way the $\SLN$ and $\PSLN$ CM systems.
Their configuration spaces are different, while the Hamiltonians are the same. Evidently,
this fact becomes important for the quantum systems.
The same is valid for the systems with non-trivial characteristic class. But now one should
consider the lattices related to the unbroken subgroups.

Finally, we should mention that in spite of apparent dissimilarity
of Hamiltonians with different characteristic classes, the corresponding integrable systems are symplectomorphic.
In particular, the MCM systems are symplectomorphic to the standard spin CM systems. The symplectomorphisms
 are provided by the so-called Symplectic
Hecke Correspondence \cite{LOZ1}. In terms of the Lax operators the
symplectomorphisms are defined by acting on them by special singular gauge transformations.
 A particular example of such transformation establishing an equivalence of  the  $\SLN$ CM system and  the $\SLN$ EA top was given in \cite{LOZ1}.
Following \cite{KW} the Symplectic Hecke Correspondence can be explained
in terms of monopole solutions of the Bogomolny equation. Details can be found in \cite{LOZ2}.

\vspace{0.3cm}
\bigskip
{\small {\bf Acknowledgments.}\\
 The work was supported by grants RFBR-09-02-00393, RFBR-09-01-92437-KEa,
NSh-3036.2008.2, RFBR-09-01-93106-NCNILa (A.Z. and A.S.),
RFBR-09-02-93105-NCNILa (M.O.)  and to the Federal Agency for
Science and Innovations of Russian Federation under contract
14.740.11.0347. The work of A.Z. was also supported by the Dynasty
fund and the President fund MK-1646.2011.1. A.L and M.O. are
grateful for hospitality to the Max Planck Institute of Mathematics,
Bonn, where the part of this work was done.}


\section{Holomorphic bundle.}

\setcounter{equation}{0}

\subsection*{Global description of holomorphic bundles}

Here we define  holomorphic bundles
over  a Riemann surface $\Si_g$ of genus $g$ following the approach developed in \cite{NS}.

Let  $\pi_1(\Si_g)$ be a fundamental group of $\Si_g$.
It has $2g$ generators $\{a_\al,b_\al\}\,$, corresponding to the
 fundamental cycles of $\Si_g$ with the relation
\beq{0.3}
\prod_{\al=1}^g [b_{\al},a_{\al}]=1\,,
\eq
where $[b_{\al},a_{\al}]= b_{\al}a_{\al}b_{\al}^{-1}a_{\al}^{-1}$ is
the group commutator.

Consider a finite-dimensional representation $\pi$ of a simple complex Lie group $G$
in a space $V$. Let $\clE_G$ is a principle $G$-bundle  over $\Si_g$. We
define a holomorphic $G$-bundle  $E=\clE_G\times_GV$
(or in more detail $E_G$ or $E_G(V)$) over $\Si_g$ using $\pi_1(\Si_g)$.
The bundle $E_G$ has the space of sections $\G(E_G)=\{s\}$,
where $s$ takes values in $V$.
Let $\rho$ be a representation of $\pi_1$ in $V$ such that $\rho(\pi_1)\subset\pi(G)$.
The bundle $E_G$ is defined by transition matrices of its sections
 around the fundamental cycles.
Let $z\in \Si_g$ be a fixed point. Then
\beq{sect}
 s(a_\al z)= \rho(a_\al)s(z)\,,~~
 s(b_\be z)= \rho(b_\be)s(z)\,.
\eq
Thus, the sections are defined by their quasi-periodicities on the fundamental cycles.
Due to (\ref{0.3}) we have
\beq{pi}
\prod_{\al=1}^g [\rho(b_{\al}),\rho(a_{\al})]=Id\,.
\eq

The $G$-bundles described in this way are topologically trivial. To consider
less trivial situation assume that $G$ has a non-trivial
center $\clZ(G)$.
Let $\zeta\in\clZ(G)$.
Replace (\ref{pi}) by
\beq{pi1}
\prod_{\al=1}^g [\rho(b_{\al}),\rho(a_{\al})]=\zeta\,.
\eq
Then the pairs $(\hat\rho(a_\al),\hat\rho(b_\be))$, satisfying (\ref{pi1}), cannot describe  transition matrices of $G$-bundle, but can serve as transition matrices of $G^{ad}=G/\clZ(G)$-bundle.
The bundle $E_G$ in this case is topologically non-trivial and
 $\zeta$ represents \emph{the characteristic class} of $E_G$. It is an obstruction to lift $G^{ad}$ bundle to $G$ bundle. We will give a formal definition in Section 44.

The transition matrices can be deformed without breaking  (\ref{pi}) or
(\ref{pi1}).  Among these deformations are the gauge transformations
 \beq{gtr}
\rho(a_\al)\to f^{-1}\rho(a_\al)f\,,~~\rho(b_\be)\to f^{-1}\rho(b_\be)f\,.
\eq
\emph{The moduli space of holomorphic bundles} $\clM_{g}$ is the space of transition
matrices defined up to the gauge transformations.
 Its dimension is independent on the characteristic class and
is equal to
\beq{mg}
\dim\,(\clM_{g})=(g-1)\dim\,(G)\,.
\eq
It means that the nonempty moduli spaces arise for the holomorphic bundles over
surfaces of genus $g>1$.

To include into the construction the surfaces with $g=0,1$
 consider a Riemann surface with $n$ marked points and attribute $E$ with
 what is called the quasi-parabolic structure at the  marked points. Let $B$ be a Borel subgroup of $G$.
We assume that the gauge transformation $f$ preserves the flag  variety $Fl=G/B$. It means that
$f\in B$ at the marked points. It follows from (\ref{fld})
 \beq{mgn}
\dim\,(\clM_{g,n})=(g-1)\dim\,(G)+n\dim\, (Fl)=(g-1)\dim\,(G)+n\sum_{j=1}^{rank\, G}(d_j-1)\,.
\eq
In the important for applications case $g=1$, $n=1$
$\dim\,(\clM_{g,1})=\dim\,(Fl)$.


\subsection*{Local description of holomorphic bundles and modification}

There exists another description of  holomorphic bundles over $\Si_g$.
Let $w_0$ be a fixed point on $\Si_g$
 and $D_{w_0}$ ($D^\times_{w_0}$) be a disc (punctured disc) with a center $w_0$
with a local coordinate $z$. Consider
a $G$-bundle $E_G=\clE_G\times_GV$ over  $\Si_g$.
It can be trivialized over  $D$ and over
$\Si_g\setminus w_0$. These two trivializations are related by a
$G$ transformation $\pi(g)$
  holomorphic in $D^\times_{w_0}$, where $D_{w_0}$ and $\Si_g\setminus w_0$ overlap.
If we consider another trivialization over $D$
then $g$ is multiplied from the right by  $h\in G$.
Likewise, a trivialization over $\Si_g\setminus w_0$ is determined up to the
multiplication on the left $g\to hg$ , where $h\in G$ is holomorphic on
 $\Si_g\setminus w_0$.
Thus, the set of isomorphism classes of
$G$-bundles are described as a double-coset
\beq{mshb}
G(\Si_g\setminus w_0)\setminus G(D^\times_{w_0}) /G(D_{w_0})\,,
\eq
where $G(U)$ denotes the group of $G$-valued holomorphic functions
on $U$.

To  define a $G$-bundle over $\Si_g$ the transition matrix $g$ should have
 a trivial monodromy around $w_0$ $g(ze^{2\pi i})=g(z)$ on the punctured disc $D^\times_{w_0}$.
 But if the monodromy is nontrivial
 $$
 g(ze^{2\pi i})=\zeta g(z)\,,~~\zeta\in\clZ(G)\,,
 $$
 then $g(z)$ is not a transition matrix. But it can be considered as
  a transition matrix for the $G^{ad}$-bundle, since  $G^{ad}=G/\clZ(G)$.
  This relation is similar to (\ref{pi1}).

  Our aim is to construct from $E$ a new bundle $\ti E$ with a non-trivial characteristic class.  This procedure is called \emph{a modification} of bundle $E$.
Smooth gauge transformations cannot change a topological type of bundles.
The modification is defined by a singular gauge transformation at some point,
say $w_0$. Since it is a local transformation we replace $\Si_g$ by a sphere
$\Si_0=\mC P^1$, where $w_0$ corresponds to the point $z=0$ on $\mC P^1$. Since $z$ is local coordinate, we
can replace
 $G(\Si_g\setminus w_0)$ in (\ref{mshb}) by the group $G(\mC((z)))$. It is the
group of Laurent series with $G$ valued coefficients. Similarly, $G(D_{w_0})$ is replaced
by the power series  $G(\mC[[z]])$. It is clear from this description of the
moduli space of bundles over $\mC P^1$ that it is a finite dimensional space.

Transform $g(z)$ by multiplication from the right on
$g(z)\to g(z)h(z)$ where $h(z)$ singular at $z=0$.
It is the singular gauge transformation mentioned above. Due to definition of
$g(z)$, $\,h(z)$ is defined up to the multiplication from the right by $f(z)\in G(\mC[[z]])$.
On the other hand, since $g(z)$ is defined up to the multiplication from the right by an
element from $G(\mC[[z]])$, $h(z)$ is element of the double coset
$$
G(\mC[[z]])\setminus G(\mC((z)))/G(\mC[[z]])\,.
$$
In particular, $h(z)$ is defined up to a conjugation. It means that as a representative
of this double coset one can take a co-character (\ref{cocharc1}) $h(z)\in t(G)$.
\beq{mtm}
g(z)\to g(z)z^{\ga}\,,~~~(z^{\ga}=\bfe(\ln\,(z\ga)))\,,
\eq
where $\ga$ belongs to the coweight lattice $(\ga=(m_1,m_2,\ldots,m_l)\in P^\vee)$ (\ref{cwl}).
The monodromy of $z^{\ga}$ is $\exp\,-(2\pi\ga)$. Since $\lan\al,\ga\ran\in\mZ$
  for any $\bfx\in\gg\,$ Ad$_{\exp\,-(2\pi\ga)}\bfx=\bfx$. Then $\exp\,-(2\pi\ga)$
 an element of $\clZ(\bG)$ (\ref{center}). If the transition matrix $g(z)$
  defining  $E$ has a trivial monodromy, the new transition matrix (\ref{mtm}) acquires a nontrivial monodromy.  In this way we come to a new bundle $\ti E$ with a non-trivial characteristic class.
The bundle $\ti E$ is called \emph{the modified bundle}.
It is defined by the new transition matrix (\ref{mtm}).
If $\ga\in Q^\vee\,$ then $\zeta=1$ and the modified bundle $\ti E$  has the same
type as $E$.

This transformation of the bundle $E$ corresponds to transformations of its sections
$\ti E$
\beq{mod1}
\G(E)\,\mapright{\Xi(\ga)} \,\G(\ti E)\,,~~
(\Xi(\ga)\sim\pi(z^{m_1},z^{m_2}\ldots z^{m_l}))\,.
\eq
We say that this modification  has a type $\ga=(m_1,m_2,\ldots,m_l)$.
Another name of the modification is \emph{the Hecke transformation}.
It acts on the characteristic classes of bundles as follows
\beq{amcc}
\Xi(\ga)\,:\,\ln\zeta(E)\to\ln\zeta(\ti E)=\ln\zeta(E)+2\pi i\ga\,,~~
\ga\in P^\vee/Q^\vee\,.
\eq

Consider the action of modification on sections (\ref{mod1})  in more details.
Let $V$ be a space of a finite-dimensional representation $\pi$ of $G$ with a
highest weight $\nu$ and $\nu_j$ $\,(j=1,\ldots,N)$ is a set of its weights
\beq{nu}
\nu_j=\nu-\sum_{\al_m\in\Pi}c_j^m\al_m\,,~~~c_j^k\in\mZ\,,~c_j^k\geq 0\,.
\eq
It means that for $\bfx\in\gH\,$ $\,\pi(\bfx)|\nu_j\ran=\lan\bfx,\nu_j\ran|\nu_j\ran$.
The weights belong to the weight diagram defined by the highest weight $\nu\in P$   of $\pi$.
The space $V$ has the weight basis
$(|\nu^{s_1}_1\ran,\ldots,|\nu^{s_N}_N\ran)$ in $V\,$, where $s_1=1,\ldots,m_1\,,\ldots$
$s_N=1,\ldots,m_N$ and $m_1$, $\ldots,m_N$ are multiplicities of weights.
Thus,  $M=\dim\,V=\sum m_j$.

Let us choose a trivialization of $E$ over $D$ by fixing this basis.
Thereby, the bundle $E$ over $D$ is represented by a sum
of $M$ line bundles $\clL_1\oplus\clL_2\oplus\ldots\oplus\clL_M$.
Cartan subgroup $\clH$ acts in this basis in a diagonal way: for
$\bfs=(|\nu^{s_1}_1\ran,\ldots,|\nu^{s_N}_N\ran)$
$$
\pi(h)\,:\,|\nu^{s_j}_j\ran\to \bfe\lan\bfx,\nu_j\ran |\nu^{s_j}_j\ran\,,~~~h=\bfe\,(\bfx)\,,~~\bfx\in\gH\,,
~~~(\bfe(x)=\exp\,(2\pi i\bfx))\,.
$$
Assume for simplicity that in (\ref{mtm}) $g(z)=1$. Then
 the modification transformation (\ref{mod1}) of the sections assumes the form
 \beq{mod}
\Xi(\ga)\,:\, |\nu^{s_j}_j\ran\to z^{\lan \ga,\nu_j\ran} |\nu^{s_j}_j\ran\,,~~j=1,\ldots,M\,.
\eq
It means that away from the point $z=0$, where the transformations are singular, the sections of $\ti E$ are the same as of $E$. But near $z=0$ they are singular with the leading terms
$|\nu^{s_j}_j\ran\sim z^{-\lan \ga,\nu_j\ran}$.

It is sufficient to consider the case when  $\ga=\varpi^\vee_i$ is a fundamental coweight and $\pi$ is
a fundamental representation $\nu=\varpi_k$.
Then from (\ref{nu}) we have
$$
z^{\lan \ga,\nu_j\ran}=z^{\lan\varpi^\vee_i,\varpi_k-\sum_{\al_m\in\Pi}c_j^m\al_m \ran}\,.
$$
The weight $\varpi_k$ can be expanded in the basis of simple roots  $\varpi_k=\sum_kA_{km}\al_m$,
where $ A_{jk}$ is the inverse Cartan matrix ($A_{jk}a_{ki}=\de_{ji}$).
Its matrix elements are rational numbers with the denominator $N=ord\,(\clZ)$.
Then from (\ref{cwl})
$$
z^{\lan \ga,\nu_j\ran}\sim z^{\frac{l}N+m}\,,~~l,m\in\mZ\,.
$$
 Note, that the branching does not happen for $G^{ad}$-bundles, because the corresponding  weights $\nu_j$ belong to the root lattice $Q$ and thereby $\lan \ga,\nu_j\ran\in\mZ$.

It is possible
to go around the branching by multiplying the sections on a scalar matrix of the form
$\di(z^{-A_{ik}},\ldots,z^{-A_{ik}})$. This matrix  no longer belongs to the representation of
$\bG$, because it has the determinant $z^{-MA_{ik}}$ $(M=\dim\,V)$. It can be checked that $MA_{ik}$ is an integer number.

If $G=\SLN$ the scalar matrix  belongs to $\GLN$. Thereby, after this transformation we come to a $\GLN$-bundle.
But this bundle is topologically non-trivial, because it has a non-trivial degree. In this way the characteristic classes for the $\SLN$-bundles
are related to another topological characteristic, namely to  degrees of the  $\GLN$-bundles. We describe below the similar construction for other simple
groups.


\section{Holomorphic bundles over elliptic curves}

\setcounter{equation}{0}

Hereinafter we consider the bundles over an elliptic curve, described as
the quotient $\Si_\tau\sim\mC/(\tau\mC\oplus\mC)$, $\,(Im\,\tau>0)$.
There are two fundamental cycles corresponding to shifts $z\to z+1$ and $z\to z+\tau$.
Let $G$ be a complex simple Lie group.
Sections of a $\bG$-bundle $E_G(V)$ over $\Si_\tau$ satisfy the quasi-periodicity conditions
(\ref{sect})
\beq{s1}
s(z+1)=\pi(\clQ)\, s(z)\,,~~~s(z+\tau)=\pi(\La)\, s(z)\,,
\eq
where $\clQ$, $\La$ take values in $G$.
A bundle $\ti E$ is equivalent to $E$ if its sections $\ti s$ are related to $s$ as
$\ti s(z)=f(z)s(z)$, where $f(z)$ is invertible operator in $V$. It follows from (\ref{s1})
that the transition operators, have the form
\beq{gtt}
\ti\clQ=f(z+1)\clQ f^{-1}(z)\,,~~~~\ti\La=f(z+\tau)\La f^{-1}(z)\,.
\eq
As we have mentioned, the moduli space $\clM_{1,n}$ is the quotient space of pairs
$(\clQ,\La)$ with respect to this action. In what follows we consider the simplest case
$n=1$, though our construction is applicable for arbitrary $n$.

The transition operators define a trivial bundle if $[\clQ,\La]=Id$.
Let  $\zeta$ be an element of $\clZ(\bar G)$. To come to a nontrivial bundle
we should find solutions
 $\La,\clQ\in\bar G$ of the equation
\beq{gce}
\La\clQ\La^{-1}\clQ^{-1}=\zeta\,.
\eq
It follows from (\ref{cG}) that the r.h.s. can be represented as $\zeta=\bfe(-\varpi^\vee)$,
where $\varpi^\vee\in P^\vee$ (\ref{cwl}). Then (\ref{gce}) takes the form
\beq{gce1}
\La\clQ\La^{-1}\clQ^{-1}=\bfe(-\varpi^\vee)\,,~~(\bfe\,(x)=\exp\,(2\pi ix))\,.
\eq

 It follows from \cite{NS} that the transition operators can be chosen as constants.  Therefore, to describe the moduli space of holomorphic bundles we should find a pair $\clQ$, $\La\in G$ satisfying
 (\ref{gce1}) and defined up to the conjugation
\beq{gtcm}
\La\to f\La f^{-1}\,,~~\clQ\to f\clQ f^{-1}\,.
\eq
Let $g=\bG$ be a simply-connected group.
Let us fix a Cartan subgroup $\clH_{\bar G}\subset \bar G$. Assume that $\clQ$
is  semisimple, and therefore   is conjugated to an element from $\clH_{\bar G}$.
We will see that by neglecting non-semisimple  transition operators we still define a big cell
in the moduli space.
Our goal is to find solutions of (\ref{gce}), where  $\clQ$ is a generic
element of a fixed Cartan subgroup $\clH\subset G$.


\subsection*{Algebraic equation}

\begin{predl}
Solutions of (\ref{gce1}) up to the conjugations have the following description.\\
$\bullet$ The  element $\La$ has the form $\La=\La^0V$, where $\La^0$
is defined uniquely by the  coweight
 $\varpi^\vee_j$ $\,(\La^0=\La^0_j)$. It is an element from the Weyl group $W$
 preserving the extended coroot system  $\Pi^{\vee ext}=\Pi^\vee\cup\al^\vee_0$, and in this way is a symmetry of the extended Dynkin diagram. $V\in\clH_{\bG}$ commutes with $\La^0$. \\
 $\bullet\bullet$ The element $\clQ$ has the form $\clQ=\clQ^0U$, where
 \beq{ka}
\clQ^0=\exp\,2\pi i\ka\,,~~~ \ka=\frac{\rho^\vee}{h}\in\gH\,,
 \eq
  where $h$  is the Coxeter number,
 $\rho^\vee=\oh\sum_{\al^\vee\in (R^\vee)^+}\al^\vee$ and $U$ commutes with $\La^0$.
 \footnote{The first statement can be found in \cite{Bou} (Proposition 5 in VI.3.2). We give another
 proof because it elucidates the proof of  the second statement.}
 \end{predl}
\emph{Proof}\\
In (\ref{ka}) $\ka$ can be chosen from a fixed Weyl chamber (\ref{wca}).
From (\ref{tb1}) and (\ref{tb}) we find  that if $\kappa_0\in Q^\vee$ then
$\bfe\,(\kappa_0)=Id$. Therefore, by shifting $\ka\to\ka+\ga$,
$\ga\in Q^\vee$, $\ka$ can be put in $ C_{alc}$ (\ref{gran}).
Rewrite (\ref{gce}) as
\beq{gce2}
\La\clQ\La^{-1}=\zeta\clQ\,,~~\zeta=\bfe\, (-\xi)\,.
\eq
Here $\La$ is defined up to multiplication from $\clH_{\bG}$ and we write it
in the form $\La^0V$, $V\in\clH_{\bG}$.

\begin{lem}
There exists a conjugation $f$ (\ref{gtcm}) such $\clQ\to\clQ$ and
$\La^0V\to\La^0 V_\la$, and $\La^0 V_\la=V_\la\La^0$.
\end{lem}
\emph{Proof}.\\
Let us take $f\in\clH_{\bG}$. Then $f$ preserves $\clQ$. It acts on the second transition operator
as
$$
\La^0V\to f\La^0f^{-1}V=\La^0(\La^0)^{-1}f\La^0f^{-1}V\,.
$$
Define $V_\la$ as  $V_\la=(\La^0)^{-1}f\La^0f^{-1}V$. Our goal is to prove that there exists such $f$ that
$V_\la$ commutes with $\La^0$. In other words, $f\La^0f^{-1}V=\La^0)^{-1}f\La^0f^{-1}V\La^0$.
Let
$V=\bfe(\bfx)$, $f=\bfe(\bfy)$, $\bfx\,,\,\bfy\in\gH$, $\la=Ad_{\La^0}$.
Then the commutativity condition takes the form
$(\la-1)\bfx=(\la^{-1}-1)\bfy+(\la-1)\bfy$.

Let $l$ be an order of $\La^0$\,, $\,((\La^0)^l=1)$.
Then a solution of this equation is given by a sum
$$
\bfy=\f1{l}\sum_{i=1}^{l}i\la^i(\bfx)\,.
$$
Thus $\La^0$ and $V$  defines $V_\la=\bfe(\bfp)$ commuting with $\La^0$, where $\bfp$
is the average along the $\la$-orbit
\beq{vla}
\bfp=\f1{l}\sum_{i=0}^{l-1}\la^i(\bfx)\,.
\eq
 $\Box$

On the next step we find $\La^0$.
Rewrite (\ref{gce1}) in the form
\beq{gce21}
\la(\ka)=\ka-\xi\,,~~\xi=\varpi^\vee_j\,,~~~\la={\rm Ad}_\La\,,
\eq
where $\ka\in C_{alc}$.
 Define a subgroup $\G_{C_{alc}}$ of the affine Weyl group $W_a'$ (\ref{q25}) $\,\G_{C_{alc}}\subset W_a'$ that preserves $C_{alc}$
(\ref{na}). It acts by permutations on its vertices (\ref{na}). Equivalently, $\G_{C_{alc}}$ acts
by  permutations of nodes of the extended Dynkin graph.
The face of $C_{alc}$ belonging to the hyperplane $\lan\al_i,x\ran=0$ contains all
vertices except $\varpi^\vee_i/n_i$. Similarly, the face belonging to the hyperplane
$(\al_0,x)=1$ contains all vertices except $0$. By this duality the permutations of vertices by
$g=(\la,\xi)\in\G_{C_{alc}}$ correspond to permutations of the faces, and in this way to permutations of the coroots $\Pi^{\vee ext}$.

Instead of  (\ref{gce21}) consider
$\la (C_{alc})+\xi=C_{alc}$.
The left hand side of this equation is a transformation $g=(\la,\xi)\in\G_{C_{alc}}$.
Let us take  $\xi=\varpi^\vee_j$, where $\varpi^\vee_j$ is a fundamental coweight that
is a vertex of $C_{alc}$ ($n_j=1$ in (\ref{na})). Remember, that only these $\varpi^\vee_j$ define nontrivial elements of the quotient $P^\vee/Q^{\vee}$.
Then we have
\beq{01}
\la_j(C_{alc})= C_{alc}-\varpi^\vee_j \equiv C'_{alc}\,.
\eq
The node $0$ of $C'_{alc}$  is an image of the node $\varpi^\vee_j$ of $C_{alc}$ after the shift.
Let us define $\la_j$.
The Weyl group $W$ action on the Weyl alcoves that contains $0$ is simple transitive.
Therefore, there exists a unique  $\la_j\in W$ such that $\la_j(C_{alc})=C'_{alc}$.
Then $(\la_j,\varpi^\vee_j)\in \G_{C_{alc}}$ defines a transformation
of $C_{alc}$, which is a permutation of its vertices (\ref{na}) such that
$\varpi^\vee_j\to 0$. Taking into account the action of $\G_{C_{alc}}$ on the extended
Dynkin graph we find
\beq{laa}
 \la_j^*(\al_k)=\left\{
 \begin{array}{ll}
                       \al_m & k\neq j \\
                       -\al_0 & k=j
                     \end{array}
 \right.
 \al_k\,,\al_m\in\Pi\,.
 \eq
Thus, taking $\xi=\varpi^\vee_j$ we find $\La_j$.

 Fixed points of the $\G_{alc}$-action are solutions of (\ref{gce21}). It will give us $\ka$ and in this way $\clQ$.
Let us prove  that a particular solution of (\ref{gce21}) is
 \beq{ka2}
 \ka=\frac{\rho^\vee}{h}\,,
 \eq
 where $h$  is the Coxeter number (\ref{co11}).
 The equation (\ref{gce21}) is equivalent to
 \beq{con}
 \lan\ka,\la_j^*(\al_k)\ran= \lan\ka,\al_k\ran-\de_{jk}\,, ~~~\al_k\in\Pi\,, k=1,\ldots,l\,.
 \eq

Since $\rho^\vee=\sum_{m=1}^l\varpi^\vee_m$ (see (\ref{arho})) for $k\neq j$
(\ref{con}) becomes a trivial identity. For $k=j$ using (\ref{mro}) we obtain
$-\f1{h}\sum_{m=1}^ln_m-\f1{h}=-1$.
It follows from  (\ref{co11}) that it is again identity.

An arbitrary solution of  (\ref{gce21}) takes the form
$$
\ka=\frac{\rho^\vee}{h}+\bfq\,,~~~\bfq\in Ker(\la_j-1)\,.
$$
In other words, the Weyl transformation $\la_j$ should preserve $\bfq$.

Thus, taking in (\ref{gce}) $\zeta=\exp \,-(2\pi i\varpi^\vee_j)$ we find solutions
$(\La_j=\La_j^0V_\la,\clQ)$, where $\La^0_j$ is a symmetry of the extended Dynkin graph corresponding
to $\varpi^\vee_j$ and
\beq{geq}
\clQ=\exp\,2\pi i(\frac{\rho^\vee}{h}+\bfq)\,.
\eq
The pair $(\bfp,\bfq)$ (\ref{vla}), belonging to the Cartan subalgebra $\gH$,  plays the role of the moduli parameters of solutions to (\ref{gce1}).
$\Box$

\begin{rem}
For $Spin(4n)\,$ there are two generators $\zeta_1$ and $\zeta_2$ of
 $\,\clZ(Spin(4n))\sim \mu_2\oplus\mu_2$  corresponding to the
 fundamental weights
$\varpi_a$, $\varpi_b$ of the left and the right spinor representations. Arguing as above we will find
two solutions $\La_a$ and $\La_b$ of (\ref{gce1}), while $\clQ$ is the same in the both cases.
\end{rem}

\bigskip
Consider a group $G$, $\,(\bG\supset G\supseteq G_{ad})$ and let $\La,\clQ\in G$.
Let us choose $\xi=\varpi$ such that it generates the group $t(G)$ of  co-characters
$t(G)=P^\vee_l$ (\ref{coch}), (\ref{tb1})
$t(G)=\varpi+Q^\vee\,,~~l\varpi\in Q^\vee$.
 Then $\zeta=\bfe\,(-\varpi)$ is a generator
of center $\clZ(G)\sim P^\vee/t(G)=\mu_l$ (see (\ref{cG})). Arguing as above
we come to
\begin{predl}
$\bullet$ The  element $\La$ is defined  by the  coweight
 $\varpi^\vee\in W$. It is a symmetry of the extended Dynkin diagram.
 $\La$ is defined up to invariant elements from $\clH_G$. \\
 $\bullet\bullet$ Let
 \beq{modv}
(\la(G)-1)\bfq=0\,,~~\bfq\in \gH\,,~\la(G)=Ad_{\La(G)}\,.
 \eq
 A general solution of (\ref{gce1}) is
\beq{ka1}
 \ka=\frac{\rho^\vee}{h}+\bfq\,,
 \eq
 \end{predl}
Therefore, the group of cocharacters $t(G)$ defines a Weyl symmetry $\La^0(G)$
of the extended Dynkin diagram $\Pi^{\vee ext}$ such that $(\La^0)^l(G)=Id$.

$\La(G)$ and $\clQ$ play the role of transition operators of  $G$-bundles over $\Si_\tau$.
A generator $\zeta=\bfe\,(\varpi)$ defines a characteristic class of the
bundles. It is an obstruction to lift $G$-bundle to $\bar G$-bundle.

\begin{rem}
If $\xi\in Q^\vee$ then $\zeta=Id$. It means that we can take $\xi=0$ as a representative of $P^\vee/ Q^\vee$.
Then $\la=1$ (see (\ref{01}) and $Ker\,(\la-1)=\gH$. In this case the bundle has
a trivial characteristic class, but has holomorphic moduli  defined by the vector $\bfq\in\gH$.
The corresponding Higgs bundle over $\Si_\tau/(z=0)$ defines elliptic spin
Calogero-Moser system.
\end{rem}


\subsection*{The moduli space}

We have described a $G$-bundle $E_G(V)$  by the transition operators
$(\La=\La^0\bfe\,(\bfp),\clQ=\bfe\,(\frac{\rho^\vee}h+\bfq)$, where $\La^0$ corresponds to the
 coweight $\varpi^\vee\in P^\vee$ .
The topological type of $E$ is defined by an
element of the quotient $P^\vee/t(G)$. Let us transform $(\La,\clQ)$ taking in (\ref{gtt})
$f=\bfe\,(-\bfq z)$. Since $f$ commutes with $\La^0$ we come to new transition operators
$\clQ=\bfe\,(\ka+\bfq)\to\clQ=\bfe\,(\ka)$,
$\La\to \La^0\bfe\,(\bfp-\bfq\tau)$.
Denote $\bfp-\bfq\tau=\ti\bfu$
\footnote{We will write $\ti\bfu$ for nontrivial bundles reserving $\bfu$ for trivial bundles.}. Then sections of $E_G(V)$  assume the quasi-periodicities
\beq{mon1}
s(z+1)=\pi(\bfe\,(\ka))\,s(z)\,,~~s(z+\tau)=\pi(\bfe\,(\ti\bfu)\La^0)\, s(z)\,.
\eq
Thus, we come to the transition operators
\beq{qala}
\clQ=\bfe\,(\ka)\,,~~~~\La=\bfe\,(\ti\bfu)\La^0\,.
\eq
Here  $\ti\bfu$ plays the role of a parameter in the moduli space.
In this subsection we describe it in details.


\subsubsection*{Trivial bundles}

Consider first the simplest case $\La=Id$ and $\bfu\in\gH$ (see Remark 3.2)).
It means, that $E$ has a trivial characteristic class.
The transition transformations $\pi(\bfe\,(\ka))$, $\pi(\bfe\,(\bfu))$ lie in
 of the Cartan subgroup $\clH_G$ of $G$.

Consider first a bundle $E_{\bG}$ for a simply-connected
group $\bar G$. Since   $t(\bar G)\sim Q^\vee$ (\ref{tb}) and due to (\ref{tb1}),
 $\bfe\,(\bfu+\ga)=\bfe\,(\bfu)$ for  $\ga\in Q^\vee$.  Taking into account that $\bfu$ lies
in a Weyl chamber we conclude that in fact   $\bfu\in C_{alc}$ as
it was already established.
Now apply the transformation $\bfe\,(\ga z)$
\beq{gtrg}
s(z)\to \pi(\bfe\,(\ga z))s(z)\,,~~\ga\in Q^\vee\,.
\eq
 The
sections  are transformed as
\beq{mon2}
s(z+1)=\pi(\bfe\,(\ka))\,s(z)\,,~~s(z+\tau)=\pi(\bfe\,(\bfu+\ga\tau))\,s(z)\,.
\eq
Thus, transition operators, defined by parameters $\bfu$ and $\bfu+\tau\ga_1+\ga_2$
$(\ga_{1,2}\in Q^\vee)$, describe  equivalent bundles.
The semidirect product of the Weyl group $W$ and the lattice $\tau Q^\vee\oplus Q^\vee$
is called \emph{the Bernstein-Schwarzman group} \cite{BS}
$$
W_{BS}=W\ltimes(\tau Q^\vee\oplus Q^\vee).
$$
Thereby,  $\bfu$ can be taken
from the fundamental domain $C^{(sc)}$ of $W_{BS}$. Thus,
\beq{csc}
\fbox{$
C^{(sc)}=\gH/W_{BS}~{\rm is~the~
moduli ~space ~of~trivial} ~\bar G-{\rm bundles}\,.$}
\eq

Consider $G^{ad}$ bundle and let $\bfe\,(\bfu)\in G^{ad}$. In this case  $\bfe\,(\ga)=1$ if $\ga\in P^\vee$ (\ref{tb}), (\ref{tb1}).
Define the group
$$
W^{ad}_{BS}=W\ltimes(\tau P^\vee\oplus P^\vee)\,.
$$
As above, we come to the similar conclusion:
\beq{cad}
\fbox{$
C^{(ad)}=\gH/W^{ad}_{BS}~{\rm is~the~
moduli ~space ~of~trivial} ~ G_{ad}-{\rm bundles}\,.$}
\eq

\bigskip

Consider a coweight  $\varpi^\vee\in P^\vee$  such that $l\varpi^\vee\in Q^\vee$, the coweight lattice  $P_l^\vee=\mZ\varpi^\vee\oplus Q^\vee$ (\ref{coch1}). Thus,  $P^\vee/ P_l^\vee\sim\mu_l$.
Consider a group $G_l$ (\ref{fgl}) and \,   generated by a coweight
$$
P_l^\vee\,,~~l\varpi^\vee\in Q^\vee\,.
$$
.
 The coweight sublattice $P_l^\vee$
 is the group of its cocharacters $t(G_l)$ (\ref{coch}). Representations of
 $G_l$ are defined by the dual to $t(G_l)$ groups of characters $\G(G_l)$ (\ref{cha}).
The dual to $P_l^\vee$ lattice $P_p\subset P$ has the form
$$
P_p=\mZ\varpi+Q\,,~~p\varpi\in Q\,.
$$

By means of $P_l^\vee$ define the affine group of the  Bernstein-Schwarzman type
$$
W^{(l)}_{BS}=W\ltimes(\tau P_l^\vee\oplus P_l^\vee)\,.
$$
Making use of the gauge transform $\bfe\,(\ga z)\in G_l$,  $\,(\ga\in P_l^\vee)$ we find
that
\beq{cl}
\fbox{$
C^{(l)}=\gH/W^{(l)}_{BS}~{\rm is~the~
moduli ~space ~of~trivial} ~ G_l-{\rm bundles}\,.$}
\eq

Consider  the dual picture and the lattice $P^\vee_p$.
It is formed by $Q^\vee$ and a coweight $\varpi^\vee$
$$
P^\vee_p=\mZ\varpi^\vee+ Q^\vee\,,~~p\varpi^\vee\in Q^\vee\,.
$$
 The lattice $P^\vee_p$ plays the role of the group  of  cocharacters for the  dual group
 $^LG_l=G_p=\bG/\mu_p$, (\ref{fgl}), while $P_l$ of defines characters of $G_p$.
 Again by means of the group
$$
W^{(p)}_{BS}=W\ltimes(P_p^\vee\oplus P_p^\vee\tau)\,.
$$
we find that
 \beq{cp}
\fbox{$
C^{(p)}=\gH/W^{(p)}_{BS}~{\rm is~the~
moduli ~space ~of~trivial} ~ G_p-{\rm bundles}\,.$}
\eq

Thus, for the $\bar G,G_l,G_p,G_{ad}$ trivial bundles we have the following
interrelations between their moduli space
\beq{msd}
\begin{array}{ccccc}
   &  & C^{(sc)} &  &  \\
   & \swarrow & \mid & \searrow &  \\
  C^{(l)} &  & \mid &  & C^{(p)} \\
   & \searrow & \downarrow & \swarrow &  \\
   &  &C^{(ad)}  &  &
\end{array}
\eq
Here arrows mean coverings. Note that  $C^{(sc)}\,,C^{(ad)}$ and as well
 $C^{(l)}\,, C^{(p)}$ are dual to each other in the sense that the defining them
 lattices are dual.

 Let $Fl$ be a flag variety located at the marked point.
In this way we have defined  a space $\ti\clM_{1,1}=(C^a,Fl)$ $\,(a=(sc),(l),(ad))$ related to the moduli space
 of trivial bundles over $\Si_\tau$ with one marked point.
 But we still have a freedom to act on $Fl$ by constant conjugations from the Cartan subgroup
 $\clH^a$.
 Thus, eventually we come to $\clM^0_{1,1}=(C^a,Fl/\clH^a)$. It has dimension of $\clM_{1,1}$ (\ref{mgn}). It is a big cell in $\clM_{1,1}$. In our construction we have excluded non-semisimple elements $\clQ$.


\subsubsection*{Nontrivial bundles}

Consider a general case $\La^0\neq Id$. It was explained above that
 $\La^0$ corresponds to some characteristic class related to $\varpi^\vee\in P^\vee\,,$ and
 $\varpi^\vee\notin Q^\vee$.
In this case $\ti\bfu\in Ker(\la-1)$, and in fact
$\,\ti\bfu\in C_{alc}\cap\ti\gH_0$, where $\ti\gH_0$ is the invariant subalgebra $\la(\ti\gH_0)=\ti\gH_0$. There is a basis in $\ti\gH_0$ defined by a system of simple coroots $\ti\Pi^\vee$ (see Section 5.4). Moreover, the corresponding root system defines a simple Lie algebra $\ti\gg_0$.

Let $\ti W$ be  the Weyl group $W$ of the root system  $\ti R=\ti R(\ti\Pi)$
\beq{iws}
\ti W=\{w\in \ti W\,|\,w(\ti R)=\ti R\,\}\,,
\eq
and
\beq{tiq}
\ti Q^\vee=\{\ga=\sum_{j=1}^pm_j\ti\al^\vee_j\,,~m_j\in\mZ\}
\eq
is the coroot lattice generated by $\ti\Pi^\vee$ (\ref{tir}).
Consider first  $E_{\bG}$ bundles.
 As above, $\bfe\,(\ti\bfu+\ga)=\bfe\,(\ti\bfu)$, $\ga\in \ti Q^\vee$.
The automorphism (\ref{gtrg}) for  $\ga\in \ti Q^\vee$ commutes with $\La$. Thus, $\ti\bfu$ and $\ti\bfu+\tau\ga_1+\ga_2$ $\,\ga_{1,2}\in\ti Q^\vee$
define equivalent $\bG$-bundles.
Consider the semidirect products
\beq{twsc}
\ti W_{BS}=\ti W\ltimes(\tau \ti Q^\vee\oplus \ti Q^\vee)\,.
\eq
The fundamental domain in $\ti\gH$ under the $\ti W_{BS}$ action
is the moduli space of $\bG$-bundles with characteristic classes defined
by $\varpi^\vee$
\beq{tsc}
\fbox{$\ti C^{sc}=\ti\gH/\ti W_{BS}
{\rm ~is~the~
moduli ~space ~of~nontrivial} ~ \bG-{\rm bundles}$}\,,
\eq

Consider $E_{G^{ad}}$-bundles.
Let $\ti\varpi_j^\vee$ be fundamental coweights
($\lan\ti\varpi_j^\vee,\ti\al_k\ran=\de_{jk}$)
and
\beq{tiq1}
\ti P^\vee=\{\ga=\sum_{j=1}^pm_j\ti\varpi^\vee_j\,,~m_j\in\mZ\}
\eq
is the coweight lattice in $\ti\gH_0$.
Define the semidirect product
\beq{twad}
\ti W^{ad}_{BS}=\ti W\ltimes(\tau\ti P^\vee\oplus \ti P^\vee)\,.
\eq
A fundamental domain under its action
\beq{tad}
 \fbox{$\ti C^{ad}=\ti\gH_0/\ti W^{ad}_{BS}
 {\rm ~is~the~
moduli ~space ~of~nontrivial} ~ G^{ad}-{\rm bundles}$}\,.
 \eq
is  a moduli space of a $G^{ad}$-bundle with characteristic class
define by $\varpi^\vee$. If $ord(\clZ(\bar{\ti G})$ is not a primitive number then we again come to the
 hierarchy of the moduli spaces similar to (\ref{msd}).

 As above the space $\clM^0_{1,1}=(C^a,Fl/\ti\clH_0)$ is a big cell in the moduli space
 of non-trivial bundles.


\section{Characteristic classes and conformal groups}
\setcounter{equation}{0}

\subsection*{Characteristic classes}

Let $\clE_G$ be is a principle $G$-bundle  over $\Si$.
 Consider a finite-dimensional representation of complex group $G$ in
 a space $V$ and let  $E_G(V)$ be the vector bundle $E_G(V)=\clE_G\times_GV$
  induced by $V$.

The first cohomology $H^1(\Si_g,G(\clO_\Si))$  of $\Si$ with coefficients in analytic sheaves define  the moduli space $\clM(G,\Si)$
of holomorphic $G$-bundles. Let $\bG$ be a simply-connected group and $G^{ad}$ be an
adjoint group.
Using (\ref{adg}) and (\ref{fgl}) we write three exact sequences
$$
\begin{array}{l}
  1\to\clZ(\bG))\to\bG(\clO_\Si)\to G^{ad}(\clO_\Si)\to 1\,, \\
    1\to\clZ_l\to\bG(\clO_\Si)\to G_l(\clO_\Si)\to 1\,, \\
    1\to\clZ(G_l)\to G_l(\clO_\Si)\to G^{ad}(\clO_\Si)\to 1\,,
\end{array}
$$
where $G_l=\bG/\clZ_l$.
Then we come to the long exact sequences
\beq{coh1}
\to H^1(\Si_g,\bG(\clO_\Si))\to H^1(\Si_g,G^{ad}(\clO_\Si))\to H^2(\Si_g,\clZ(\bG))\sim\clZ(\bG))\to 0 \,,
\eq
\beq{coh2}
\to H^1(\Si_g,\bG(\clO_\Si))\to H^1(\Si_g,G_l(\clO_\Si))\to H^2(\Si_g,\clZ_l)\sim\mu_l\to 0\,,
\eq
\beq{coh3}
\to H^1(\Si_g,G_l(\clO_\Si)\to H^1(\Si_g,G^{ad}(\clO_\Si))\to H^2(\Si_g,\clZ(G_l))\sim\mu_p\to 0\,.
\eq
The elements from $H^2$ are obstructions to lift bundles, namely
$$
\begin{array}{c}
\zeta(E_{G^{ad}})\in H^2(\Si_g,\clZ(\bG)) -{\rm ~obstructions ~to~ lift~}
 E_{G^{ad}}-{\rm bundle~to~}E_{\bG}-{\rm bundle}\,,\\
  \zeta(E_{G_l})\in H^2(\Si_g,\clZ_l) -{\rm ~obstructions ~to~ lift~}
 E_{G_l}-{\rm bundle~to~}E_{\bG}-{\rm bundle}\,, \\
    \zeta^\vee(E_{G^{ad}})\in H^2(\Si_g,\clZ(G_l)) -{\rm ~obstructions ~to~ lift~}
 E_{G^{ad}}-{\rm bundle~to~}E_{G^l}-{\rm bundle}\,.
  \end{array}
$$
\begin{defi}
Images of $H^1(\Si_g,G(\clO_\Si))$ in $H^2(\Si_g,\clZ)$ are called the characteristic classes
$\zeta(E_G)$ of $G$-bundles.
\end{defi}
Since
$\clZ_l\to\clZ(\bG)\to\clZ(G_l)$.
we have the following relations between these characteristic classes
$\zeta^\vee(E_{G^{ad}})=\zeta(E_{G^{ad}})\,mod\,\clZ_l$,
and the characteristic class $\zeta(E_{G_l})$ coincides with $\zeta(E_{G^{ad}})$
as an obstruction to lift a $E_{G_l}$-bundle, treated as a $E_{G^{ad}}$-bundle to a $E_{\bG}$- bundle.

Consider a particular case $\bG=\SLN$, $G^{ad}=\PSLN$. Then the elements
$\zeta\in \clZ(\SLN)\sim\mu_N$ are obstructions to lift $\PSLN$-bundles
to $\SLN$-bundles.
They represent the characteristic classes of $\PSLN$-bundles.
On the other hand, the exact sequence
\beq{0.6}
1\to \clO^*\to\GLN\to {\rm PGL}(N,\mC)\to Id
\eq
gives rise to the exact sequence of cohomology
\beq{0.7}
H^1(\Si_g,\GLN)\to H^1(\Si_g,{\rm PGL}(N,\mC))\to H^2(\Si_g,\clO^*)\,.
\eq
The Brauer group $H^2(\Si,\clO^*)$ vanishes and, therefore, there are
no obstructions to lift ${\rm PGL}(N,\mC)\sim{\rm PSL}(N,\mC)$-bundles to $\GLN$-bundles.
A topological characteristic of a $\GLN$-bundle is the degree of its determinant bundle.
In following subsections we will construct an analog of $\GLN$  for other simple groups.
We call them the \emph{conformal groups}. The main goal is to
relate the characteristic classes to the degrees of some line bundles connected to
the conformal groups.


\subsection*{Conformal groups}

Here we introduce an analog of the group $\GLN$ for other simple groups apart from $\SLN$.
Let
\beq{phi1}
\phi\,:\, \clZ(\bG)\hookrightarrow \left(\mC^*\right)^r
\eq
 be an embedding of the
 center $\clZ(\bG)$ into algebraic torus $\left(\mC^*\right)^r$ of
  minimal dimension ($r=1$ for a cyclic center and $r=2$ for
  $\mu_2\times\mu_2$).
 Note that any two embeddings are
 conjugate from the left: $\phi_1=A\phi_2$ for some automorphism
 $A$ of the torus $\left(\mC^*\right)^r$.
 It is not true for $\clZ(\SLN)=\mu_N$. But for other groups we deal with
 $\mu_2,\mu_3,\mu_4$ or $\mu_2\times\mu_2$.
In these cases  nontrivial  roots of unity coincide or they are
 inverse to each other. In the latter case $A\,:\,x\to x^{-1}$.

 Consider the "anti-diagonal" embedding
 $ \clZ(\bG)\to \bG\times \left(\mC^*\right)^r\,,$
$\,\zeta\mapsto (\zeta,\phi(\zeta)^{-1})\,,$ $\,\zeta\in \clZ(\bG)$.
 The image of this map is a normal
subgroup since $\clZ$ is the center of $\bG$.
\begin{defi}
The quotient
$$
C\bG=\left(\bG\times \left(\mC^*\right)^r\right)/\clZ(\bG)
$$
is called  the conformal version of $\bG$.
\end{defi}
In the similar way the conformal version can be defined for any $G$ with a non-trivial center.
If the center of $G$ is trivial as for $G^{ad}$ then $CG=G\times\mC^*$.

The group $C\bG$ does not depend on embedding  in $\mC^r$
   due to above remark about conjugacy of $\phi$'s.
 We have a natural inclusion $\bG\subset C\bG$.

 Consider the quotient  torus
 $Z^{\vee}=\left(\mC^*\right)^r/\clZ(\bG)\sim\left(\mC^*\right)^r$.
  The last isomorphism is defined by $\lambda\to \lambda^N$ for cyclic center
 and $(\lambda_1,\lambda_2)\to (\lambda_1^2,\lambda_2^2)$ for $D_{even }$.
  The sequence
 \beq{bgcg}
 1\to \bG\to C\bG\to Z^{\vee}\to 1
 \eq
 is the analogue of
$$
1\to \SLN\to \GLN\to \mC^* \to 1\,.
$$

 On the other hand, we have embedding $\left(\mC^*\right)^r\to C\bG$ with the quotient $C\bG/\left(\mC^*\right)^r=G^{ad}$. Then the sequence
 \beq{cgad}
 1\to\left(\mC^*\right)^r\to C\bG\to G^{ad}\to 1
 \eq
 is similar  to the sequence
 $$
 1\to\Bbb C^*\to \GLN\to \PGLN\to 1\,.
 $$

Let $\pi$ be an irreducible  representation of $\bG$ and  $\chi$ is a character of the torus $\left(\mC^*\right)^r$. It follows from (\ref{bgcg})
that an irreducible  representation $\ti\pi$ of $C\bG$ is defined as
 \beq{tipi}
\ti\pi=\pi\boxtimes\chi((\mC^*)^r)\,,~{~\rm such~that}~\pi|_{\clZ(\bG)}=\chi\phi\,,~~~(\phi~(\ref{phi1}))\,.
 \eq

 Assume for the simplicity that  $\pi$ is a fundamental representation. It means that  the highest weight $\nu$ of $\pi$ is a fundamental weight.
Let $\varpi^\vee$ be
 a fundamental coweight generating $\clZ(\bG)$ for $r=1$. In other words,
 $\zeta=\bfe(\varpi^\vee)$ is a generator of $\clZ(\bG)$ ($\zeta^N=1$, $\,N=$ord$(\clZ(\bG))$.
 Then $\pi|_{\clZ(\bG)}$ acts as a scalar
$\bfe\lan\varpi^\vee,\nu\ran$. The highest weight can be expanded in the basis of
simple roots $\nu=\sum_{\al\in\Pi}c^\nu_\al\al$. Then the coefficients $c^\nu_\al$
are rows of the inverse Cartan matrix.
They have the form $k/N$, where $k$ is an integer. Therefore the scalar
\beq{scal}
\bfe\lan\varpi^\vee,\nu\ran=\bfe\,
\Bigl(\sum_{\al\in\Pi}c^\nu_\al\de_{\lan\varpi^\vee,\al\ran}\Bigr)
\eq
is a root of unity. On the other hand, let $\chi_m(\mC^*)=w^m$ ($w\in\mC^*$) be a character of $\mC^*$,  and
$\phi(\zeta)=\bfe\,(l/N)$.
In terms of weights the definition of $\ti\pi$ (\ref{tipi}) takes the form
$\bfe\lan\varpi^\vee,\nu\ran=\bfe\,\Bigl(\frac{ml}N\Bigr)$.
It follows from this construction that characters of $C\bG$ are defined by
the weight lattice $P$ and the integer lattice $\mZ$ with an additional restriction
$$
\chi_{(\ga,m)}(\bfx,w)=\exp\,2\pi i\lan\ga,\bfx\ran w^m\,,~~
\lan\ga,\varpi^\vee\ran=\frac{ml}N+j\,,~~\ga\in P\,,~~m,j\in\mZ\,,~~\bfx\in\gH\,.
$$
The case $D_{even}$ $(r=2)$ can be considered in the similar way.

\begin{rem}
Simple groups can be defined as subgroups of $GL(V)$ preserving
some multi-linear forms in $V$. For examples, in the fundamental representations
these forms are  symmetric forms for $SO$,
 antisymmetric forms for $Sp$, a trilinear form for $E_6$ and a form of
 fourth order for $E_7$.
 In a generic situation $G$ is defined as a subgroup of $GL(V)$ preserving a
 three tensor in $V^*\otimes V^*\otimes V$ \cite{GP}.
 The conformal versions of these groups can be
 alternatively defined as transformations preserving the forms up to dilatations.
 We prefer to use here the algebraic construction, but this approach justifies the name "conformal version".
 \end{rem}

The conformal versions can be also defined in terms of exact representations of $\bG$.
 Let $V$ be such a representation and assume that $\clZ(\bG)$
 is a cyclic group.
  Then $C\bG$ is a subgroup of $GL(V)$ generated by $G$ and dilatations
 $\mC^*$. The character $\det\,V$ is equal to $\lambda^{\dim\,(V)}$, where $\la$ is equal to (\ref{scal}) for fundamental representations.

 For $D_{even }$ we use two representations, f.e.
 the left and right spinors $Spin^{L,R}$. The conformal group
 $CSpin_{4k}$ is a subgroup
 of $GL(Spin^L\oplus Spin^R)$ generated by $Spin_{4k}$ and $\mC^*\times\mC^*$,
 where the first factor $\mC^*$ acts by dilatations on $Spin^L$ and
the second factor acts on $Spin^R$. The character
$\det\,Spin^L$ ($\det\,Spin^R$) is equal to $\lambda_1^{\dim\,(Spin_{4k}^L)}$
($\lambda_2^{\dim\,(Spin_{4k}^R)}$), $\,(\dim\,(Spin_{4k}^{L,R})=2^{2k-1})$.


\subsection*{Characteristic classes and degrees of vector bundles}

From the exact sequence (\ref{cgad}) and vanishing of the second cohomology of
 a curve $H^2(\Si,\clO^*)=0$ with coefficients in analytic sheaf
 we get that
 any $G^{ad}({\cal O})$-bundle (even topological non-trivial with
 $\zeta(G^{ad}({\cal O})\neq 0$)  can be lifted to a  $C\bG ({\cal O})$-bundle.

 Let $V$ be an exact representation either irreducible or the sum
$Spin^L\oplus Spin^R$ for $D_{2k}$. Then from (\ref{phi1})
one has an embedding of $\clZ(\bG)$ to the automorphisms of $V$
\beq{phv}
\phi_V\,:\,\clZ(\bG)\hookrightarrow
\left(\mC^*\right)^r={\rm Aut}_{\bG}(V)\,.
\eq
In particular case, when $V$ is a space of a fundamental representation
 the center acts on $V$ by multiplication on (\ref{scal}).

 Let ${\cal E}_{C\bG}$ be a principal $C\bG({\cal O})$-bundle.
Denote by $E( V)={\cal E}\otimes_{C\bG}V$ (or $E(Spin^{L,R})$) a vector bundle
 induced by a representation $V$ ($Spin^{L,R}$ for $D_{even}$).

\begin{theor}
\footnote{For $G=\GLN$ this theorem was proved in \cite{NS}}
Let $E_{ad}=E(Ad)$ be the adjoint bundle with the  characteristic class
 ${\zeta}(E_{ad})$. The image of $\zeta(E_{ad})$
 under $\phi_V$ (\ref{phv}) is
 $$
 \phi_V(\zeta(E_{ad}))=\left\{
 \begin{array}{l}
  \exp\bigl(-2\pi i\,{\rm deg}\,(E(V))/{ \dim V}\bigr)\,, \\
   \exp\bigl(-2\pi i{\rm deg}\,(E(Spin_{4k}^{L,R}))/2^{2k-1}\bigr)\,.
 \end{array}
 \right.
 $$
\end{theor}

\emph{ Proof.} Consider the commutative diagram
$$
\begin{CD}
  @.1@. 1@. @.\\
@. @AAA@AAA@.@.\\
  1 @>>>Z^{\vee}({\cal O}_\Si)@>\sim>> Z^{\vee}({\cal O}_\Si)@>>> 1@.\\
@. @AA[N]A@AA A@AAA@.\\
  1 @>>>\left({\cal O}_\Si^*\right)^r@>>>C\bG({\cal O}_\Si)@>>>G^{ad}({\cal O}_\Si)@>>>1\\
@. @AAA@AAA@AAA@.\\
 1@>>>\clZ(\bG)@>>>\bG({\cal O}_\Si)@>>>G^{ad}({\cal O}_\Si)@>>>1\\
@. @AAA@AAA@AAA@.\\
@.1@. 1@.1 @.\\
\end{CD}
$$
and corresponding diagram of \^{C}ech cochains.
Let $\psi$ be a 1-cocycle with values in $G^{ad}({\cal O}_\Si)$.
Consider its preimage as a cocycle with values in $C\bG ({\cal O}_\Si)$.
Due to definition of $C\bG$ this cocycle is a pair of cochains  $(\Psi,\nu)$
with values in $\bG({\cal O}_\Si)$ and $\left({\cal O}_{\Si}^* \right)^r$ such
that $\phi_V(d\Psi)d\nu=1\in \left({\cal O}^* \right)^r$, where $d$ is the
\^{C}ech coboundary operator. The cohomology class of $d\Psi$ by definition is the characteristic class ${\bf c}$, so $\phi_V$ of it is opposite to the class of
$d\nu$:
$\,\phi_V(\zeta(E_{ad}))=(d\nu)^{-1}$.
Since $\nu$ acts in $V$ as a scalar $\nu^{{\rm dim}V}$, it is a one-cocycle as a
determinant of this action. It represents the
determinant of the bundle $E(V)$. In this way $\nu$ is a
preimage of the cocycle $\nu^{{\rm dim}V}$ under the taking $N=\dim\,(V)$ power
${\cal O}^*\stackrel{[N]}{\rightarrow}{\cal O}^*\,,$ $\,\nu\to
\nu^{N}\,,$   $\,N=\dim\,(V)$.

Consider the long exact sequence
$$
1\to\mu_{N} \to{\cal O}_\Si^*\stackrel{[N]}{\rightarrow}{\cal O}_\Si^*\to 1\,,
~~~(\mu_{N}=\mZ/N\mZ)\,.
 $$
It  induces the map $H^1(\Si,{\cal O}_\Si^*)\ {\rightarrow}H^2(\Si,\mu_N)$.
 The cocycle $d\nu$ lies in the cohomology class which is an image
of the class of $\det\,E(V)=\nu^N$ under the
coboundary map $H^1(\Sigma, {\cal O}^*)\to H^2(\Sigma, \mu_N)$.  Denote it by
 ${\rm Inv}_N=$Image$(\det\,E(V))$. Thus, by the definition, the class of $d\nu$ equals to ${\rm Inv}_{N}(\det\,E(V))=
{\rm Inv}_{N}(\zeta_1(E(V)))$.

The statement of the theorem follows from  the following
proposition
\begin{predl}
 Let $\gamma$ be a 1-cocycle with values in
${\cal O}^*$. Then ${\rm Inv}_N(\gamma)=\exp\left(\frac1N2\pi
i\,{\rm deg}(\gamma)\right)$.
\end{predl}
\emph{Proof}\\
 Consider the diagram
$$
\begin{CD}
0@>>>\mu_N@>>>{\cal O}_\Si^*@>[N]>>{\cal O}_\Si^*@>>>0\\
@. @.@AA\exp A@AA\exp A @.\\
 @.0 @>>>{\cal O}_{\Si}@>\times N >>{\cal O}_{\Si}@>>>0 \\
 @. @.@AAA @AAA  @.\\
 @. @.2\pi i \Bbb Z@>\times N >>2\pi i \Bbb Z@. \\
\end{CD}
$$

Let $\gamma$ be a 1-cocycle of ${\cal O}_{\Si}^*$. By definition its
image in $H^2(X,\mu_N)$ is equal to the coboundary of 1-cochain
$\gamma^{1/N}$ of ${\cal O}_{\Si}^*$, $(\gamma^{1/N})^N=\gamma $. Let
$\log(\gamma)$ be a preimage of the cycle $\gamma$ under
exponential map; $\log(\gamma)$ is a 1-cochain of ${\cal O}_{\Si}$ and
its coboundary equals to degree of $\gamma$ times $2\pi i$. As the
multiplication by $N$ is invertible on ${\cal O}_{\Si}$, the cochain
$\frac1N\log(\gamma)$ is well-defined, due to commutativity of the
diagram we can choose $\exp\left(\frac1N\log(\gamma)\right)$ as
$\gamma^{1/N}$. Hence, the image of $\gamma$ in $H^2(X,\mu_N)$
equals to coboundary of $\exp\left(\frac1N\log(\gamma)\right)$ equals
exponential of coboundary of  $ \frac1N\log(\gamma) $ equals
exponential of degree of $\gamma$ times $\frac{2\pi i}N$.

The case $r=2$,   can be analyzed in the same way. The theorem is
proved.
$\Box$

\bigskip

Let as above $\varpi^\vee$ be a fundamental coweight generating a center $\clZ(\bG)$
and $\nu$ is weight of representation of $\bG$ in $V$.
Then it follows from Theorem 4.1 and (\ref{scal} ) that
\beq{cde}
deg\,(E(V))=\dim\,(V)(\lan\varpi^\vee,\nu\ran+k)\,,~~~k\in\mZ\,.
\eq
Then for the fundamental representations of $\bG$ we have the following realization
of this formula.
\vspace{3mm}
\begin{center}
\begin{tabular}{|c|l|l|l| }
  \hline
   $\bG$ &$\nu$, &$V$ & deg$\,(E(V))$ \\
 \hline
SL$(n,\mC)$ &  $\varpi^\vee_1$ & $\underline{n}$ &$-1+kn$ \\
Spin$_{2n+1}(\mC)$&  $\varpi^\vee_n$ & $\underline{2^n}$ & $2^{n-1}(1+2k)$ \\
Sp$_n(\mC)$& $\varpi^\vee_1$  & $\underline{2n}$ &  $n(1+2k)$ \\
Spin$^{L,R}_{4n}(\mC)$& $\varpi^\vee_{n,n-1}$ & $\underline{2^{2n-1}}$ &$2^{2n-2}(1+2k)$ \\
Spin$_{4n+2}(\mC) $& $\varpi^\vee_n$  & $\underline{2^n}$ & $2^{n-2}(1+4k)$   \\
$E_6(\mC)$ &  $\varpi^\vee_1  $ & $\underline{27}$ & $9(1+3k)$  \\
$E_7(\mC)$ &  $\varpi^\vee_1$ & $\underline{56}$ & $28(1+2k)$  \\
  \hline
\end{tabular}
$$
(k\in\mZ)
$$
\vspace{5mm}
\textbf{Table 3.} Degrees of
 bundles for conformal groups.
\end{center}
It follows from our considerations that replacing the transition matrix
$$
\La\to\ti\La=\bfe\,(\lan\varpi^\vee,\nu\ran( z+\frac{\tau}2))\La
$$
defines the bundle of conformal group $CG$ of degree (\ref{cde}).


\section{GS-basis in simple Lie algebras}
\setcounter{equation}{0}

We pass from the Chevalley basis (\ref{CBA}) to a new basis that is more convenient
to define bundles corresponding to nontrivial characteristic classes.
We call it \emph{the generalized sin basis} (GS-basis), because for $A_{n}$ case and
degree one bundles it coincides with the sin-algebra basis (see, for example, \cite{FFZ}).

Let us take an element $\zeta\in\clZ(\bar G)$ of order $l$ and the corresponding $\La^0\in W$ from
(\ref{gce}).  Then $\La^0$ generates a cyclic group  $\mu_l=(\La^0,(\La^0)^2,\ldots,(\La^0)^l=1)$ isomorphic
to a subgroup of $\clZ(\bar G)$. Note that $l$ is  a divisor of ord$(\clZ(\bar G))$.
Consider the action of  $\La^0$ on $\gg$. Since $(\La^0)^l=Id$ we have a $l$-periodic
gradation
\beq{gra}
\gg=\oplus_{a=0}^{l-1}\gg_a\,, ~~\la(\gg_a)=\om^a\gg_a\,,~~\om=\exp\,\frac{2\pi i}l\,,~~~\la=Ad_{\La^0}\,,
\eq
\beq{gra1}
[\gg_a,\gg_b]=\gg_{a+b}\,,~(mod \,l)\,,
\eq
where $\gg_0$ is a subalgebra $\gg_0\subset\gg$ and the subspaces
$\gg_a$ are its representations.

Since  $\La^0\in W$
it preserves the root system $R$.
 Define the quotient set $\clT_l=R/\mu_l$.
Then $R$ is represented  as a union of $\mu_l$-orbits
$R=\cup_{\clT_l}\clO$.
We denote by  $\clO(\babe)$  an orbit starting from the root $\be$
$$
\clO(\babe)=\{\be\,,\la(\be)\,,\ldots,\la^{l-1}(\be)\}\,, ~~\babe\in \clT_l\,.
$$
The number of elements in an orbit $\clO$ (the length of $\clO$) is $l/p_\al=l_\al$, where $p_\al$ is a divisors of $l$.
Let $\nu_\al$ be a number of orbits $\clO_{\baal}$ of the length $l_\al$.
Then $\sharp\, R=\sum\nu_\al l_\al$.
Note, that if $\clO(\babe)$ has length $l_\be$ $\,(l_\be\neq 1)$, then the elements
$\la^k\be$ and $\la^{k+l_\be}\be$ coincide.


\subsection*{Basis in $\gL$ (\ref{CD})}

Transform first the root basis $\clE=\{E_\be\,,~\be\in R\}$ in $\gL$.
Define  an  orbit in $\clE$
$$
\clE_{\babe}=\{E_\be\,,E_{\la(\be)}\,,\ldots\,,E_{\la^{l-1}(\be)}\}
$$
 corresponding to $\clO(\babe)$. Again
$\clE=\cup_{\babe\in\clT_l}\clE_{\babe}$.

For  $\clO(\babe)$ define the set of integers
\beq{dc}
J_{p_\al}=\{a=mp_\al\,|\,m\in\mZ\,, ~~a~ {\rm is ~ defined~}mod\,l\,\}\,, ~~~(p_\al=l/l_{\al})\,.
\eq
"The Fourier transform" of the root basis on the orbit $\clO(\babe)$ is defined as
\beq{ft}
\gt^a_{\babe}=\f1{\sqrt{l}}\,\sum_{m=0}^{l-1}\om^{ma}E_{\la^m(\be)}\,,~~
\om=\exp\,\frac{2\pi i}{l}\,,~~a\in J_\be\,.
\eq
This transformation is invertible
$E_{\la^k(\be)}=\f1{\sqrt{l}}\sum_{a\in J_l}\om^{-ka}\gt^a_{\babe}$,
and therefore there is the one-to-one  map
$\clE_{\be} \leftrightarrow \{\gt^a_{\babe}\,,~a\in J_\be\}$.
In this way we have defined the new basis
\beq{fbl}
\{\gt^a_{\babe}\,,~(a\in J_l\,,~\babe\in\clT_l)\}\,.
\eq

Since $\la (E_\al)=E_{\la(\al)}$  we have for $\La\bfe(\ti\bfu)$ $\,(\ti\bfu\in\ti\gH_0)$
\beq{nlt}
Ad_{\La}(\gt^a_{\babe})=\bfe(\lan\ti\bfu,\be\ran-\frac{a}l)\gt^a_{\babe}\,,~~~~\bfe(x)=\exp\,(2\pi ix)\,.
\eq
It means that
 $\,\gt^a_{\babe}$ $\,(\babe\in\clT_l)$ is a part of basis in $\gg_{l-a}$ (\ref{gra}).
 Moreover,
  \beq{nqt}
Ad_{\clQ}(\gt^a_{\babe})=\bfe(\lan\ka,\be\ran)\gt^a_{\babe}\,.
\eq
 This relations follows from  (\ref{gce2}) and (\ref{geq}).
 We also  take into account that $\clQ$ and $\La$
commute in the adjoint representation and
$\bfe\,(x)E_\al\bfe\,(-x)=\bfe\,\lan x,\al\ran E_\al$ for $x\in\ti\gH_0$.

Picking another element $\La'$ generating a subgroup $\clZ_{l'_1}$ $(l'\neq l)$
we come to another set of orbits and to another basis. We have as many types of bases
as many of non-isomorphic subgroups in $\clZ(\bar G)$.


\subsubsection*{The Killing form}

 Consider two orbits  $\clO(\baal)$ and $\clO(\babe)$, passing through $E_\al$
 and $E_\be$. Assume that there exists such integer $r$
that $\al=-\la^r(\be)$.
It implies that elements of two orbits are related as
$\la^n(\al)=-\la^m(\be)$ if $m-n=r$.  In other words, $-\be\in\clO(\baal)$.
 In particular, it means that orbits have the same length.
It follows from (\ref{ft}) and (\ref{are6}) that
\beq{kfl}
(\gt^{c_1}_{\baal},\gt^{c_2}_{\babe})= \de_{\al,-\la^r(\be)}
\de^{(c_1+c_2,0\,\,(mod\,l))}\om^{-rc_1}\frac{2p_\al}{(\al,\al)}\,,
\eq
where $p_\al=l/l_\al$, and $l_\al$ is the length of $\clO(\baal)$.  In particular,
$(\gt_{\baal}^a,\gt_{-\baal}^{-a})=\frac{2p_\al}{(\al,\al)}$.

In what follows we need a dual basis $\gT_{\baal}^b$
\beq{dbt}
(\gT_{\baal_1}^{b_1},\gt^{b_2}_{\baal_2})=
\de^{(b_1+b_2,0\,\,(mod\,l))}\de_{\baal_1,-\baal_2}\,,
~~~\gT_{\baal}^b=\gt^{-b}_{-\baal}\frac{(\al,\al)}{2p_\al}\,.
\eq
The Killing form in this basis is inverse to (\ref{kfl})
$$
(\gT_{\baal_1}^{a_1},\gT_{\baal_2}^{a_2})=
\de_{\al_1,-\la^r(\al_2)}
\de^{(a_1+a_2,0\,\,(mod\,l))}\om^{ra_1}\frac{(\al_1,\al_1)}{2p_{\al_1}}\,.
$$
In particular,
\beq{kf2}
(\gT_{\al}^{a},\gT_{-\al}^{-a})=
\frac{(\al,\al)}{2p_\al}\,.
\eq



\subsection*{A basis in the Cartan subalgebra}

Almost the same construction exists in $\gH$.
Again let $\La^0$ generates the group $\mu_l$. Since  $\La^0$ preserves
the  extended Dynkin diagram, its action preserves  the extended coroot system
 $\Pi^{\vee ext}=\Pi^\vee\cup \al^\vee_0$ in $\gH$.
 Consider the quotient $\clK_l=\Pi^{\vee ext}/\mu_l$.
Define an orbit $\clH(\baal)$ of length $l_\al=l/p_\al$   in $\Pi^{\vee ext}$
passing through $H_\al\in\Pi^{\vee ext}$
$$
\clH(\baal)=\{H_\al\,,H_{\la(\al)}\,,\ldots,H_{\la^{l-1}(\al)}\}\,,
~~~\baal\in\clK_l=\Pi^{\vee ext}/\mu_l\,.
$$
The set $\Pi^{\vee ext}$ is a union of $\clH(\baal)$
$$
(\Pi^{\vee})^{ext}=\cup_{\baal\in\clK_l}\clH(\baal)\,.
$$

Define "the Fourier transform"
\beq{fth}
\gh^c_{\baal}=\f1{\sqrt{l}}
\sum_{m=0}^{l-1}\om^{mc}H_{\la^m(\al)}\,,~~\om=\exp\,\frac{2\pi i}{l}\,,
~~c\in J_\al ~(\ref{dc})\,.
\eq

The basis $\gh^c_{\baal}\,$, $(c\in J_\al,~\baal\in\clK_l)$
 is over-complete in $\gH$.
Namely, let $\clH(\baal_0)$ be an orbit passing through
the minimal coroot $\{H_{\al_0},H_{\la(\al_0)},\ldots,H_{\la^{l-1}(\al_0)}\}$.
Then the element $\gh^0_{\bar{\al}_0}$ is a linear combination of elements
$\gh^0_{-\baal}\,$, $(\al\in\Pi)$ and we should exclude it from the basis.
We replace the basis $\Pi^\vee$ in $\gH$ by
\beq{cb}
\gh^c_{\baal}\,,~(c\in J_\al)\,,~~\,,~~
\left\{
\begin{array}{ll}
  \baal\in\ti\clK_l=\clK_l\setminus\clH(\baal_0)\,, & c=0 \\
  \baal\in \clK_l\,,  & c\neq 0\,.
\end{array}
\right.
\eq
As before there is a one-to-one map $\Pi^\vee\leftrightarrow\{\gh^c_{\baal}\}$.

The elements $(\gh^a_{\baal},\gt^a_{\baal})$ form GS basis in $\gg_{l-a}$ (\ref{gra}).


\subsubsection*{The Killing form}

The Killing form in the basis (\ref{cb}) can be found from (\ref{kilh})
\beq{clA}
(\gh^{a}_{\baal},\gh^{b}_{\babe})=\de^{(a+b,0\,(mod\,l))}
\clA^a_{\al,\be}\,,~~~
\clA^a_{\al,\be}=\frac{2}{(\be,\be)}\sum _{s=0}^{l-1}\om^{-sa}
a_{\be,\la^s(\al)}\,,
\eq
where $a_{\al,\be}$ is the Cartan matrix (\ref{CMa}).

The dual basis is generated by elements $\gH^a_{\baal}$
\beq{dbh}
(\gH^a_{\baal}, \gh^b_{\babe}) =\de^{(a+b,0\,(mod\,l))}\de_{\al,\be}\,,~~~  \gH^a_{\baal}=\sum_{\be\in\Pi}(\clA^a_{\al,\be})^{-1}\gh^{-a}_{\babe}\,,
~~\gh^{a}_{\babe}=\sum_{\al\in\Pi}(\clA^{-a}_{\al,\be})\gH^{-a}_{\baal}
\eq
The Killing form in the dual basis takes the form
\beq{kfdbh}
(\gH^{a_1}_{\baal_1},\gH^{a_2}_{\baal_2})=\de^{(a_1+a_2,0\,(mod\,l))}
(\clA^{a_1}_{\baal_1,\baal_2})^{-1}\,.
\eq

\bigskip

In summary, we have defined the GS-basis in $\gg$
\beq{GSB}
\{\gt^{a}_{\babe},\gh^c_{\baal}\,,~~(a,\babe,c,\baal)~
{\rm are~ defined ~in ~(\ref{fbl}),~ (\ref{cb})}\}\,,
\eq
and the dual basis
\beq{DGSB}
\{\gT^{a}_{\babe},\gH^c_{\baal}\,,~~(a,\babe,c,\baal)~
{\rm are~ defined ~in ~(\ref{dbt}),~ (\ref{dbh})}\}\,,
\eq
along with the Killing forms.


\subsection*{Commutation relations \label{commrell}}

The commutation relations in the GS basis can be found from the
commutation relations in the Chevalley basis (\ref{cbcr}). Taking
into account the invariance of the structure constants with
respect to the Weyl group action $C_{\la\al,\la\be}=C_{\al,\be}$
it is not difficult to derive the  commutation relations in the GS
basis using its definition in the Chevalley basis (\ref{ft}),
(\ref{fth}). In the case of root-root commutators we come to the
following relations
 \beqn{com1}
 [\gt^{a}_{\alpha},\gt^{b}_{\beta}]\,=\left\{
\begin{array}{ll}
\frac{1}{\sqrt{l}}\,\sum\limits_{s=0}^{l-1}\, \omega^{bs} \,
C_{\alpha,\, \lambda^s\beta}\,\gt^{a+b}_{\alpha+\lambda^s\beta},&
\alpha\neq \,-\lambda^{s} \beta\\ &
\\
\frac{p_{\alpha}}{\sqrt{l}}\,\omega^{s\,b}\,\gh^{a+b}_{\alpha}&\alpha=
\,-\lambda^{s} \beta
\end{array}\right.
  \eqn
 The Cartan-root commutators are:
\beqn{com2}
\begin{array}{l}
\left[\gh^{\,k}_{\,\alpha}, \gt^{\,m}_{\,\beta}\right] =
\frac{1}{\sqrt{l}}\,\sum\limits_{s=0}^{l-1}\,\omega^{-ks}\,\frac{2(\alpha,
\lambda^{s}\beta) }{(\alpha,\alpha)}\,\gt^{k+m}_{\,\beta}\\
 \left[\gH^{\,k}_{\,\alpha},\gt^{\,m}_{\,\beta}
\right] = \frac{1}{\sqrt{l}}\,\sum\limits_{s=0}^{l-1}\,\omega^{-
ks }\,\frac{(\alpha,\alpha)}{2}\,({\hat \alpha}, \lambda^{s}\beta)
\,\gt^{k+m}_{\,\beta}
\end{array}
 \eqn
Here we denote by ${\hat \alpha}$ the dual to the simple roots
elements in the Cartan subalgebra: \beqn{dr} ({\hat \alpha_{i}},
\beta_{j})=\delta_{ij} \eqn In Section \ref{clrll}, for explicit
computations with Lax operators and $r$-matrices, it will be much
more convenient to use the following normalized basis for Cartan
subalgebra:
 \beqn{g2}
 \bar{\gh}^{\,k}_{\,\alpha}=\frac{(\alpha,\alpha)}{2}\,\gh^{\,k}_{\,\alpha},\
 \ \ \ \ \bar{\gH}^{\,k}_{\,\alpha}=\frac{2}{(\alpha,\alpha)}\,\gH^{\,k}_{\,\alpha}
 \eqn
This reparametrization leads to the following commutation
relations: \beqn{crel2}
\begin{array}{l}
 \left[\bar{\gh}^{\,k}_{\,\alpha},
\gt^{\,m}_{\,\beta}\right] =
\frac{1}{\sqrt{l}}\,\sum\limits_{s=0}^{l-1}\,\omega^{-ks}\,(\alpha,
\lambda^{s}\beta)\,\gt^{k+m}_{\,\beta}\\
 \left[\bar{\gH}^{\,k}_{\,\alpha},\gt^{\,m}_{\,\beta}
\right] = \frac{1}{\sqrt{l}}\,\sum\limits_{s=0}^{l-1}\,\omega^{-
ks }\,({\hat \alpha}, \lambda^{s}\beta) \,\gt^{k+m}_{\,\beta}
\end{array}
 \eqn
The following simple formula expresses the decomposition of Cartan
element in the basis of simple roots: \beqn{decsr}
\bar{\gh}_{\beta}^{k}=\sum\limits_{\alpha\in
\Pi}\,(\hat{\alpha},\beta)\,\bar{\gh}_{\alpha}^{k},\ \ \ \beta \in
R \eqn the connection of dual bases is clear from the following
expression: \beqn{connbas}  \sum\limits_{\beta\in \Pi}\,
(\hat{\alpha}, \beta) \, \bar{\gh}_{\beta}^{k}
=\sum\limits_{\beta\in \Pi}\, (\alpha, \beta) \,
\bar{\gH}_{\beta}^{k}  \eqn The Cartan elements have the following
"sign"-property: \beqn{sprop}
\bar{\gh}^{k}_{-\alpha}=-\bar{h}^{k}_{\alpha},\ \
\bar{\gH}^{k}_{-\alpha}=-\bar{\gH}^{k}_{\alpha},\ \
\bar{S}^{\gh,k}_{-\alpha}=-\bar{S}^{\gh,k}_{\alpha},\ \
\bar{S}^{\gH,k}_{-\alpha}=-\bar{S}^{\gH,k}_{\alpha} \eqn



 From the definition of GS-basis we simply find:
\beqn{tran1} \gt^{\,k}_{\,\lambda^{s} \alpha}=
\omega^{-ks}\,\gt^{\,k}_{\,\alpha},\ \ \
\gh^{\,k}_{\,\lambda^{s}\alpha}=
\omega^{-ks}\,\gh^{\,k}_{\,\alpha},\ \ \
\gH^{\,k}_{\,\lambda^{s}\alpha}=
\omega^{-ks}\,\gH^{\,k}_{\,\alpha} \eqn The same identities we
suppose for classical variables (see Section \ref{clrll} ) :
\beqn{tran2} S^{\gL \,k}_{\,\lambda^{s} \alpha}=
\omega^{-ks}\,S^{\gL\,k}_{\,\alpha},\ \ \ S^{\gh
\,k}_{\,\lambda^{s}\alpha}= \omega^{-ks}\,S^{\gh \,k}_{\,\alpha},\
\ \ S^{\gH\,k}_{\,\lambda^{s}\alpha}=
\omega^{-ks}\,S^{\gH\,k}_{\,\alpha} \eqn


\subsection*{Invariant subalgebra}

Consider the invariant subalgebra  $\gg_0$.
It is generated by the basis $(\gt^0_{\babe}\,,\,\gh^0_{\baal})$ (\ref{GSB}).
In particular,  $\{\gh^0_{\baal}\}$ (\ref{fth}),  (\ref{cb}) form a basis in the Cartan subalgebra
$\ti{\gH}_0\subset\gH$ ($\dim\,\ti{\gH}_0=p<n$).

We  pass from $\{\gh^0_{\baal}\}$ to  a special basis  in  $\ti{\gH}_0$
\beq{tir}
\ti\Pi^\vee=\{\ti{\al_k}^\vee\,|\,k=1,\ldots,p\}\, .
\eq
It is constructed in the following way.
Consider a subsystem of simple coroots
 \beq{dpi1}
 \Pi_1^\vee= \Pi^{ext\vee}\setminus\clO(\baal_0^\vee)
 \eq
 (see (\ref{cb})).
In other words, $\Pi_1^\vee$ is a subset of simple coroots that does not
contain simple coroots from the orbit passing through $\al_0$.
 For $A_{N-1}$, $B_n$, $E_6$ and $E_7$ the coroot basis $\ti\Pi^\vee$ (\ref{tir})  is a result of
 an averaging along the $\la$ orbits in  $\Pi_1^\vee$
\beq{invc}
\ti\al^\vee=\sum_{m=1}^{l-1}H_{\la^m(\al)}\,,~~~H_{\al}\in\Pi_1^\vee\,.
\eq
In the $C_n$ and $D_n$ cases this construction is valid for almost all coroots except the last on the Dynkin diagram (see Remark 10.1 below).
Consider the dual vectors $\ti\Pi=\{\ti{\al_k}\,|\,k=1,\ldots,p\,,~ \lan\ti{\al_k},\ti{\al_k}^\vee\ran=2\}$
in $\ti{\gH}_0^*$.
\begin{predl}
The set of vectors in $\ti\gH_0^*$
\beq{tiPi}
\ti\Pi=\{\ti{\al_k}\,|\,k=1,\ldots,p\}\,,
\eq
is a system of simple roots of a simple Lie subalgebra  $\ti\gg_0\subset\gg_0$
defined by the root system $\ti R=\ti R(\ti\Pi)$ and the Cartan matrix
 $\lan\ti{\al_k},\ti{\al_j}^\vee\ran$.
\end{predl}

We will check this statement \cite{LOSZII} case by case.

 Let  $ R_1= R_1(\Pi_1)$ be a subset of roots generated by simple roots $\Pi_1=\Pi^{ext}\setminus\clO(\al_0)$. It is invariant under $\la$ action.
 The root system $\ti R$ of  $\ti\gg_0$ corresponds to the $\la$ invariant set of $R_1$.
 Consider the complementary set of roots $R\setminus R_1$
 and  the set of orbits
 \beq{clt'}
 \clT'_l=(R\setminus R_1)/\mu_l\,.
  \eq
  It is a subset of all orbits  $\clT_l=R/\mu_l$.
 Therefore, $\clT_l=\ti R\cup\clT'_l$.
 The $\la$-invariant subalgebra $\gg_0$ contains the subspace
\beq{gd2}
V=\{\sum_{\babe\in \clT'_l}a_{\babe}\gt^0_{\babe}\,,~~a_{\babe}\in\mC\}\,.
\eq
 Then $\gg_0$ is a sum of $\ti{\gg}_0$ and $V$
\beq{gd}
\gg_0=\ti{\gg}_0\oplus V\,.
\eq
The components of this decomposition are orthogonal with respect to the Killing form (\ref{clA}), and $V$ is a representation of $\ti{\gg}_0$
We find below the explicit forms of $\gg_0$ for all simple algebras from our list.

Let $\gH'$ be a subalgebra of $\gH$ with the basis $\gh_{\baal}^c\,$ $c\neq 0$ (\ref{fth}) and $\ti\gH$ is a Cartan subalgebra of $\ti\gg_0$.
Then
\beq{deco}
\gH=\ti\gH_0\oplus\gH'\,.
\eq

We summarize the information about invariant subalgebras in Table 3.

\bigskip
\begin{center}
\begin{tabular}{|c|c|c|c|c|c|c|}
  \hline
  &             &         &                  &          &   & \\
  $\Pi$  & $\clZ(\bG)$ &  $\varpi_j^\vee$  &$\Pi_1$ & $l=$ord$\,(\La)$ & $\ti\gg_0$ &  $\gg_0$ \\
    &             &         &                  &          &  &  \\
  \hline
  1& 2& 3& 4& 5& 6& 7 \\
   \hline
  $A_{N-1}\,,~(N=pl)$ &$\mu_N$&$\varpi_{N-1}^\vee$& $\cup_{1}^l A_{p-1}$ & $N/p$ &  ${\bf sl_p}$ &  ${\bf sl_p}\oplus_{j=1}^{l-1}{\bf gl_p}$\\
  $B_n$ & $\mu_2$&$\varpi_{n}^\vee$ & ${\bf so_{2n-1}}$  & 2& ${\bf so(2n-1)}$ & $ {\bf so(2n)}$ \\
  $C_{2l}\,,~(l>1)$ &$\mu_2$ &$\varpi_{2l}^\vee$ &  $A_{2l-1}$ & 2 &  ${\bf so(2l)}$ & ${\bf gl_{2l}}$ \\
  $C_{2l+1}$ &$\mu_2$  & $\varpi_{2l+1}^\vee$ &$A_{2l}$ & 2 & ${\bf so(2l+1)}$ &  ${\bf gl_{2l+1}}$ \\
  $D_{2l+1}\,,~(l>1)$ &$\mu_4$  & $\varpi_{2l+1}^\vee$ &$A_{2l-2}$ & 4 & ${\bf so(2l-1)}$& ${\bf so(2l)}\oplus{\bf so(2l)}\oplus\underline{1}$  \\
  $D_{2l+1}\,,~(l>1)$ &$\mu_4$  &$\varpi_{1}^\vee$ & $D_{2l}$ & 2 & ${\bf so(4l-1)}$& ${\bf so(4l)}\oplus\underline{1}$  \\
  $D_{2l}\,,~(l>2)$ &$\mu_2\oplus\mu_2$ &$\varpi_{2l}^\vee$ & $A_{2l-1}$ & 2 & ${\bf so(2l)}$ &  ${\bf so(2l)}\oplus{\bf so(2l)}$ \\
  $D_{2l}\,,~(l>2)$ &$\mu_2\oplus\mu_2$ & $\varpi_{1}^\vee$ &$D_{2l-1}$ & 2 & ${\bf so(4l-3)}$&  ${\bf so(4l-2)}\oplus\underline{1}$ \\
  $E_6$ &$\mu_3$&$\varpi_{1}^\vee$ & $D_4$ & 3 & ${\bf g_2}$ & ${\bf so(8)}\oplus 2\cdot\underline{1}$\\
  $E_7$ &$\mu_2$&$\varpi_{7}^\vee$ & ${\bf e_6}$ & 2 & ${\bf f_4}$ &  ${\bf e_6}\oplus\underline{1}$\\
  \hline
\end{tabular}
\bigskip

\textbf{Table 4}\\
Invariant subalgebras $\ti\gg_0=\gg_{\ti\Pi}$ and $\gg_0$ of simple Lie algebras.\\
The coweights generating central elements are displaced in column 3.
\end{center}


\bigskip

In  the invariant simple algebra $\ti{\gg}_0$ instead of
 the basis $(\gh^0_{\baal},\gt^0_{\babe})$  we use the Chevalley basis
and incorporate it in the GS-basis
\beq{hal}
\{\gh^0_{\baal}\,, \gt^0_{\babe}\}\,\to\,\{\ti\gg_0=(H_{\ti\al}\,,\ti\al\in\ti\Pi\,,~E_{\ti\be}\,,\ti\be\in\ti R)\,,~
V=(\gt^0_{\babe}\,,\babe\in\clT')\}\,.
\eq

\begin{rem}
For any $\xi\in Q^\vee$ a solution of (\ref{01}) is $\La=Id$.
In this case $\ti\gg_0=\gg$ and GS-basis is the Chevalley basis.
\end{rem}


\subsection*{The GS  basis from a canonical basis in $\gH$}

Let $(e_1,e_2,\ldots,e_n)$ be a canonical basis in $\gH$,
$((e_j,e_k)=\de_{jk})$.
\footnote{For $A_n$ and $E_6$ root systems it is convenient to choose canonical
bases in $\gH\oplus\mC$.}
Since $\La$ preserves $\gH$ we can
consider the action of $\mu_l$ on the canonical basis.
Define an orbit  of length $l_s=l/p_s$
passing through $e_s$
$\clO(s)=\{e_s,\la(e_{s}),\ldots,\la^{(l-1)}e_{s)}\}$.

The Fourier transform along $\clO(s)$  takes the form
\beq{ahk2}
\gh_{s}^c=\f1{\sqrt{l}}\sum_{m=0}^{l-1}\om^{mc}\la^m(e_{s})\,,~~
c\in J_{p_s}\,, ~~\om=\exp\,(\frac{2\pi i}{l})\,,
\eq
where $J_{p_s}=\{c=mp_s\, mod(l)\,|\,m\in\mZ\}$.
Consider the quotient $\clC_l=(e_1,e_2,\ldots,e_n)/\mu_l$.
Then we can pass from the canonical basis to the GS basis
$$
(e_1,e_2,\ldots,e_n)\longleftrightarrow \{\gh_{s}^c\,,~s\in \clC_l\}\,.
$$


The Killing form is read of from (\ref{ahk2})
\beq{kfcb1}
(\gh_{s_1}^{c_1},\gh_{s_2}^{c_2})=\de_{(s_1,s_2)}\de^{(c_1,-c_2)}\,.
\eq
Then the dual generators are
\beq{kfcb}
\gH_{s}^{c}=\gh_{s}^{-c}\,.
\eq

The commutation relations in $\gg$ in this form of  GS basis take the form
\beq{crh1}
[\gh^{k_1}_{s},\gt^{k_2}_{\babe}]=
\f1{\sqrt{l}}\sum_{r=0}^{l-1}\om^{-rk_1}\lan\la^r(\be),e_s\ran
\gt^{k_1+k_2}_{\babe}\,,
 \eq
$$
[\gt^{k_1}_{\baal},\gt^{k_2}_{\babe}]=\f1{p_\al\sqrt{l}}\om^{rk_2}\sum_s
(\al^\vee,e_s)\gh^{k_1+k_2}_{s}\,,~~ {\rm if~}\al= -\la^r(\be){\rm ~for~ some~} r\,.
$$
We obtain the last relation from (\ref{ft}) and from the expansion
$\gh^{k}_\al=\sum_s(\al^\vee,e_s)\gh^k_s$.


\section{General description of systems with non-trivial characteristic classes}
\setcounter{equation}{0}


\subsection*{The Lax operators and Symplectic Hecke Correspondence}

 Consider a meromorphic section $\Phi$ of the adjoint bundle $End\,E_G\otimes K$, where $K$ is a canonical class.
 The pair $(\Phi,E_G)$ is called the Higgs bundle over $\Si$. The set of these
pairs  is a cotangent bundle to the space of holomorphic bundles equipped with a canonical
symplectic structure. Evidently, the gauge transformations (\ref{gtr}) can be lifted to the Higgs bundle as canonical transformations with respect to the symplectic form.
The hamiltonian reduction of the Higgs bundle under this action leads to integrable systems
\cite{Hi}, and the Higgs field become the Lax operator $L$.
 The moduli space of a
Higgs bundle becomes a phase space of an integrable systems.
This construction is valid for curves with marked points. In this case we deal with the Higgs
bundle with quasi-parabolic structures at the marked points. It implies that the Lax operators
have first order poles at the marked poles with residues belong to generic coadjoint orbits $\clO$.
The coadjoint orbits are affine spaces over the flag varieties mentioned at Section 2. The dimension of  the moduli space of the Higgs bundle is twice of $\dim\,\clM_{g,n}$
(\ref{mgn})
$$
\dim\,(T^*\clM_{g,n})=2(g-1)\dim\,(G)+n\dim\, (\clO)\,.
$$
The commuting integrals are generated by the quantities $(L^{d_j})$, where $d_j$ are degrees of invariant polynomials.
The integrals  are coefficients of expansion $(L^{d_j})$ in the basis of functions on $\Si$ with prescribed singularities at the marked points.

For $g=1$, $n=1$ the phase space has dimension of a coadjoint orbit $2\sum_{j=1}^{rank G}(d_j-1)$.
In this case $L$ satisfies the  conditions
\beq{fm}
L(z+1)=\clQ L(z)\clQ^{-1}\,,~~~
L(z+\tau)=\La L(z)\La^{-1}\,,
\eq
where $\clQ$ and $\La$ are solutions of (\ref{gce}), and
\beq{repl}
\bp L(z)=\bfS\de(z,\bz)\,.
\eq
In other words $Res|_{z=0}\,L(z)=\bfS$. These conditions fix $L$.

To make dependence on the characteristic class $\zeta(E_G)$ explicit  we will write
 $L(z)^{\varpi^\vee_j}$,  if the Lax matrix satisfies the quasi-periodicity conditions  with $\La=\La_{\varpi^\vee_j}$, $\clQ_{\varpi^\vee_j}$
 where $\La_{\varpi^\vee_j}$ $\clQ_{\varpi^\vee_j}$
 are solutions of (\ref{gce}) with
$\zeta=\bfe(-\varpi^\vee_j)$, $\,\varpi^\vee_j\in P^\vee$.

The modification $\Xi(\ga)$  of $E_G$ changes the characteristic class (\ref{amcc}).
It acts on $L^{\varpi^\vee_j}$ as follows
\beq{mhb}
L^{\varpi^\vee_j}\Xi(\ga) =\Xi(\ga)L^{\varpi^\vee_j+\ga} \,.
\eq
In this form it was introduced in \cite{LOZ1}. It is called \emph{the Symplectic
Hecke Correspondence}, because it acts as a symplectomorphism on phase spaces
of Hitchin integrable systems. In implies that the Higgs systems corresponding
to the bundles with different characteristic classes are in fact symplectomorphic. In particular,
they are symplectomorphic to the standard CM systems.

The action (\ref{mhb}) allows one to write down condition on $\Xi(\ga)$.
Since $L^{\varpi^\vee_j}$ has a simple pole at $z=0$
the modified Lax matrix  $L^{\varpi^\vee_j+\ga}$ should have also a
 simple pole at $z=0$.
Decompose $L^{\varpi^\vee_j}$ and $L^{\varpi^\vee_j+\ga}$ in the Chevalley basis (\ref{CD}), (\ref{CBA})
$$
L^{\varpi^\vee_j}=L_{\gH}(z)+\sum_{\al\in R}L_\al(z)E_\al\,,~~~
L^{\varpi^\vee_j+\ga}=\ti L_{\gH}(z)+\sum_{\al\in R}\ti L_\al(z)E_\al\,.
$$
Expand $\al$ in the basis of simple roots (\ref{ale1})
 $\al=\sum_{j=1}^lf_j^\al\al_j$ and $\ga$
in the basis of fundamental coweights $\ga=\sum_{j=1}^lm_j\varpi_j^\vee$.
Assume  that
$\lan\ga,\al_j\ran\geq 0$ for simple $\al_j$.
In other words $\ga$ is a dominant coweight.
  Then
$\lan\ga,\al\ran=\sum_{j=1}^lm_jn_j^\al$ is an integer number, positive for $\al\in R^+$ and negative for $\al\in R^-$.
From (\ref{mod1}) and (\ref{mhb})  we find
\beq{shc}
 L^{\varpi^\vee_j+\ga}_{\gH}(z)=L^{\varpi^\vee_j}_{\gH}(z)\,,~~~
 L^{\varpi^\vee_j+\ga}_\al(z)=z^{\lan\ga,\al\ran}L^{\varpi^\vee_j}_\al(z)\,.
\eq
 In a neighborhood of $z=0$ $L_\al(z)$ should have the form
\beq{cbl}
L^{\varpi^\vee_j}_\al(z)=a_{\lan\ga,\al\ran}z^{-\lan\ga,\al\ran}+
a_{\lan\ga,\al\ran+1}z^{-\lan\ga,\al\ran+1}+
\ldots\,,~~~(\al\in R^-)\,,
\eq
otherwise the transformed Lax operator becomes singular.
 It means that the type of the modification $\ga$
is not arbitrary, but depends on the local behavior of the Lax operator.
It allows one to find the dimension of the space of the Hecke transformation.
We do not need it here.

Now consider on a global behavior of $L(z)$ (\ref{fm}). Then we find that
 $\Xi(\ga)$ should intertwine the the quasi-periodicity conditions
$$
\Xi(\ga,z+1)\clQ_{\varpi^\vee_j}=\clQ_{\varpi^\vee_j+\ga}\Xi(\ga,z)\,,~~
\Xi(\ga,z+\tau)\La_{\varpi^\vee_j}=\La_{\varpi^\vee_j+\ga}\Xi(\ga,z)\,.
$$
Solutions of these equation do not belong to the fixed Cartan subgroup $\clH\subset G$ as in (\ref{mod1}). So it is not easy to find a consistent behavior of
 $\Xi(\ga)$ and $L^{\varpi^\vee_j}$ at $z=0$ as in (\ref{cbl}).
 For $G=\SLN$, $\ga=\varpi^\vee_1$ and special residue of $L$ it was done
 in \cite{LOZ1}.


\subsection*{The Lax matrix. Explicit form}

Assume that $L$ has a residue at $z=0$ taking values in a coadjoint orbit $\clO\subset \gg^*$.
\beq{res}
Res\,L|_{z=0}=\bfS=\sum_{\al\in\Pi}\oh(\al,\al)S^\gH_{\al}
\sum_{\be\in\Pi}a^{-1}_{\al,\be}H_{\be}+\sum_{\be\in R}S^\gL_{\be} \frac{(\be,\be)}{2} E_{-\be}
\eq
$$
=\sum_{j=1}^nS_je_j+\sum_{\be\in R}S^\gL_{\be}\frac{(\be,\be)}{2} E_{-\be}\,.
$$
We identify $\gg^*$ and $\gg$ by means the Killing form (\ref{kilh}),
 (\ref{are6}).
Then  the coordinates are linear functionals on $\gg$
 $ S^\gH_{\al}=(\bfS,H_\al)\,,$ or $S_j=(\bfS,e_j))\,,$ $\, S^\gL_{\be}=(\bfS,E_\be)$.

The Poisson brackets for
$S^\gH_{\al},$ $\,(S_j)$, $\,S^\gL_{\be}$ have the same structure constants as $\gg$
(\ref{cbcr}).
To define a generic orbit $\clO$ we fix the Casimir functions $C_j(\bfS)$.

Rewrite (\ref{res}) in the dual GS-basis.
We use the gradation (\ref{gra}) to define the Lax matrix
\beq{lmg}
L(z)=\sum_{a=0}^{l-1}L_a(z)\,,
\eq
where the zero component is decomposed according with  (\ref{gd})
$L_0(z)=\ti L_0(z)+L'_0(z)$.
Then
$$
\bfS=Res\,L|_{z=0}=Res\,\ti L_0|_{z=0}+Res\,L_0'|_{z=0}+\sum_{a=1}^{l-1}Res\,L_a|_{z=0}=
\ti\bfS_0+\bfS'_0+\sum_{a=1}^{l-1}\bfS_a\,,
$$
where
$$
\ti\bfS_0=\sum_{\ti\al,\ti\be\in\ti\Pi}
\ti S^{\gH}_{\ti\al}\frac{(\al,\al)(\be,\be)}{4(\ti\al,\ti\be)}H_{\ti\be}+
\sum_{\ti\be\in\ti R}\ti S^{\gL}_{\ti\be}\frac{(\ti\be,\ti\be)}{2} E_{-\ti\be}\,,
$$
\beq{res1}
\bfS'_0=
\sum_{\babe\in \clT'_l}S^{'}_{\babe}\gT^0_{\babe}\,,~~~
\bfS_a=\sum_{\baal\in\clK_l}
S^{\gH,a}_{\baal}\gH^{l-a}_{\baal}
+\sum_{\babe\in\clT_l}S^{\gL,a}_{\babe}\gT^{l-a}_{-\babe}\,,
\eq
(see (\ref{clt'}), (\ref{gd2})).
Again, the coordinates
$$
S^{\gH,a}_{\baal}=(\bfS,\gh^a_{\baal})\,,~~S^{'}_{\babe}=(\bfS,\gt^0_{\babe})\,,~~
S^{\gL,a}_{\babe}=(\bfS,\gt^a_{\babe})\,~~\ti S^{\gH}_{\ti\al}=(\bfS,H_{\ti\al})\,,~~S^{\gL}_{\ti\be}=(\bfS, E_{\ti\be})
$$
have the structure constants of the Poisson brackets as in
(\ref{com1}), (\ref{com2}), (\ref{crh1}). We can pass from one
data to another by the Fourier transform introduced above. We
rewrite $\ti\bfS_0$ in terms of a canonical basis
$(e_1,\ldots,e_{\ti n})$ in the invariant Cartan algebra
$\ti\gh_0$ \beq{res0} \ti\bfS_0=\sum_{j=1}^{\ti n} \ti
S^{\gH}_{j}e_j+ \sum_{\ti\be\in\ti R}\ti
S^{\gL}_{\ti\be}\frac{(\ti\be,\ti\be)}{2} E_{-\ti\be}\,. \eq

It follows from  (\ref{nlt}), (\ref{nqt})  and from definition of the dual basis (\ref{dbt}), (\ref{dbh})  that
$$
Ad_{\La}(\gT^c_{\babe})=\bfe\,(\frac{c}{l}-\lan\ti\bfu,\be\ran)\gT^c_{\babe}\,,~~
Ad_{\La}(\gH^c_{\babe})=\bfe\,(\frac{c}{l})\gH^c_{\babe}\,,~~
(\bfe\,(x)=\exp\,(2\pi ix))\,.
$$
In addition,  we have
\beq{qc}
Ad_{\clQ}(\gH^c_{\babe})=\gH^c_{\babe}\,,~~
Ad_{\clQ} (H_{\ti\al})=H_{\ti\al}\,,
\eq
\beq{qr}
Ad_{\clQ}(\gT^c_{\babe})=\bfe\,(-\lan\ka,\be\ran)\gT^c_{\babe}\,,~~
Ad_{\clQ}( E_{\ti\al})=\bfe\,\lan\ka,\ti\al\ran E_{\ti\al}\,.
\eq
Using  (\ref{arho1}) we obtain   $\lan\ka,\al\ran=f_\al/h$. Then the last relation
assumes the form
\beq{qr1}
Ad_{\clQ}(\gT^c_{\babe})=\bfe\,(-f_\be/h)\gT^c_{\babe}\,,~~
Ad_{\clQ}( E_{\ti\al})=\bfe\,(f_\al/h) E_{\ti\al}\,.
\eq

There are also the evident relations:
$$
Ad_{\La}( E_{\ti\al})=\bfe\,(\lan\ti\bfu,\ti\al\ran)E_{\ti\al}\,,
~~Ad_{\La}( H_{\ti\al})=H_{\ti\al}\,,~~\ti\bfu\in\ti\gH\,.
$$




The quasi-periodicity conditions and the existence of pole at $z=0$ dictate the form
of the components for  $a\neq 0$. We  define matrix element of Lax operator
using $\phi(u,z)$ (\ref{phiz}).
Let
\beq{kfi}
\varphi^a_\be(\bfx,z)=\bfe\,(\lan\ka,\be\ran z)\phi(\lan\bfx+\ka\tau,\be\ran+\frac{a}l,z)=\bfe\,( zf_\be/h)\phi(\lan\bfx,\be\ran+\tau f_\be/h+\frac{a}l,z)\,.
\eq
The last equality follows from the identity $\lan\ka,\al\ran=\f1{h}\lan\rho^\vee,\al\ran=f_\al/h$ (see (\ref{arho1})).
It follows from (\ref{qpp1}) that
$\varphi^a_\be(\bfx,z+1)=\bfe\,(\lan\ka,\be\ran)\varphi^a_\be(\bfx,z)$,
$~\varphi^a_\be(\bfx,z+\tau)=\bfe\,(-\lan x,\be\ran-\frac{a}l)\varphi^a_\be(\bfx,z)$.

Then from (\ref{fm}) we find
\beq{LM}
L_a(z)=\sum_{\baal\in\clK_l}S^{\gH,a}_{\baal}\phi(\frac{a}{l},z)\gH^{l-a}_{\baal}
+
\sum_{\babe\in \clT_l}S^{\gL,a}_{\babe}\varphi^{a}_{\be}(-\ti\bfu,z)\gT^{l-a}_{-\babe}\,,
\eq
and
$ L'_0(z)=\sum_{\baal\in \clT'_l}S^{'}_{\baal}\varphi^0_{\al}(-\ti\bfu,z)\gT^0_{-\baal}$.

In the canonical basis in $\gH$ (\ref{ahk2}) $L_a(z)$ takes the form
$$
L_a(z)=\sum_{s\in\clC_l}S^{\gH,a}_{s}\phi(\frac{a}{l},z)\gh^{-a}_{s}
+
\sum_{\babe\in \clT_l}S^{\gL,a}_{\babe}\varphi^{a}_{\be}(-\ti\bfu,z)\gT^{l-a}_{-\babe}\,.
$$
It follows from (\ref{kfi}), (\ref{resp}) and (\ref{qpp}), and that $L_a(z)$ has the required quasi-periodicities and the residues.

We replace the basis on the dual basis using (\ref{dbt}) and (\ref{kfcb}) and finally  obtain
\beq{lcbk}
L_a(z)=\sum_{s\in\clC_l}S^{\gH,a}_{s}\phi(\frac{a}{l},z)\gh^{-a}_{s}
+
\sum_{\babe\in \clT_l}S^{\gL,a}_{\babe}\varphi^{a}_{\be}(-\ti\bfu,z)\gt^{a}_{\babe}\frac{(\be,\be)}{2p_\be}\,,
\eq
\beq{LM0p}
 L'_0(z)=
\sum_{\baal\in \clT'_l}S^{'}_{\baal}
\varphi^0_\al(-\ti\bfu,z)\gt^0_{\baal}\frac{(\al,\al)}{2p_\al}\,.
\eq

Consider the invariant subalgebra  $\ti{\gg}_0$. For $\ti{\gg}_0$ we
write down the Lax matrix in the Chevalley basis.
Let $p\leq n$ be a rank of $\ti{\gg}_0$, $(e_1,\ldots,e_{p})$ is a canonical
basis in $\ti\gH_0$ and $E_{\ti\al}$ are generators of the root subspaces.
The matrix elements of  $\ti L_0$ are constructed by means of
$\varphi^0_\be$ (\ref{kfi}) and the Eisenstein functions (\ref{e1f}):
\beq{tLM0}
\ti L_0(z)=\sum_{j=1}^{p}(v_{j}+ \ti S^{\gH}_{j}E_1(z))e_j+
\sum_{\ti\be\in\ti R}\ti  S^{\gL}_{\ti\be}\varphi^0_{\ti\al} (-\ti\bfu,z)E_{\ti\be}\,.
\eq
Here $\ti\bfv=(v_{1},\dots,v_{p})$ are momenta vector, dual to
$\ti\bfu=(u_{1},\dots,u_{p})$.
The Lax operator (\ref{tLM0}) differs from the standard Lax operator related to $\ti\gg_0$\\
$\ti L_0(z)=\sum_{j=1}^{p}(v_{j}+ \ti S^{\gH}_{j}E_1(z))e_j+
\sum_{\ti\al\in\ti R} \ti S^{\gL}_{\ti\al}\phi (\lan-\ti\bfu,{\ti\al}\ran,z)E_{\ti\al}$.
It is gauge equivalent to the previous one after $\ti\bfu\to\ti\bfu+\ka$.
 For this reason we call $\ti\gg_0$ \emph{the unbroken subalgebra}.
The operator $\ti L_0(z)$ has the needed residue (see (\ref{resp}) and (\ref{rze})).
However, the Cartan term containing $E_1(z)$ breaks the quasi-periodicities
(see (\ref{qpz})), because
there are no double-periodic functions with one pole on $\Si_\tau$.
To go around this problem we use the Poisson reduction procedure.


\subsection*{The Lax matrix. Poisson reduction}

 The Lax element we have defined  depends on
the spin variables representing an orbit ${\clO}$,
 on the moduli vector $\ti\bfu$ in the moduli space described in Section 5
and the dual covector $\ti\bfv$. It is a Poisson manifold $\bfP$ with
the canonical brackets for $\ti\bfv$,  $\ti\bfu$ and the Poisson-Lie brackets
for $\bfS$.
\beq{pmu}
\bfP=T^*C\times\clO=\{\ti\bfv\,,\ti\bfu\,,\bfS\}\,,~~\ti\bfu\in C\,,~~
\bfS\in\clO\,.
\eq
It has dimension $\dim\,(\clO)+2\dim\,(\ti\gH_0)$

Consider  the Poisson algebra $\clA=C^\infty(\bfP)$ of smooth function
on $\bfP$.
Let $\ep\in\ti\gH_0$ and $\ga$ is a small contour around $z=0$. Consider the following function
$\mu_\ep=\oint_\ga(\ep,L(\ti\bfv,\ti\bfu,\bfS))=(\ep,\bfS_0^\gH)$,
$~\bfS_0^\gH=\sum_{j=1}^{p} S^{\gH}_{j}e_j$.
It generates the vector field on $\bfP~$
$V_\ep\,:\,L(\ti\bfv,\ti\bfu,\bfS)\to\{\mu_\ep,L(\ti\bfv,\ti\bfu,\bfS)\}=
[\ep,L(\ti\bfv,\ti\bfu,\bfS)]$.

Let $\clA^{inv}$ be an invariant subalgebra of $\clA$ under $V_\ep$ action.
Then
$I=\{\mu_\ep F(\ti\bfv,\ti\bfu,\bfS)\,|\,F(\ti\bfv,\ti\bfu,\bfS)\in\clA \}$
is the Poisson ideal in $\clA^{inv}$. The reduced Poisson algebra is
 the factor-algebra
$$
\clA^{red}=\clA^{inv}/I=\clA//\ti\clH_0\,,~~(\ti\clH_0=\exp\,\ti\gH_0)\,.
$$
The reduced Poisson manifold $\bfP^{red}$ is defined by
 the moment constraint $ S^{\gH}_{0,\ti\al}=0$
(or $S^{\gH}_{0,s}=0$) and $\dim\,\ti\gH$ gauge fixing constraints on the
spin variables that we do not specify
$$
\bfP^{red}=\bfP//\ti\clH_0=\bfP(S^{\gH}_{0,s}=0)/\ti\clH_0\,,~~~
\dim\,(\bfP^{red})=\dim\,(\bfP)-2\dim\,(\ti\clH_0)=\dim\,(\clO)\,.
$$
 Thus, after the reduction we come to the
Poisson manifold that has dimension of the coadjoint orbit $\clO$,
but the Poisson structure on $\bfP^{red}$ is not the Lie-Poisson structure.
The Poisson brackets on  $\bfP^{red}$ are the Dirac brackets \cite{Di}.

Due to the moment constraints we come to the Lax operator
 that has the correct periodicity. It depends on variables
$\{\ti\bfv\,,\ti\bfu\,,\bfS\}\in\bfP^{red}$
\beq{tLM1}
\ti L_0(z)=\sum_{j=1}^{\ti n}v_{j}e_j+
\sum_{\ti\be\in\ti R} S^{\gL}_{\ti\be}\varphi^0_{\ti\al} (\ti\bfu,z)E_{\ti\be}\,.
\eq
 Here $S^{\gL}_{\ti\be}$ are not free due to the gauge fixing.


\subsection*{Hamiltonians}

Let us define commuting integrals. For this purpose consider
 the ring of invariant polynomials on $\gg$. It is generated $n$ homogeneous polynomials $P_1,P_2,\ldots,P_n$ of degrees $d_1=2,\ldots,d_n=h$.
It follows from the general approach \cite{Hi} that  $P_j(L(z))$ generate commuting integrals. They are double
 periodic meromorphic functions on $\Si_\tau$ and thereby can be expanded in the basis of elliptic functions
$$
\f1{m_j}P_j(L(z))=I_{j,0}+I_{j,2}E_2(z)+\ldots+I_{j,j}E_j(z)\,.
$$
The coefficients $I_{j,k}$
$\,(0\leq k\leq m_j\,,~k\neq 1)$ become commuting independent
integrals. The highest order coefficients $I_{jj}$ are Casimir functions fixing the orbits. The coefficient $I_{j,1}$ vanishes, because there is no double periodic
functions with one simple pole.
 The number of rest  coefficients is equal to $\oh\dim\,(\clO)$.

Consider the second order in the spin variables Hamiltonian $H=I_{1,0}$ coming from the expansion
$\f1{2}P_1(L(z))=H+I_{2,2}E_2(z)$.
We represent it in the form
\beq{hami}
H=\ti H_0 +H'+\sum_{a=1}^M H_a\,,~~M=\Bigl[\frac{l}2\Bigr]\,.
\eq
Due to the orthogonality of $L_a$ and $L_b$ $(a\neq -b$ mod$\,\,l$) with respect to the Killing form the Hamiltonians
$\ti H$, $H'$ and $H_k$ are determined by pairing of the corresponding
Lax operators
$$
\ti H_0 = \oh(\ti L_0(z),\ti L_0(z))|_{const}\,,~~
H' = \oh( L'_0(z), L'_0(z))|_{const}\,,~~
H_a = \oh( L_a(z), L_{l-a}(z))|_{const}\,
$$
To calculate the Hamiltonians we use (\ref{kf2}), (\ref{kfcb1})
(\ref{wpphi}),
 Then we come to the following expressions
 \footnote{In what follows we shall use the standard CM Hamiltonians, where the coordinate
vector is shifted $\ti\bfu\to\ti\bfu+\ka\tau$.}
\beq{h0}
\ti H_0=\oh\sum_{s=1}^{\ti n}v^2_{s} -\sum_{\ti\be\in\ti R}
\f1{(\ti\be,\ti\be)}\ti S^{\gL}_{\ti\be}\ti S^{\gL}_{-\ti\be}
E_2(\lan\ti\bfu-\ka\tau,\ti\be\ran)\,.
\eq
As it was noted above $\ti H_0$ is the elliptic CM Hamiltonian related to $\ti\gg_0$.

Using (\ref{kfl}) we find
\beq{hprim}
H'=
-\sum_{\baal,\babe\in \clT'_l,}\de_{ \be=-\la^r(\al)}
\frac{(\be,\be)}{p_\be}S^{'}_{\baal}S^{'}_{\babe}
E_2(\lan\ti\bfu-\ka\tau,\be\ran)\,.
\eq

Similarly, from (\ref{kfdbh}), (\ref{kf2}) and (\ref{clA})
$$
H_a=
-\oh\sum_{s\in\clC_l}E_2\Bigl(\frac{a}{l}\Bigr)
S^{\gH,a}_{s}S^{\gH,l-a}_{s}
$$
\beq{hk}
-\sum_{\baal,\babe\in\gT_l}\de_{\al,-\la^r(\be)}\om^{-ar}\frac{(\al,\al)}{p_\al}
S^{\gL,a}_{\babe}S^{\gL,l-a}_{\baal}E_2(\lan\ti\bfu-\ka\tau,\be\ran)\,.
\eq
The Hamiltonians $H'$, $H_a$ are Hamiltonians of EA tops with the inertia
tensors depending on $\ti\bfu$.

-

 On the reduced space $\bfP^{red}$ the equations of motion corresponding to integrals
 $I_{jk}$ acquire the Lax form
 $ \p_{t_{jk}}L=[L,M_{jk}]$.
 The operator $M_{jk}$ is reconstructed from $L$ and  defined below $r$-matrix \cite{BV}.


\section{The proof of the classical $RLL$-relation \label{clrll}}
\begin{predl}

$\bullet$ The r-matrix  and the Lax operator described above
define the Poisson brackets on the reduced phase space
$\textbf{P}^{\textrm{red}}$ via $RLL$-equation: \beqn{RLLe}
 \left\{ L(z)\otimes 1,\,1\otimes L(w)
\right\} \,=\left[L(z)\otimes 1+1\otimes L(w),
r(z,w)\right]-\frac{\sqrt{l}}{2}\,\sum\limits_{k=0}^{l-1}\sum\limits_{\alpha\in\,R}\,
|\alpha|^2 \,\partial_{1}\,\varphi^{k}_{\alpha}(z-w)
\,\bar{S}^{\gh\,0}_{\,\alpha}\,t^{\,k}_{\,\alpha}\otimes
t^{-k}_{-\alpha}\ \ \ \eqn These brackets have the following
explicit form:
\end{predl}
\beqn{PB}
\begin{array}{|l|}
\hline
\\
 \left\{ S^{\gL, a}_{\alpha},S^{\gL,b}_{\beta}
\right\}\,=\left\{
\begin{array}{ll}
\frac{1}{\sqrt{l}}\,\sum\limits_{s=0}^{l-1}\, \omega^{bs} \,
C_{\alpha,\, \lambda^s\beta}\,S^{\gL,
a+b}_{\alpha+\lambda^s\beta},&
\alpha\neq \,-\lambda^{s} \beta\\ & \\
\frac{p_{\alpha}}{\sqrt{l}}\,\omega^{s\,b}\,S^{\gh,a+b}_{\alpha}&\alpha=
\,-\lambda^{s} \beta
\end{array}\right.\\
\\
\left\{\bar{S}^{\gh,k}_{\,\alpha}, S^{\gL ,m}_{\,\beta}\right\} =
\frac{1}{\sqrt{l}}\,\sum\limits_{s=0}^{l-1}\,\omega^{-ks}\,(\alpha,
\lambda^{s}\beta)\,S^{\gL, k+m}_{\,\beta}\\
\\
 \left\{\bar{S}^{\gH\,k}_{\,\alpha},S^{\gL, m}_{\,\beta}
\right\} = \frac{1}{\sqrt{l}}\,\sum\limits_{s=0}^{l-1}\,\omega^{-
ks }\,({\hat \alpha}, \lambda^{s}\beta) \,S^{\gL, k+m}_{\,\beta}\\
\\
\left\{\bar{v}^{\gH}_{\alpha}, u_{\beta}\right\}=\frac{1}{\sqrt{l}}\,\sum\limits_{s=0}^{l-1}\,( \hat{\alpha},\lambda^{s}\beta )\\
\\
\left\{v_{\alpha}, S^{\gL,a}_{\alpha}\right\}\,=\left\{v_{\alpha},
S^{\gH,a}_{\alpha}\right\}\,=\left\{v_{\alpha},
S^{\gh,a}_{\alpha}\right\}\,=0\\
\\
\left\{u_{\alpha}, S^{\gL,a}_{\alpha}\right\}\,=\left\{u_{\alpha},
S^{\gH,a}_{\alpha}\right\}\,=\left\{u_{\alpha},
S^{\gh,a}_{\alpha}\right\}\,=0\\
\\
 \hline

\end{array}
  \eqn

It will be convenient to subdivide the Lax operator and the
classical $r$-matrix  into Cartan and root parts:
 \beqn{LR}
 \begin{array}{l}
  L(z)=L_{R}(z)\,+L_{\gH}(z)+L_{\gH}^{0}(z) \\
  \\
r(z,w)=r_{R}(z,w)+r_{\gH}(z,w)
\end{array}
 \eqn
 where
\beqn{r}
\begin{array}{l}
 L_{R}(z)=\frac{1}{2}\,\sum\limits_{a=\,0}^{l-1} \,
\sum\limits_{\beta\,\in\, R}\,| \beta |^2\,
\varphi^{\,a}_{\,\beta} (z) \,
S^{\gL,-a}_{-\beta} \, \gt^{\,a}_{\,\beta}\\
\\
 L_{\gH}(z)=\sum\limits_{a=\,1}^{l-1}
\sum\limits_{\alpha\,\in\,\Pi} \, \varphi^{\,a}_{\,0}(z) \,
S^{\gH,-a}_{\alpha}\, \gh^{\,a}_{\,\alpha},\ \ L_{\gH}^{0}(z)=
\sum\limits_{\alpha\,\in\,\Pi} \,
\Big(v_{\,\alpha}^{\gH}+E_{1}(z)\,S^{\gH,0}_{\alpha}\Big) \, \,
\gh^{\,0}_{\,\alpha}  \end{array} \eqn and:
 \beqn{Rma}  r_{R}(z,w)=\frac{1}{2}\,\sum\limits_{a=\,0}^{l-1}\, \sum\limits_{\alpha\,\in\,R}\,
|\alpha|^2\,\varphi^{\,a}_{\,\alpha}(z-w) \,\gt^{\,a}_{\,\alpha}
\otimes \gt^{-a}_{-\alpha},\ \ r_{\gH}(z,w)=
\sum\limits_{a=\,0}^{l-1}\,\sum_{\alpha\,\in\,\Pi}\,\varphi^{\,a}_{\,0}(z-w)\,
\gH^{\,a}_{\,\alpha}\otimes\gh^{-a}_{\,\alpha}
 \eqn\\

Note that in these formulae the summation spreads over all roots
but not over orbits. Obviously these definitions of Lax operators
and $r$-matrices are equivalent if one takes into account
identities (\ref{tran1}), (\ref{tran2}) and (\ref{erb}) The proof
of the proposition of this section is purely technical calculation
and to simplify intermediate expressions we will use normalized
Cartan generators defined in Section \ref{commrell}. In terms of
these generators the Cartan part of Lax operator and $r$-matrix
take the form:
 \beqn{cp2}
 \begin{array}{l}
r_{\gH}(z,w)=\sum\limits_{a=\,0}^{l-1}\,\sum\limits_{\alpha\,\in\,\Pi}\,\varphi^{\,a}_{\,0}(z-w)\,
\gH^{-a}_{\,\alpha}\otimes\gh^{\,a}_{\,\alpha}=\sum\limits_{a=\,0}^{l-1}\,\sum\limits_{\alpha\,\in\,\Pi}\,
\varphi^{\,a}_{\,0}(z-w)\,
\bar{\gH}^{\,-a}_{\,\alpha}\otimes\bar{\gh}^{\,a}_{\,\alpha}\\
L_{\gH}(z)=\sum\limits_{a=\,1}^{l-1}
\sum\limits_{\alpha\,\in\,\Pi} \, \varphi^{\,a}_{\,0}(z) \,
\bar{S}^{\gH,-a}_{\alpha}\, \bar{\gh}^{\,a}_{\,\alpha}, \ \ \
L_{\gH}^{0}(z)= \sum\limits_{\alpha\,\in\,\Pi} \,
\Big(v_{\,\alpha}^{\gH}+E_{1}(z)\,\bar{ S}^{\gH,0}_{\alpha}\Big)
\, \, \bar{\gh}^{\,0}_{\,\alpha}
\end{array}
 \eqn\\
 \textit{proof}\\

 Non-vanishing terms in the left hand side of (\ref{RLLe}) are:
 \beqn{proof1}\nonumber
  \left\{ L(z)\otimes 1,\,1\otimes L(w)
\right\}=  \underbrace{\left\{ L_{R}(z)\otimes 1,\,1\otimes
L_{R}(w) \right\}}_{\gt^{k}_{\alpha}\otimes\gt^{m}_{\beta}}+
\underbrace{\left\{ L_{R}(z)\otimes 1,\,1\otimes L_{\gH}(w)
\right\}}_{\gt^{k}_{\alpha}\otimes  \bar{\gh}^{m}_{\beta} }+\\
\underbrace{\left\{ L_{R}(z)\otimes 1,\,1\otimes L_{\gH}^{0}(w)
\right\}}_{\gt^{k}_{\alpha}\otimes\bar{\gh}^{0}_{\beta}}+\nonumber
\underbrace{\left\{ L_{\gH}(z)\otimes 1,\,1\otimes L_{R}(w)
\right\}}_{\bar{\gh}^{k}_{\alpha}\otimes\gt^{m}_{\beta}}+
\underbrace{\left\{ L_{\bar{\gH}}^{0}(z)\otimes 1,\,1\otimes
L_{R}(w) \right\}}_{\bar{\gh}^{0}_{\alpha}\otimes\gt^{m}_{\beta}}
 \eqn
where we denoted by $\gt^{k}_{\alpha}\otimes\gt^{m}_{\beta}$,
$\gt^{k}_{\alpha}\otimes \bar{\gh}^{m}_{\beta}$ etc. the
corresponding tensor structure of the brackets. Our strategy is to
check the equation (\ref{RLLe}) for all possible tensor structures
using the commutation relations (\ref{com1}) and Poisson brackets
(\ref{PB}). From the previous expression we see that we should
consider only three structures:
$\gt^{k}_{\alpha}\otimes\gt^{m}_{\beta}$,
$\bar{\gh}^{k}_{\alpha}\otimes\gt^{m}_{\beta}$ and
$\bar{\gh}^{0}_{\alpha}\otimes\gt^{m}_{\beta}$. The computations
for the rest terms $\gt^{k}_{\alpha}\otimes \bar{\gh}^{m}_{\beta}$
and $\gt^{k}_{\alpha}\otimes\bar{\gh}^{0}_{\beta}$ are symmetric
to $\bar{\gh}^{k}_{\alpha}\otimes\gt^{m}_{\beta}$ and
$\bar{\gh}^{0}_{\alpha}\otimes\gt^{m}_{\beta}$ cases respectively.

The commutator $[\gt^{k}_{\alpha}, \gt^{m}_{\beta}]$ depends on
whether $\alpha\in \cal{O}(-\beta)$ or not, therefore, it will be
convenient to consider the case of tensor structure
$\gt^{k}_{\alpha}\otimes\gt^{-k}_{-\alpha}$ separately. The table
below describes the terms in the right hand side of the equation
(\ref{RLLe}) getting contributions to the corresponding tensor
structures: \beqn{proof2} \begin{array}{|c|c|}\hline & \\
\gt^{k}_{\alpha}\otimes\gt^{m}_{\beta} & \left[L_{R}(z) \otimes1 ,
r_{R}(z,w)\right],\ \ [1\otimes L_{R}(w)  , r_{R}(z,w)]\\  \hline & \\
\bar{\gh}^{k}_{\alpha}\otimes\gt^{m}_{\beta}& \left[L_{R}(z)
\otimes1 ,
r_{R}(z,w)\right],\ \ [1\otimes L_{R}(w)  , r_{R}(z,w)]\\  \hline & \\
\bar{\gh}^{0}_{\alpha}\otimes\gt^{m}_{\beta} &\left[L_{R}(z)
\otimes1 ,
r_{R}(z,w)\right],\ \ [1\otimes L_{R}(w)  , r_{\gH}^{0}(z,w)]\\  \hline & \\
\gt^{k}_{\alpha}\otimes\gt^{-k}_{-\alpha}&\left[L_{\gH}^{0}(z)
\otimes1 , r_{R}(z,w)\right],\ \ [1\otimes L_{\gH}^{0}(w), r_{R}(z,w)] \\
 & \\
\hline
\end{array}\eqn
Therefore we prove  (\ref{RLLe}) by checking four equations for
every tensor structure:
\begin{small}
 \beqn{prone}
  \Big\{ L_{R}(z)\otimes1,
1\otimes L_{R}(w)
\Big\}\Big|_{\gt^{k}_{\alpha}\otimes\gt^{m}_{\beta}}-\Big[
L_{R}(z)\otimes 1 , r_{R}(z,w)
\Big]\Big|_{\gt^{k}_{\alpha}\otimes\gt^{m}_{\beta}}-\Big[\,1\otimes
L_{R}(w)\,, \,r(w,z)
\,\Big]\Big|_{\gt^{k}_{\alpha}\otimes\gt^{m}_{\beta}}=0\ \ \ \eqn
\end{small}

 for
for generic $m$, $k$, $\alpha$ and $\beta$, \beqn{prtwo}
\Big\{L_{\gH}(z)\otimes1, 1\otimes L_{R}(w)
\Big\}\Big|_{\bar{\gh}^{k}_{\alpha}\otimes\gt^{m}_{\beta}}- \Big[
L_{R}(z)\otimes1,\, r_{R}(z,w)
\Big]\Big|_{\bar{\gh}^{k}_{\alpha}\otimes\gt^{m}_{\beta}} -
\Big[1\otimes L_{R}(w),
r_{\gH}(z,w)\Big]\Big|_{\bar{\gh}^{k}_{\alpha}\otimes\gt^{m}_{\beta}}=0\
\ \ \eqn for $k\neq0$,
 \beqn{prthree} \Big\{ L_{\gH}^{0}(z)\otimes1,\,
1\otimes L_{R}(w)
\Big\}\Big|_{\bar{\gh}^0_{\alpha}\otimes\gt^{m}_{\beta}}- \Big[
L_{R}(z)\otimes1,\, r_{R}(z,w)
\Big]\Big|_{\bar{\gh}^0_{\alpha}\otimes\gt^{m}_{\beta}} -
\Big[1\otimes L_{R}(w),
r_{\gH}(z,w)\Big]\Big|_{\bar{\gh}^0_{\alpha}\otimes\gt^{m}_{\beta}}=0\
\ \ \eqn and
 \beqn{prfour}\nonumber  \Big\{L_{R}(z)\otimes1,\,1\otimes
L_{R}(w)\Big\}\Big|_{\gt^{k}_{\alpha}\otimes\gt^{-k}_{-\alpha}}-\Big[
L_{\gH}^{0}(z)\otimes1, r_{R}(z,w)\Big]-\Big[ 1\otimes
L_{\gH}^{0}(w), r_{R}(z,w)\Big]=\\
=-\ \frac{\sqrt{l}}{2}\,\sum\limits_{k=0}^{l-1}\,
\sum\limits_{\alpha\,\in\,R} \,
|\alpha|^2\,\partial_{1}\varphi^{k}_{\alpha}(z-w)\,\bar{S}^{\gh,0}_{\alpha}\,
\gt^{\,k}_{\,\alpha} \otimes \gt^{-k}_{-\alpha}  \eqn In the next
subsections we check these identities using only commutation
relations (\ref{PB}) and (\ref{com1}).
\subsection*{ $\gt^{\,k}_{\,\alpha}\otimes \gt^{\,m}_{\,\beta}$ -terms for generic $\alpha$ and $\beta$ }

\beqn{pr1}\nonumber \Big\{ L_{R}(z)\otimes1, 1\otimes L_{R}(w)
\Big\}=\,\left\{ \frac{1}{2}\,\sum\limits_{k=\,0}^{l-1} \,
\sum\limits_{\alpha\,\in\, R}\,
|\alpha|^2\,\varphi^{\,k}_{\,\alpha} (z) \, S^{\gL,-k}_{-\alpha}
\, \gt^{\,k}_{\,\alpha}\otimes1\, ,\,
\frac{1}{2}\,\sum\limits_{m=\,0}^{l-1} \,
\sum\limits_{\beta\,\in\, R}\,|\beta|^2 \, \varphi^{\,m}_{\,\beta}
(w) \,
S^{\gL,-m}_{-\beta} \, 1\otimes\gt^{\,m}_{\,\beta} \right\}=\\
\nonumber =\frac{1}{4}\,
\sum\limits_{k,m=\,0}^{l-1}\sum\limits_{\alpha,\beta\,\in\,
R}\,|\alpha|^2|\beta|^2\, \varphi^{\,k}_{\,\alpha} (z) \,
\varphi^{\,m}_{\,\beta} (w) \, \left\{ S^{\gL,-k}_{-\alpha} ,
S^{\gL,-m}_{-\beta}\right\} \,
\gt^{\,k}_{\,\alpha}\otimes\gt^{\,m}_{\,\beta}\,
\stackrel{(\ref{PB})}{=}\\ \nonumber
\stackrel{(\ref{PB})}{=}\frac{1}{4\sqrt{l}}\,\sum\limits_{k,m,s=\,0}^{l-1}\sum\limits_{\alpha,\beta\,\in\,
R}\,|\alpha|^2|\beta|^2\, \omega^{-ms}  \varphi^{\,k}_{\,\alpha}
(z) \, \varphi^{\,m}_{\,\beta} (w)\,C_{-\alpha,-\lambda^s \beta }
\, S^{\gL,-k-m}_{-\alpha-\lambda^{s} \beta} \,
\gt^{\,k}_{\,\alpha}\otimes\gt^{\,m}_{\,\beta}\,\ \ \ \ \ \ \
 \eqn

 \beqn{pr2}
 \begin{array}{|c|}
 \hline
 \\
 \Big\{ L_{R}(z)\otimes1, 1\otimes L_{R}(w) \Big\}\Big|_{\gt^{k}_{\alpha}\otimes \gt^{m}_{\beta}}=\,\frac{1}{4 \sqrt{l}}\,\sum\limits_{k,m,s=\,0}^{l-1}\sum\limits_{\alpha,\beta\,\in\,
R}\, |\alpha|^2|\beta|^2 \,\omega^{-ms}  \varphi^{\,k}_{\,\alpha}
(z) \, \varphi^{\,m}_{\,\beta} (w)\,C_{-\alpha,-\lambda^s \beta }
\, S^{\gL,-k-m}_{-\alpha-\lambda^{s} \beta} \,
\gt^{\,k}_{\,\alpha}\otimes\gt^{\,m}_{\,\beta}\\
\\
\hline
 \end{array}\ \ \ \
 \eqn

 \beqn{pr3} \nonumber \Big[ L_{R}(z)\otimes 1 , r_{R}(z,w) \Big]\,=\,\left[\frac{1}{2}\,\sum\limits_{b\,=\,0}^{l-1} \,
\sum\limits_{\gamma\,\in\, R}\, |\gamma|^2
\varphi^{\,b}_{\,\gamma} (z) \, S^{\gL,-b}_{-\gamma} \,
\gt^{\,b}_{\,\gamma}\otimes1,\,
\frac{1}{2}\sum\limits_{m\,=\,0}^{l-1}\,
\sum\limits_{\beta\,\in\,R}\,
|\beta|^2\,\varphi^{-m}_{-\beta}(z-w) \,\gt^{-m}_{-\beta} \otimes
\gt^{m}_{\beta}\right] =\\ \nonumber=
\frac{1}{4}\,\sum\limits_{b,m\,=\,0}^{l-1} \,
\sum\limits_{\beta,\gamma\,\in\, R}\,|\gamma|^2|\beta|^2\,
\varphi^{\,b}_{\,\gamma} (z) \,\varphi^{-m}_{-\beta}(z-w) \,
S^{\gL,-b}_{-\gamma} \, [\,\gt^{\,b}_{\,\gamma},
\gt^{-m}_{-\beta}] \otimes
\gt^{m}_{\beta}\,\stackrel {(\ref{com1})}{=}\,\\
\nonumber\,\stackrel {(\ref{com1})}{=}\,\frac{1}{4\sqrt{l}}\,
\sum\limits_{b,m,s\,=\,0}^{l-1} \,
\sum\limits_{\beta,\gamma\,\in\, R}\,|\gamma|^2|\beta|^2
\varphi^{\,b}_{\,\gamma} (z) \,\varphi^{-m}_{-\beta}(z-w) \,
S^{\gL,-b}_{-\gamma} \,\omega^{-ms}\,C_{\gamma,-\lambda^s\beta}\,
 \gt^{\,b-m}_{\,\gamma-\lambda^{s}\beta} \otimes
 \gt^{m}_{\beta}=\\ \nonumber=\frac{1}{4\sqrt{l}}\sum\limits_{k,m,s\,=\,0}^{l-1} \,
\sum\limits_{\alpha,\beta\,\in\, R}
\,|\alpha+\lambda^s\beta|^2|\beta|^2\,\omega^{-ms}\,
\varphi^{\,k+m}_{\,\alpha+\lambda^s\beta} (z)
\,\varphi^{-m}_{-\beta}(z-w) \,
S^{\gL,-k-m}_{-\alpha-\lambda^s\beta}
\,C_{\alpha+\lambda^s\beta,-\lambda^s\beta}\,
 \gt^{\,k}_{\,\alpha} \otimes
 \gt^{m}_{\beta}\,\stackrel {(\ref{scp}),(\ref{erb})}{=}\\
 \nonumber
\stackrel
{(\ref{scp}),(\ref{erb})}{=}-\frac{1}{4\sqrt{l}}\sum\limits_{k,m,s\,=\,0}^{l-1}
\, \sum\limits_{\alpha,\beta\,\in\,
R}\,|\alpha|^2|\beta|^2\,\omega^{-ms}\,
\varphi^{\,k+m}_{\,\alpha+\beta} (z) \,\varphi^{-m}_{-\beta}(z-w)
\, S^{\gL,-k-m}_{-\alpha-\lambda^s\beta}
\,C_{-\alpha,-\lambda^s\beta}\,
 \gt^{\,k}_{\,\alpha} \otimes
 \gt^{m}_{\beta}\,
  \eqn
  Therefore we get:
\beqn{pr4}  \begin{array}{|l|} \hline\\
 \Big[ L_{R}(z)\otimes 1 ,
r_{R}(z,w)
\Big]\Big|_{\gt^{k}_{\alpha}\otimes\gt^{m}_{\beta}}\,=\\\,
-\frac{1}{4\sqrt{l}}\sum\limits_{k,m,s\,=\,0}^{l-1} \,
\sum\limits_{\alpha,\beta\,\in\,
R}\,|\alpha|^2|\beta|^2\,\omega^{-ms}\,
\varphi^{\,k+m}_{\,\alpha+\beta} (z) \,\varphi^{-m}_{-\beta}(z-w)
\, S^{\gL,-k-m}_{-\alpha-\lambda^s\beta}
\,C_{-\alpha,-\lambda^s\beta}\,
 \gt^{\,k}_{\,\alpha} \otimes
 \gt^{m}_{\beta}\\
 \\ \hline \end{array}\ \ \
  \eqn
The last term: \beqn{pr5}
 \nonumber \Big[\,1\otimes L_{R}(w)\,,
\,r(w,z) \,\Big]= \left[ \frac{1}{2}\,\sum\limits_{p\,=\,0}^{l-1}
\, \sum\limits_{\gamma\,\in\, R}\,|\gamma|^2\,
\varphi^{\,p}_{\,\gamma} (w) \, S^{\gL,-p}_{-\gamma} \,
1\otimes\gt^{\,p}_{\,\gamma}\,,\,
\frac{1}{2}\,\sum\limits_{k\,=\,0}^{l-1}\,
\sum\limits_{\alpha\,\in\,R}\,|\alpha|^2
\varphi^{\,k}_{\,\alpha}(z-w) \,\gt^{\,k}_{\,\alpha} \otimes
\gt^{-k}_{-\alpha} \right]=\\
\nonumber=\frac{1}{4}\,\sum\limits_{k,p\,=\,0}^{l-1} \,
\sum\limits_{\alpha,\gamma\,\in\, R}\,|\alpha|^2|\gamma|^2
\varphi^{\,p}_{\,\gamma} (w) \,\varphi^{\,k}_{\,\alpha}(z-w) \,
S^{\gL,-p}_{-\gamma} \, \gt^{\,k}_{\,\alpha} \otimes[
\gt^{\,p}_{\,\gamma}
,\gt^{-k}_{-\alpha}]\,\stackrel{(\ref{com1})}{=}\,\\
\nonumber
 \stackrel{(\ref{com1})}{=}\,\frac{1}{4\sqrt{l}}
\sum\limits_{k,p,s\,=\,0}^{l-1} \,
\sum\limits_{\alpha,\gamma\,\in\, R}\,|\alpha|^2|\gamma|^2
\varphi^{\,p}_{\,\gamma} (w) \,\varphi^{\,k}_{\,\alpha}(z-w) \,
S^{\gL,-p}_{-\gamma} \, \omega^{-ks}\, C_{\gamma,
-\lambda^s\alpha} \gt^{\,k}_{\,\alpha} \otimes
\gt^{p-k}_{\gamma-\lambda^s\alpha}=
\\  \nonumber
=\frac{1}{4 \sqrt{l}}\sum\limits_{k,m,s\,=\,0}^{l-1} \,
\sum\limits_{\alpha,\beta\,\in\,
R}\,|\alpha|^2|\beta+\lambda^s\alpha|^2\,
\varphi^{\,k+m}_{\,\beta+\lambda^s\alpha} (w)
\,\varphi^{\,k}_{\,\alpha}(z-w) \,
S^{\gL,-k-m}_{-\beta-\lambda^s\alpha} \, \omega^{-ks}\,
C_{\beta+\lambda^s\alpha, -\lambda^s\alpha} \gt^{\,k}_{\,\alpha}
\otimes \gt^{m}_{\beta}\,\stackrel{(\ref{tran2})}{=}\\  \nonumber
\stackrel{(\ref{tran2})}{=}\frac{1}{4\sqrt{l}}\,\sum\limits_{k,m,s\,=\,0}^{l-1}
\, \sum\limits_{\alpha,\beta\,\in\,
R}\,|\alpha|^2|\beta+\lambda^s\alpha|^2\,
\varphi^{\,k+m}_{\,\beta+\lambda^s\alpha} (w)
\,\varphi^{\,k}_{\,\alpha}(z-w) \,
S^{\gL,-k-m}_{-\alpha-\lambda^{-s} \beta} \, \omega^{ms}\,
C_{\beta+\lambda^s\alpha, -\lambda^s\alpha} \gt^{\,k}_{\,\alpha}
\otimes \gt^{m}_{\beta}\,=\\
\nonumber=\frac{1}{4\sqrt{l}}\sum\limits_{k,m,s\,=\,0}^{l-1} \,
\sum\limits_{\alpha,\beta\,\in\,
R}\,|\alpha|^2|\beta+\lambda^{-s}\alpha|^2\,
\varphi^{\,k+m}_{\,\beta+\lambda^{-s}\alpha} (w)
\,\varphi^{\,k}_{\,\alpha}(z-w) \,
S^{\gL,-k-m}_{-\alpha-\lambda^{s} \beta} \, \omega^{-ms}\,
C_{\beta+\lambda^{-s}\alpha, -\lambda^{-s}\alpha}
\gt^{\,k}_{\,\alpha} \otimes
\gt^{m}_{\beta}\,\stackrel{(\ref{scp}),(\ref{erb})}{=}\\
\nonumber
\stackrel{(\ref{scp}),(\ref{erb})}{=}\frac{1}{4\sqrt{l}}\,
\sum\limits_{k,m,s\,=\,0}^{l-1} \,
\sum\limits_{\alpha,\beta\,\in\, R}\,|\alpha|^2|\beta|^2
\varphi^{\,k+m}_{\,\beta+\alpha} (w)
\,\varphi^{\,k}_{\,\alpha}(z-w) \,
S^{\gL,-k-m}_{-\alpha-\lambda^{s} \beta} \, \omega^{-ms}\,
C_{-\alpha, -\lambda^{s}\beta}\, \gt^{\,k}_{\,\alpha} \otimes
\gt^{m}_{\beta}
 \eqn
therefore: \beqn{pr6} \begin{array}{|l|}\hline \\
 \Big[\,1\otimes
L_{R}(w)\,, \,r(w,z) \,\Big]\Big|_{\gt^{k}_{\alpha}\otimes
\gt^{m}_{\beta}}=\\
=\frac{1}{4\sqrt{l}}\,\sum\limits_{k,m,s\,=\,0}^{l-1} \,
\sum\limits_{\alpha,\beta\,\in\, R}\,|\alpha|^2|\beta|^2
\varphi^{\,k+m}_{\,\beta+\alpha} (w)
\,\varphi^{\,k}_{\,\alpha}(z-w) \,
S^{\gL,-k-m}_{-\alpha-\lambda^{s} \beta} \, \omega^{-ms}\,
C_{-\alpha, -\lambda^{s}\beta}\, \gt^{\,k}_{\,\alpha} \otimes
\gt^{m}_{\beta}\\
\\
\hline
 \end{array}\ \ \ \eqn
Summing up (\ref{pr2}), (\ref{pr4}) and (\ref{pr6}) and using Fay
identity (\ref{fay1})  we obtain:
$$
\Big\{ L_{R}(z)\otimes1, 1\otimes L_{R}(w)
\Big\}\Big|_{\gt^{k}_{\alpha}\otimes \gt^{m}_{\beta}}-\Big[
L_{R}(z)\otimes 1 , r_{R}(z,w) \Big]\Big|_{\gt^{k}_{\alpha}\otimes
\gt^{m}_{\beta}}-\Big[\,1\otimes L_{R}(w)\,, \,r(w,z)
\,\Big]\Big|_{\gt^{k}_{\alpha}\otimes \gt^{m}_{\beta}}=0
$$


\subsection*{ $\bar{\gh}^{\,k}_{\,\alpha}\otimes \gt^{\,m}_{\,\beta}$-terms for $k\neq 0$ }

Let us consider the following Poisson bracket:  \beqn{t1}
\nonumber \Big\{L_{\gH}(z)\otimes1, 1\otimes L_{R}(w) \Big\}\,=\,
\left\{\sum\limits_{k=\,0}^{l-1} \sum\limits_{\alpha\,\in\,\Pi} \,
\varphi^{\,k}_{\,0}(z) \, \bar{S}^{\gH,-k}_{\alpha}\,
\bar{\gh}^{\,k}_{\,\alpha}\otimes 1,
\frac{1}{2}\,\sum\limits_{m=\,0}^{l-1} \,
\sum\limits_{\beta\,\in\, R}\,| \beta |^2\,
\varphi^{\,m}_{\,\beta} (w) \, S^{\gL,-m}_{-\beta} \,
1\otimes\gt^{\,m}_{\,\beta}
 \right\}=\ \ \ \\
 \nonumber
 =\frac{1}{2}\,\sum\limits_{k,m=\,0}^{l-1}
\sum\limits_{\beta\,\in\,R}\,\sum\limits_{\alpha\in\,\Pi} \,|
\beta |^2\, \varphi^{\,k}_{\,0}(z) \varphi^{\,m}_{\,\beta} (w) \,
\left\{\bar{S}^{\gH,-k}_{\alpha},\, S^{\gL,-m}_{-\beta}\right\} \,
\bar{\gh}^{\,k}_{\,\alpha}\otimes
 \gt^{\,m}_{\,\beta}\stackrel{(\ref{PB})}{=}\ \ \ \\
 \nonumber
 \stackrel{(\ref{PB})}{=}\,-\frac{1}{2\,\sqrt{l}}\,\sum\limits_{k,m,s=\,0}^{l-1}
\sum\limits_{\beta\,\in\,R}\,\sum\limits_{\alpha\in\,\Pi} \,|
\beta |^2\, \varphi^{\,k}_{\,0}(z) \varphi^{\,m}_{\,\beta} (w)
\,\omega^{ks}\,(\hat{\alpha},\lambda^{s}\beta)\,
S^{\gL,-k-m}_{-\beta} \bar{\gh}^{\,k}_{\,\alpha}\otimes
 \gt^{\,m}_{\,\beta}\ \ \ \ \ \
\eqn and therefore: \beqn{ht1}
\begin{array}{|c|}
\hline\\
\Big\{L_{\gH}(z)\otimes1, 1\otimes L_{R}(w)
\Big\}\Big|_{\gh^{k}_{\alpha}\otimes
\gt^{m}_{\beta}}\,=\,-\frac{1}{2\,\sqrt{l}}\,\sum\limits_{k,m,s=\,0}^{l-1}
\sum\limits_{\beta\,\in\,R}\,\sum\limits_{\alpha\in\,\Pi}  \,|
\beta |^2\, \varphi^{\,k}_{\,0}(z) \varphi^{\,m}_{\,\beta} (w)
\,\omega^{ks}\,(\hat{\alpha},\lambda^{s}\beta)\,
S^{\gL,-k-m}_{-\beta} \bar{\gh}^{\,k}_{\,\alpha}\otimes
 \gt^{\,m}_{\,\beta}
 \\
 \\
 \hline
\end{array} \ \ \ \
\eqn In the second part we need to compute the Cartan part of the
commutator: \beqn{t2} \nonumber \Big[ L_{R}(z)\otimes1,\,
r_{R}(z,w)
\Big]\Big|_{\bar{\gh}^{k}_{\alpha}\otimes\gt^{m}_{\beta} } =
\left[\frac{1}{2}\,\sum\limits_{p=\,0}^{l-1} \,
\sum\limits_{\gamma\,\in\, R}\,| \gamma |^2\,
\varphi^{\,p}_{\,\gamma} (z) \, S^{\gL,-p}_{-\gamma} \,
\gt^{\,p}_{\,\gamma}\otimes1,\,
\frac{1}{2}\,\sum\limits_{m=\,0}^{l-1}\,
\sum\limits_{\beta\,\in\,R}\,
|\beta|^2\,\varphi^{-m}_{-\beta}(z-w) \,\gt^{\,-m}_{\,-\beta}
\otimes \gt^{m}_{\beta}
\right]\Big|_{\bar{\gh}^{k}_{\alpha}\otimes\gt^{m}_{\beta} }=\ \
\\\nonumber =\left(\frac{1}{4}\,\sum\limits_{p,m=\,0}^{l-1} \,
\sum\limits_{\gamma,\beta\,\in\, R}\,| \gamma |^2\,|\beta|^2\,
\varphi^{\,p}_{\,\gamma} (z) \, \varphi^{-m}_{-\beta}(z-w)\,
S^{\gL,-p}_{-\gamma} \,
\left[\gt^{\,p}_{\,\gamma},\gt^{\,-m}_{\,-\beta}\right]\otimes\gt^{m}_{\beta}\right)
\Big|_{\bar{\gh}^{k}_{\alpha}\otimes\gt^{m}_{\beta}
}\ \ \ \ \ \
 \eqn
 To separate the needed part from the expression in the bracket we
 must take into account only those terms for which $\gamma\in {\cal O}
 (\beta)$. For this purpose it convenient to make use of the
 "delta"-symbol of the orbit:
$$
 \delta\Big(\alpha\in\,{\cal
O}(\beta)\Big)=\left\{\begin{array}{ll} 1 &
\textrm{if}\ \ \alpha\in\,{\cal O}(\beta)\\
0 & \textrm{if}\ \ \alpha\notin\,{\cal O}(\beta)
\end{array}\right.
$$
Using this notation we rewrite the last expression in the form:
\beqn{tt1} \nonumber\frac{1}{4}\,\sum\limits_{p,m=\,0}^{l-1} \,
\sum\limits_{\gamma,\beta\,\in\, R}\, \delta\Big( \gamma\in {\cal
O}(\beta) \Big) \, | \gamma |^2\,|\beta|^2\,
\varphi^{\,p}_{\,\gamma} (z) \, \varphi^{-m}_{-\beta}(z-w)\,
S^{\gL,-p}_{-\gamma} \,
\left[\gt^{\,p}_{\,\gamma},\gt^{\,-m}_{\,-\beta}\right]\otimes\gt^{m}_{\beta}\stackrel{(\ref{com1})}{=}\\
\nonumber
\stackrel{(\ref{com1})}{=}\frac{1}{2\sqrt{l}}\,\sum\limits_{p,m=\,0}^{l-1}
\, \sum\limits_{\gamma,\beta\,\in\, R}\, \delta\Big( \gamma\in
{\cal O}(\beta) \Big) \, p_{\gamma} \omega^{-m\theta(\gamma,
\beta) } |\beta|^2\, \varphi^{\,p}_{\,\gamma} (z) \,
\varphi^{-m}_{-\beta}(z-w)\, S^{\gL,-p}_{-\gamma} \,
\bar{\gh}^{\,p-m}_{\gamma}\otimes\gt^{m}_{\beta}=\\
\nonumber =\frac{1}{2\sqrt{l}}\,\sum\limits_{k,m=\,0}^{l-1} \,
\sum\limits_{\gamma,\beta\,\in\, R}\, \delta\Big( \gamma\in {\cal
O}(\beta) \Big) \, p_{\gamma} \omega^{-m\theta(\gamma, \beta) }
\,|\beta|^2\, \varphi^{\,m+k}_{\,\gamma} (z) \,
\varphi^{-m}_{-\beta}(z-w)\, S^{\gL,-m-k}_{-\gamma} \,
\bar{\gh}^{\,k}_{\gamma}\otimes\gt^{m}_{\beta}\stackrel{(\ref{tran2})}{=}\\
\nonumber
\stackrel{(\ref{tran2})}{=}\frac{1}{2\sqrt{l}}\,\sum\limits_{k,m=\,0}^{l-1}
\, \sum\limits_{\gamma,\beta\,\in\, R}\, \delta\Big( \gamma\in
{\cal O}(\beta) \Big) \, p_{\gamma} \omega^{k \theta(\gamma,
\beta) } |\beta|^2\, \varphi^{\,m+k}_{\,\gamma} (z) \,
\varphi^{-m}_{-\beta}(z-w)\, S^{\gL,-m-k}_{-\beta} \,
\bar{\gh}^{\,k}_{\gamma}\otimes\gt^{m}_{\beta}\stackrel{(\ref{decsr})}{=}\\
\nonumber
\stackrel{(\ref{decsr})}{=}\frac{1}{2\sqrt{l}}\,\sum\limits_{k,m=\,0}^{l-1}
\, \sum\limits_{\gamma,\beta\,\in\, R}\,
\sum\limits_{\alpha\in\Pi} \,(\hat{\alpha}, \gamma)\,  \delta\Big(
\gamma\in {\cal O}(\beta) \Big) \, p_{\gamma} \omega^{k
\theta(\gamma, \beta) }\,|\beta|^2\, \varphi^{\,m+k}_{\,\gamma}
(z) \, \varphi^{-m}_{-\beta}(z-w)\, S^{\gL,-m-k}_{-\beta} \,
\bar{\gh}^{\,k}_{\alpha}\otimes\gt^{m}_{\beta}=\\
\nonumber =
\frac{1}{2\sqrt{l}}\,\sum\limits_{s=0}^{l_{\beta}-1}\sum\limits_{k,m,=\,0}^{l-1}
\, \sum\limits_{\beta\,\in\, R}\, \sum\limits_{\alpha\in\Pi}
\,(\hat{\alpha}, \lambda^{s}\beta)\, p_{\beta} \omega^{k s }
|\beta|^2\, \varphi^{\,m+k}_{\,\lambda^{s}\beta} (z) \,
\varphi^{-m}_{-\beta}(z-w)\, S^{\gL,-m-k}_{-\beta} \,
\bar{\gh}^{\,k}_{\alpha}\otimes\gt^{m}_{\beta}\stackrel{(\ref{erb})}{=}\\
\nonumber \stackrel{(\ref{erb})}{=}
\frac{1}{2\sqrt{l}}\,\sum\limits_{k,m,s=\,0}^{l-1} \,
\sum\limits_{\beta\,\in\, R}\, \sum\limits_{\alpha\in\Pi}
\,(\hat{\alpha}, \lambda^{s}\beta)\,\omega^{k s } |\beta|^2\,
\varphi^{\,m+k}_{\,\beta} (z) \, \varphi^{-m}_{-\beta}(z-w)\,
S^{\gL,-m-k}_{-\beta} \,
\bar{\gh}^{\,k}_{\alpha}\otimes\gt^{m}_{\beta}
 \eqn
 here we use the denotation $\theta(\alpha,\beta)=s$ if
 $\alpha=-\lambda^{s}\beta$.
 Finally we obtain:
 \beqn{ht2} \begin{array}{|c|}\hline \\  \Big[ L_{R}(z)\otimes1,\,
r_{R}(z,w)
\Big]\Big|_{\bar{\gh}^{k}_{\alpha}\otimes\gt^{m}_{\beta} } =
\frac{1}{2\sqrt{l}}\,\sum\limits_{k,m,s=\,0}^{l-1} \,
\sum\limits_{\beta\,\in\, R}\, \sum\limits_{\alpha\in\Pi}
\,(\hat{\alpha}, \lambda^{s}\beta)\,\omega^{k s } |\beta|^2\,
\varphi^{\,m+k}_{\,\beta} (z) \, \varphi^{-m}_{-\beta}(z-w)\,
S^{\gL,-m-k}_{-\beta} \,
\bar{\gh}^{\,k}_{\alpha}\otimes\gt^{m}_{\beta} \\
\\ \hline  \end{array}\ \ \ \eqn
For the last, third term we have: \beqn{tt3}\nonumber
\Big[1\otimes L_{R}(w),
r_{\gH}(z,w)\Big]\Big|_{\bar{\gh}^{k}_{\alpha}\otimes\gt^{m}_{\beta}}=
\left[\frac{1}{2}\,\sum\limits_{a=\,0}^{l-1}
\, \sum\limits_{\beta\,\in\, R}\,| \beta |^2\,
\varphi^{\,a}_{\,\beta} (w) \, S^{\gL,-a}_{-\beta} \,
1\otimes\gt^{\,a}_{\,\beta},\,
\sum\limits_{k=\,0}^{l-1}\,\sum_{\alpha\,\in\,\Pi}\,\varphi^{\,k}_{\,0}(z-w)\,
\bar{\gH}^{\,k}_{\,\alpha}\otimes\bar{\gh}^{-k}_{\,\alpha}
\right]\Big|_{\bar{\gh}^{k}_{\alpha}\otimes\gt^{m}_{\beta}}=\\
\nonumber = \frac{1}{2}\,\sum\limits_{a,k=\,0}^{l-1} \,
\sum\limits_{\beta\in\, R}\,\sum\limits_{\alpha\in\Pi}  | \beta
|^2\, \varphi^{\,a}_{\,\beta} (w) \,\varphi^{\,k}_{\,0}(z-w)\,
S^{\gL,-a}_{-\beta}\, \bar{\gH}^{\,k}_{\,\alpha}\otimes
[\gt^{\,a}_{\,\beta},\bar{\gh}^{-k}_{\,\alpha}]
\stackrel{(\ref{crel2})}{=}\\
\nonumber
\stackrel{(\ref{crel2})}{=}-\frac{1}{2\sqrt{l}}\,\sum\limits_{a,k,s=\,0}^{l-1}\sum\limits_{\beta\in\,
R}\,\sum\limits_{\alpha\in\Pi} \omega^{\,s
k}\,(\alpha,\lambda^{s}\beta)\,| \beta |^2\,
\varphi^{\,a}_{\,\beta} (w) \,\varphi^{\,k}_{\,0}(z-w)\,
S^{\gL,-a}_{-\beta}\, \bar{\gH}^{\,k}_{\,\alpha}\otimes
\gt^{a-k}_{\beta}=\\
\nonumber
=-\frac{1}{2\sqrt{l}}\,\sum\limits_{m,k,s=\,0}^{l-1}\sum\limits_{\beta\in\,
R}\,\sum\limits_{\alpha\in\Pi} \omega^{\,s
k}\,(\alpha,\lambda^{s}\beta)\,| \beta |^2\,
\varphi^{\,k+m}_{\,\beta} (w) \,\varphi^{\,k}_{\,0}(z-w)\,
S^{\gL,-k-m}_{-\beta}\, \bar{\gH}^{\,k}_{\,\alpha}\otimes
\gt^{m}_{\beta}\stackrel{(\ref{connbas})}{=}\\
\nonumber \stackrel{(\ref{connbas})}{=}
-\frac{1}{2\sqrt{l}}\,\sum\limits_{m,k,s=\,0}^{l-1}\sum\limits_{\beta\in\,
R}\,\sum\limits_{\alpha\in\Pi} \omega^{\,s
k}\,(\hat{\alpha},\lambda^{s}\beta)\,| \beta |^2\,
\varphi^{\,k+m}_{\,\beta} (w) \,\varphi^{\,k}_{\,0}(z-w)\,
S^{\gL,-k-m}_{-\beta}\, \bar{\gh}^{\,k}_{\,\alpha}\otimes
\gt^{m}_{\beta}  \eqn and we get: \beqn{ht3}
\begin{array}{|c|}\hline \\ \Big[1\otimes
L_{R}(w),
r_{\gH}(z,w)\Big]\Big|_{\bar{\gh}^{k}_{\alpha}\otimes\gt^{m}_{\beta}}=
-\frac{1}{2\sqrt{l}}\,\sum\limits_{m,k,s=\,0}^{l-1}\sum\limits_{\beta\in\,
R}\,\sum\limits_{\alpha\in\Pi} \omega^{\,s
k}\,(\hat{\alpha},\lambda^{s}\beta)\,| \beta |^2\,
\varphi^{\,k+m}_{\,\beta} (w) \,\varphi^{\,k}_{\,0}(z-w)\,
S^{\gL,-k-m}_{-\beta}\, \bar{\gh}^{\,k}_{\,\alpha}\otimes
\gt^{m}_{\beta} \\
\\ \hline \end{array} \ \ \ \eqn
Summing up (\ref{ht1}), (\ref{ht2}), (\ref{ht3}) and using Fay
identity (\ref{fay1}) we obtain:
$$
\Big\{L_{\gH}(z)\otimes1, 1\otimes L_{R}(w)
\Big\}\Big|_{\bar{\gh}^{k}_{\alpha}\otimes\gt^{m}_{\beta}}- \Big[
L_{R}(z)\otimes1,\, r_{R}(z,w)
\Big]\Big|_{\bar{\gh}^{k}_{\alpha}\otimes\gt^{m}_{\beta}} -
\Big[1\otimes L_{R}(w),
r_{\gH}(z,w)\Big]\Big|_{\bar{\gh}^{k}_{\alpha}\otimes\gt^{m}_{\beta}}=0
$$


\subsection*{$\gh^{\,0}_{\,\alpha}\otimes \gt^{\,m}_{\,\beta}$-terms}

\beqn{hnt1}\nonumber \Big\{ L_{\gH}^{0}(z)\otimes1,\, 1\otimes
L_{R}(w) \Big\} = \left\{ \sum\limits_{\alpha\,\in\,\Pi} \,
\Big(\bar{v}_{\,\alpha}^{\gH}+E_{1}(z)\,\bar{S}^{\gH,0}_{\alpha}\Big)
\, \, \bar{\gh}^{\,0}_{\,\alpha}\otimes1,\,
\frac{1}{2}\,\sum\limits_{m=\,0}^{l-1} \,
\sum\limits_{\beta\,\in\, R}\,| \beta |^2\,
\varphi^{\,m}_{\,\beta} (w) \, S^{\gL,-m}_{-\beta} \,
1\otimes\gt^{\,m}_{\,\beta}   \right\}=\\\nonumber = \frac{1}{2}\,
\sum\limits_{\alpha\,\in\,\Pi} \,\sum\limits_{m=\,0}^{l-1} \,
\sum\limits_{\beta\,\in\, R}\,
  | \beta |^2\, \Big(\left\{\bar{v}_{\,\alpha}^{\gH},\varphi^{\,m}_{\,\beta} (w)\right\}\,S^{\gL,-m}_{-\beta}+E_{1}(z)\,\varphi^{\,m}_{\,\beta} (w)
   \left\{\bar{S}^{\gH,0}_{\alpha},\, S^{\gL,-m}_{-\beta}\right\} \Big)
\,
\bar{\gh}^{\,0}_{\,\alpha}\otimes\gt^{\,m}_{\,\beta}\stackrel{(\ref{PB})}{=}\\\nonumber
\stackrel{(\ref{PB})}{=}\frac{1}{2}\,
\sum\limits_{\alpha\,\in\,\Pi} \,\sum\limits_{m=\,0}^{l-1} \,
\sum\limits_{\beta\,\in\, R}\,
  | \beta |^2\, \Big( S^{\gL,-m}_{-\beta}\,\partial_{1}\varphi^{m}_{\beta}(w)\,
  \frac{1}{\sqrt{l}}\,\sum\limits_{s=0}^{l-1}\,(\hat{\alpha},\lambda^s\beta )
- \frac{1}{\sqrt{l}}\,E_{1}(z)\,\varphi^{\,m}_{\,\beta} (w)
   \, S^{\gL,-m}_{-\beta}\sum\limits_{s=0}^{l-1}\,(\hat{\alpha},\lambda^s\beta )  \Big)
\, \bar{\gh}^{\,0}_{\,\alpha}\otimes\gt^{\,m}_{\,\beta} \eqn
therefore we have: \beqn{hngt1}\begin{array}{|l|}\hline\\
  \Big\{
L_{\gH}^{0}(z)\otimes1,\, 1\otimes L_{R}(w)
\Big\}\Big|_{\bar{\gh}^{0}_{\alpha}\otimes\gt^{m}_{\beta}}=\\
=\frac{1}{2\sqrt{l}}\, \sum\limits_{\alpha\,\in\,\Pi}
\,\sum\limits_{m,s=\,0}^{l-1} \, \sum\limits_{\beta\,\in\, R}\,
  | \beta |^2\,  S^{\gL,-m}_{-\beta}\, (\hat{\alpha},\lambda^s\beta ) \Big(\partial_{1}\varphi^{m}_{\beta}(w)\,
- E_{1}(z)\,\varphi^{\,m}_{\,\beta} (w)  \Big) \,
\bar{\gh}^{\,0}_{\,\alpha}\otimes\gt^{\,m}_{\,\beta}\\
\\
\hline
\end{array}\ \ \ \eqn
The calculation of the second term $\Big[ L_{R}(z)\otimes1,\,
r_{R}(z,w) \Big]\big|_{\gh^{0}\otimes\gt }$ is analogous to the
computation of (\ref{tt1}), and to get the answer we need to put
$k=0$ in the expression (\ref{tt1}): \beqn{hngt2}
\begin{array}{|c|}\hline \\  \Big[ L_{R}(z)\otimes1,\,
r_{R}(z,w)
\Big]\Big|_{\bar{\gh}^{0}_{\alpha}\otimes\gt^{m}_{\beta}} =
\frac{1}{2\sqrt{l}}\,\sum\limits_{m,s=\,0}^{l-1} \,
\sum\limits_{\beta\,\in\, R}\, \sum\limits_{\alpha\in\Pi}
\,(\hat{\alpha}, \lambda^{s}\beta)|\beta|^2\,
\varphi^{\,m}_{\,\beta} (z) \, \varphi^{-m}_{-\beta}(z-w)\,
S^{\gL,-m}_{-\beta} \,
\bar{\gh}^{\,0}_{\alpha}\otimes\gt^{m}_{\beta} \\
\\ \hline  \end{array}\ \ \ \eqn
Analogously, we obtain the answer for $\Big[1\otimes L_{R}(w),
r_{\gH}(z,w)\Big]\Big|_{\bar{\gh}^{0}_{\alpha}\otimes\gt^{m}_{\beta}}$
substituting into (\ref{ht3}) $k=0$, and replacing
$\varphi^{0}_{0}(z-w)\rightarrow E_{1}(z-w)$: \beqn{hngt3}
\begin{array}{|c|}\hline \\ \Big[1\otimes
L_{R}(w),
r_{\gH}(z,w)\Big]\Big|_{\bar{\gh}^{0}_{\alpha}\otimes\gt^{m}_{\beta}}=
-\frac{1}{2\sqrt{l}}\,\sum\limits_{m,s=\,0}^{l-1}\sum\limits_{\beta\in\,
R}\,\sum\limits_{\alpha\in\Pi}(\hat{\alpha},\lambda^{s}\beta)\,|
\beta |^2\, \varphi^{\,m}_{\,\beta} (w) \,E_{1}(z-w)\,
S^{\gL,-m}_{-\beta}\, \bar{\gh}^{\,0}_{\,\alpha}\otimes
\gt^{m}_{\beta} \\
\\ \hline \end{array} \ \ \ \eqn
Using the second Fay identity (\ref{fay2}) we finally find:
$$
\Big\{ L_{\gH}^{0}(z)\otimes1,\, 1\otimes L_{R}(w)
\Big\}\Big|_{\bar{\gh}^{0}_{\alpha}\otimes\gt^{m}_{\beta}}- \Big[
L_{R}(z)\otimes1,\, r_{R}(z,w)
\Big]\Big|_{\bar{\gh}^{0}_{\alpha}\otimes\gt^{m}_{\beta}} -
\Big[1\otimes L_{R}(w),
r_{\gH}(z,w)\Big]\Big|_{\bar{\gh}^{0}_{\alpha}\otimes\gt^{m}_{\beta}}=0
$$


\subsection*{$\gt^{k}_{\alpha}\otimes\gt^{-k}_{-\alpha}$-terms}

\beqn{gtgt1}\nonumber \Big\{L_{R}(z)\otimes1,\,1\otimes
L_{R}(w)\Big\}\Big|_{\gt^{k}_{\alpha}\otimes\gt^{-k}_{-\alpha}}=\\\nonumber
\left\{ \frac{1}{2}\,\sum\limits_{k=\,0}^{l-1} \,
\sum\limits_{\alpha\,\in\, R}\,| \alpha |^2\,
\varphi^{\,k}_{\,\alpha} (z) \, S^{\gL,-k}_{-\alpha} \,
\gt^{\,k}_{\,\alpha}\otimes1,\,
\frac{1}{2}\,\sum\limits_{m=\,0}^{l-1} \,
\sum\limits_{\beta\,\in\, R}\,| \beta |^2\,
\varphi^{\,m}_{\,\beta} (z) \, S^{\gL,-m}_{-\beta} \,
1\otimes\gt^{\,m}_{\,\beta}\right\}\Big|_{\gt^{k}_{\alpha}\otimes\gt^{-k}_{-\alpha}}=\\
\nonumber =\left(\frac{1}{4}\,\sum\limits_{k,m=\,0}^{l-1} \,
\sum\limits_{\alpha,\beta\,\in\, R}\, | \alpha |^2\,| \beta |^2\,
\varphi^{\,k}_{\,\alpha}(z) \varphi^{\,m}_{\,\beta} (w) \,
\left\{S^{\gL,-k}_{-\alpha}S^{\gL,-m}_{-\beta}\right\} \
\gt^{\,k}_{\,\alpha}\otimes\gt^{\,m}_{\,\beta}\right)\Big|_{\gt^{k}_{\alpha}\otimes\gt^{-k}_{-\alpha}}
\eqn To separate the needed part we must take into account only
those terms for which $m=-k$, and $\beta~\in\cal{O}~(-\alpha)$,
therefore we put $m=-k$, $\beta=- \lambda^{s}\alpha$, and take the
sum over $ s\in \{0,...,l_{\alpha}-1\}$: \beqn{gtgt2} \nonumber
\frac{1}{4}\,\sum\limits_{k\,=0}^{l-1}
\,\sum\limits_{s\,=0}^{l_{\alpha}-1} \sum\limits_{\alpha\in\, R}\,
| \alpha |^4\,|\, \varphi^{\,k}_{\,\alpha}(z)
\varphi^{\,-k}_{\,-\lambda^s\alpha} (w) \,
\left\{S^{\gL,-k}_{-\alpha}S^{\gL,k}_{\lambda^s\alpha}\right\} \
\gt^{\,k}_{\,\alpha}\otimes\gt^{\,-k}_{\,-\lambda^s\alpha}\stackrel{(\ref{PB}),(\ref{erb}),(\ref{tran1})}{=}\\
\nonumber \stackrel{(\ref{PB}),(\ref{erb}),(\ref{tran1})}{=}
\frac{1}{4\sqrt{l}}\,\sum\limits_{k\,=0}^{l-1}
\,\sum\limits_{s\,=0}^{l_{\alpha}-1} \sum\limits_{\alpha\in\, R}\,
| \alpha |^4\,|\, \varphi^{\,k}_{\,\alpha}(z)
\varphi^{\,-k}_{\,-\alpha} (w) \, p_{\alpha}\,S^{\gh, 0}_{-\alpha}
\,
\gt^{\,k}_{\,\alpha}\otimes\gt^{\,-k}_{\,-\alpha}=\\
\nonumber =\frac{\sqrt{l}}{2}\,\sum\limits_{k\,=0}^{l-1}
\sum\limits_{\alpha\in\, R}\, | \alpha |^2\,|\,
\varphi^{\,k}_{\,\alpha}(z) \varphi^{\,-k}_{\,-\alpha}
(w)\,\bar{S}^{\gh, 0}_{-\alpha} \,
\gt^{\,k}_{\,\alpha}\otimes\gt^{\,-k}_{\,-\alpha}\stackrel{(\ref{sprop})}{=}\\
\nonumber
\stackrel{(\ref{sprop})}{=}-\frac{\sqrt{l}}{2}\,\sum\limits_{k\,=0}^{l-1}
\sum\limits_{\alpha\in\, R}\, | \alpha |^2\,|\,
\varphi^{\,k}_{\,\alpha}(z) \varphi^{\,-k}_{\,-\alpha}
(w)\,\bar{S}^{\gh, 0}_{\alpha} \,
\gt^{\,k}_{\,\alpha}\otimes\gt^{\,-k}_{\,-\alpha}\eqn Therefore we
have: \beqn{gtgt3}\begin{array}{|c|}\hline\\
\Big\{L_{R}(z)\otimes1,\,1\otimes
L_{R}(w)\Big\}\Big|_{\gt^{k}_{\alpha}\otimes\gt^{-k}_{-\alpha}}=-\frac{\sqrt{l}}{2}\,\sum\limits_{k\,=0}^{l-1}
\sum\limits_{\alpha\in\, R}\, | \alpha |^2\,|\,
\varphi^{\,k}_{\,\alpha}(z) \varphi^{\,-k}_{\,-\alpha}
(w)\,\bar{S}^{\gh, 0}_{\alpha} \,
\gt^{\,k}_{\,\alpha}\otimes\gt^{\,-k}_{\,-\alpha}
\\
\\
\hline
\end{array}\eqn
The second term: \beqn{gtgt4}\nonumber \Big[
L_{\gH}^{0}(z)\otimes1,
r_{R}(z,w)\Big]=\left[\sum\limits_{\gamma\,\in\,\Pi} \,
\Big(\bar{v}_{\,\gamma}^{\gH}+E_{1}(z)\,\bar{S}^{\gH,0}_{\gamma}\Big)
\, \, \bar{\gh}^{\,0}_{\,\gamma}\otimes1,\,
\frac{1}{2}\,\sum\limits_{k=\,0}^{l-1}\,
\sum\limits_{\alpha\,\in\,R}\,
|\alpha|^2\,\varphi^{\,k}_{\,\alpha}(z-w) \,\gt^{\,k}_{\,\alpha}
\otimes \gt^{-k}_{-\alpha} \right]=\\\nonumber
=\frac{1}{2}\,\sum\limits_{k=\,0}^{l-1}\,
\sum\limits_{\alpha\,\in\,R}\,\sum\limits_{\gamma\,\in\,\Pi} \,
\Big(\bar{v}_{\,\gamma}^{\gH}+E_{1}(z)\,\bar{S}^{\gH,0}_{\gamma}\Big)\,
|\alpha|^2\,\varphi^{\,k}_{\,\alpha}(z-w)
\,\left[\bar{\gh}^{\,0}_{\,\gamma},\gt^{\,k}_{\,\alpha}\right]
\otimes \gt^{-k}_{-\alpha}\stackrel{(\ref{crel2})}{=}\\\nonumber
\stackrel{(\ref{crel2})}{=}\frac{1}{2\sqrt{l}}\,\sum\limits_{k,s=\,0}^{l-1}\,
\sum\limits_{\alpha\,\in\,R}\,\sum\limits_{\gamma\,\in\,\Pi} \,
\Big(\bar{v}_{\,\gamma}^{\gH}+E_{1}(z)\,\bar{S}^{\gH,0}_{\gamma}\Big)\,
|\alpha|^2\,\varphi^{\,k}_{\,\alpha}(z-w) \,(\gamma,
\lambda^{s}\alpha)\,\gt^{\,k}_{\,\alpha} \otimes
\gt^{-k}_{-\alpha}
 \eqn
 and we obtain:
 \beqn{gtgt5}\begin{array}{|c|}\hline\\
 \Big[
L_{\gH}^{0}(z)\otimes1,
r_{R}(z,w)\Big]=\frac{1}{2\sqrt{l}}\,\sum\limits_{k,s=\,0}^{l-1}\,
\sum\limits_{\alpha\,\in\,R}\,\sum\limits_{\gamma\,\in\,\Pi} \,
\Big(\bar{v}_{\,\gamma}^{\gH}+E_{1}(z)\,\bar{S}^{\gH,0}_{\gamma}\Big)\,
|\alpha|^2\,\varphi^{\,k}_{\,\alpha}(z-w) \,(\gamma,
\lambda^{s}\alpha)\,\gt^{\,k}_{\,\alpha} \otimes
\gt^{-k}_{-\alpha}
\\
\\
\hline
\end{array}\ \ \ \eqn
the last  term:
 \beqn{gtgt6}\nonumber \Big[
1\otimes L_{\gH}^{0}(w),
r_{R}(z,w)\Big]=\left[\sum\limits_{\gamma\,\in\,\Pi} \,
\Big(\bar{v}_{\,\gamma}^{\gH}+E_{1}(w)\,\bar{S}^{\gH,0}_{\gamma}\Big)
\, 1\otimes\bar{\gh}^{\,0}_{\,\gamma},\,
\frac{1}{2}\,\sum\limits_{k=\,0}^{l-1}\,
\sum\limits_{\alpha\,\in\,R}\,
|\alpha|^2\,\varphi^{\,k}_{\,\alpha}(z-w) \,\gt^{\,k}_{\,\alpha}
\otimes \gt^{-k}_{-\alpha} \right] \eqn Obviously the result
differs from the previous expression (\ref{gtgt6}) only by sign
and by substitution $E_{1}(z)\rightarrow E_{1}(w)$, therefore:
\beqn{gtgt7}\begin{array}{|c|}\hline\\   \Big[ 1\otimes
L_{\gH}^{0}(w),
r_{R}(z,w)\Big]=-\,\frac{1}{2\sqrt{l}}\,\sum\limits_{k,s=\,0}^{l-1}\,
\sum\limits_{\alpha\,\in\,R}\,\sum\limits_{\gamma\,\in\,\Pi} \,
\Big(\bar{v}_{\,\gamma}^{\gH}+E_{1}(w)\,\bar{S}^{\gH,0}_{\gamma}\Big)\,
|\alpha|^2\,\varphi^{\,k}_{\,\alpha}(z-w) \,(\gamma,
\lambda^{s}\alpha)\,\gt^{\,k}_{\,\alpha} \otimes
\gt^{-k}_{-\alpha}\\
\\
\hline
\end{array}\ \ \ \eqn
The sum of the last two expressions (\ref{gtgt6}), (\ref{gtgt7}):
\beqn{gtgt8}\nonumber \Big[ L_{\gH}^{0}(z)\otimes1,
r_{R}(z,w)\Big]+\Big[ 1\otimes L_{\gH}^{0}(w),
r_{R}(z,w)\Big]=\\\nonumber
\frac{1}{2\sqrt{l}}\,\sum\limits_{k,s=\,0}^{l-1}\,
\sum\limits_{\alpha\,\in\,R}\,\sum\limits_{\gamma\,\in\,\Pi} \,
\Big( E_{1}(z)-E_{1}(w) \Big)\,\bar{S}^{\gH,0}_{\gamma}\,
|\alpha|^2\,\varphi^{\,k}_{\,\alpha}(z-w) \,(\gamma,
\lambda^{s}\alpha)\,\gt^{\,k}_{\,\alpha} \otimes
\gt^{-k}_{-\alpha}\stackrel{(\ref{decsr})}{=}\\\nonumber
\stackrel{(\ref{decsr})}{=}
\frac{1}{2\sqrt{l}}\,\sum\limits_{k,s=\,0}^{l-1}\,
\sum\limits_{\alpha\,\in\,R} \, \Big( E_{1}(z)-E_{1}(w)
\Big)\,\bar{S}^{\gh,0}_{\lambda^{s}\alpha}\,
|\alpha|^2\,\varphi^{\,k}_{\,\alpha}(z-w) \,\gt^{\,k}_{\,\alpha}
\otimes \gt^{-k}_{-\alpha}\stackrel{(\ref{tran2})}{=}\\\nonumber
\stackrel{(\ref{tran2})}{=}\frac{\sqrt{l}}{2}\,\sum\limits_{k=0}^{l-1}\,
\sum\limits_{\alpha\,\in\,R} \, |\alpha|^2\,\Big(
E_{1}(z)-E_{1}(w)
\Big)\,\varphi^{\,k}_{\,\alpha}(z-w)\,\bar{S}^{\gh,0}_{\alpha}\,
\gt^{\,k}_{\,\alpha} \otimes \gt^{-k}_{-\alpha}
 \eqn
 and finally:
 \beqn{gtgt9}\nonumber  \Big\{L_{R}(z)\otimes1,\,1\otimes
L_{R}(w)\Big\}\Big|_{\gt^{k}_{\alpha}\otimes\gt^{-k}_{-\alpha}}-\Big[
L_{\gH}^{0}(z)\otimes1, r_{R}(z,w)\Big]-\Big[ 1\otimes
L_{\gH}^{0}(w), r_{R}(z,w)\Big]=\\
\nonumber \frac{\sqrt{l}}{2}\,\sum\limits_{k=0}^{l-1}\,
\sum\limits_{\alpha\,\in\,R} \, |\alpha|^2\,\Big(
-\varphi^{\,k}_{\,\alpha}(z)\,\varphi^{-k}_{-\alpha}(w)- \left(
E_{1}(z)-E_{1}(w)
\right)\,\varphi^{\,k}_{\,\alpha}(z-w)\,\Big)\,\bar{S}^{\gh,0}_{\alpha}\,
\gt^{\,k}_{\,\alpha} \otimes
\gt^{-k}_{-\alpha}=\\
\nonumber =-\frac{\sqrt{l}}{2}\,\sum\limits_{k=0}^{l-1}\,
\sum\limits_{\alpha\,\in\,R} \,
|\alpha|^2\,\partial_{1}\varphi^{k}_{\alpha}(z-w)\,\bar{S}^{\gh,0}_{\alpha}\,
\gt^{\,k}_{\,\alpha} \otimes \gt^{-k}_{-\alpha}  \eqn The last
step here is by the second Fay identity (\ref{fay2}).


\subsection*{Classical Yang-Baxter equation}

\begin{predl}
The $r$-matrix (\ref{Rma}) satisfies the classical dynamical
Yang-Baxter equation: \beq{dfg5}
[r_{12}(z,w),r_{13}(z,x)]+[r_{12}(z,w),r_{23}(w,x)]+[r_{13}(z,x),r_{23}(w,x)]-
\eq
$$
\sqrt{l}\sum\limits_{k=0}^{l-1}\sum\limits_{\alpha\in\,R}\frac{|\al|^2}{2}
\gt^{\,k}_{\,\al}\otimes \gt^{\,-k}_{\,-\al}\otimes
\bar{\gh}_{\al}^{0}\,\p_1\vf_\al^k(z-w)-
\frac{|\al|^2}{2}\gt^{\,k}_{\,\al}\otimes
\bar{\gh}_{\al}^{0}\otimes
\gt^{\,-k}_{\,-\al}\,\p_1\vf_\al^k(z-x)+
$$
$$
\frac{|\al|^2}{2}\bar{\gh}_{\al}^{0}\otimes
\gt^{\,k}_{\,\al}\otimes\gt^{\,-k}_{\,-\al}\,\p_1\vf_\al^k(w-x)=0
$$
\end{predl}

\emph{Proof}

Let us examine the "non-dynamical" part (the upper line) of
(\ref{dfg5}):

\underline{off Cartan part}:

\beq{dfg6}
\sum\limits_{k,\,n=0}^{l-1}\sum\limits_{\alpha,\be\in\,R}
\vf_\al^k(z-w)\vf_\be^n(z-x) [\gt_\al^k,\gt_\be^n]\otimes
\gt_{-\al}^{-k}\otimes \gt_{-\be}^{-n} +\eq
$$
\vf_\al^k(z-w)\vf_\be^n(w-x) \gt_\al^k\otimes
[\gt_{-\al}^{-k},\gt_\be^n]\otimes \gt_{-\be}^{-n}+
$$
$$
\vf_\al^k(z-x)\vf_\be^n(w-x) \gt_\al^k\otimes \gt_\be^n\otimes
[\gt_{-\al}^{-k},\gt_{-\be}^{-n}]=
$$
\beq{dfg7}
\sum\limits_{k,\,n,s=0}^{l-1}\sum\limits_{\alpha,\be\in\,R}\frac{|\al|^2|\be|^2}{4\sqrt{l}}
\vf_\al^k(z-w)\vf_\be^n(z-x)\om^{ns}C_{\al,\,\lambda^s\be}\,
\gt_{\al+\lambda^s(\be)}^{k+n}\otimes \gt_{-\al}^{-k}\otimes
\gt_{-\be}^{-n}+ \eq
$$
\frac{|\al|^2|\be|^2}{4\sqrt{l}}\vf_\al^k(z-w)\vf_\be^n(w-x)\om^{ns}C_{-\al,\,\lambda^s\be}\,
\gt_\al^k\otimes \gt_{-\al+\lambda^s(\be)}^{-k+n}\otimes
\gt_{-\be}^{-n}+
$$
$$
\frac{|\al|^2|\be|^2}{4\sqrt{l}}\vf_\al^k(z-x)\vf_\be^n(w-x)\om^{-ns}C_{-\al,-\lambda^s\be}\,
\gt_\al^k\otimes \gt_\be^n\otimes
\gt_{-\al+\lambda^s(-\be)}^{-k-n}
$$
Making shifts $k\rightarrow k+n$ and $\al\rightarrow
\al+\lambda^s(\be)$ in the second line of (\ref{dfg7}) we have:
$$
\frac{|\al+\la^s\be|^2|\be|^2}{4\sqrt{l}}
\vf_{\al+\lambda^s(\be)}^{k+n}(z-w)\vf_\be^n(w-x)\om^{ns}
C_{-\al-\lambda^s\be,\lambda^s\be}\,
\gt_{\al+\lambda^s(\be)}^{k+n}\otimes \gt_{-\al}^{-k}\otimes
\gt_{-\be}^{-n}
$$

Similarly, one should make the following substitutions in the
third line: $\be\rightarrow -\be$, $n\rightarrow -n$,
$k\leftrightarrow n$, $\al\rightarrow \al+\lambda^s\be$,
$n\rightarrow n+k$, $\al\leftrightarrow\be$,
$\al\rightarrow\la^{-s}\al$, $\be\rightarrow\lambda^s\be$.
Finally:
$$
\frac{|\al+\la^s\be|^2|\al|^2}{4\sqrt{l}}
\vf^{k+n}_{\al+\lambda^s(\be)}(z-x)\vf^{-k}_{-\lambda^{-s}(\al)}(w-x)\om^{ks}
C_{-\al-\lambda^s\be,\al}\, \gt^{k+n}_{\al+\lambda^s\be}\otimes
\gt^{-k}_{-\lambda^{-s}\al} \otimes \gt^{-n}_{-\lambda^s\be}
$$
Using the property $\gt^k_{\lambda^s\al}=\om^{-ks}\gt^k_\al$ we
have:
$$
\frac{|\al+\la^s\be|^2|\al|^2}{4\sqrt{l}}
\vf^{k+n}_{\al+\lambda^s(\be)}(z-x)\vf^{-k}_{-\lambda^{-s}(\al)}(w-x)\om^{ns}
C_{-\al-\lambda^s\be,\al}\, \gt^{k+n}_{\al+\lambda^s\be}\otimes
\gt^{-k}_{-\al} \otimes \gt^{-n}_{-\be}
$$

By the definition of the structure constants $C_{\al,\be}$ it is
easy to show that
$C_{-\al-\lambda^s\be,\lambda^s\be}=-\frac{|\al|^2}{|\al+\la^s\be|}C_{\al,\lambda^s\be}$
and
$C_{-\al-\lambda^s\be,\,\al}=\frac{|\be|^2}{|\al+\la^s\be|}C_{\al,\lambda^s\be}$.

Now we can combine all three lines and get a common multiple for
$$
\frac{|\al|^2|\be|^2}{4\sqrt{l}}\om^{ns}C_{\al,\lambda^s\be}\,
\gt_{\al+\lambda^s(\be)}^{k+n}\otimes \gt_{-\al}^{-k}\otimes
\gt_{-\be}^{-n}
$$
The multiple is
$$
\vf_\al^k(z-w)\vf_\be^n(z-x)-\vf_{\al+\lambda^s(\be)}^{k+n}(z-w)\vf_\be^n(w-x)+
\vf^{k+n}_{\al+\lambda^s(\be)}(z-x)\vf^{-k}_{-\lambda^{-s}(\al)}(w-x)
$$
It vanishes due to Fay identity. The proof of this fact is direct.

\underline{Cartan part}:

Consider terms with Cartan elements in the third component of the
tensor product (the other components can be obtained by the cyclic
permutations). There are two origins for this type terms: 1. a
direct appearance from $[r_{12}(z,w),r_{13}(z,x)]$ and
$[r_{12}(z,w),r_{23}(w,x)]$ 2. appearance from
$[r_{13}(z,x),r_{23}(w,x)]$ due to Cartan part of the commutator
$[\gt_{-\al}^{-k},\gt_{-\be}^{-n}]$:

\beq{dfg8} \sum\limits_{k,\,n=0}^{l-1}\,\,\sum\limits_{\alpha\in\,
R,\be\in\,\Pi}\frac{|\al|^2}{2\sqrt{l}} \vf_\al^k(z-w)\vf_0^n(z-x)
[\gt_\al^k,\gH_{\be}^{\,n}]\otimes \gt_{-\al}^{-k}\otimes
\gh_{\,\be}^{-n} +\eq
$$
\sum\limits_{k,\,n=0}^{l-1}\,\,\sum\limits_{\alpha\in\,
R,\be\in\,\Pi}\frac{|\al|^2}{2\sqrt{l}} \vf_\al^k(z-w)\vf_0^n(w-x)
\gt_\al^k\otimes [\gt_{-\al}^{-k},\gH_{\be}^{\,n}]\otimes
\gh_{\,\be}^{-n}+
$$
$$
\sum\limits_{k,\,n=0}^{l-1}\,\,\sum\limits_{\alpha,\be=-\la^r\al\in\,R}
\frac{|\al|^2|\be|^2}{4\sqrt{l}} \vf_\al^k(z-x)\vf_\be^n(w-x)
\gt_\al^k\otimes \gt_\be^n\otimes
[\gt_{-\al}^{-k},\gt_{-\be}^{-n}]=
$$

\beq{dfg9}
-\sum\limits_{k,\,n,s=0}^{l-1}\,\,\sum\limits_{\alpha\in\,
R,\be\in\,\Pi}\frac{|\al|^2|\be|^2}{4\sqrt{l}}
\vf_\al^{k-n}(z-w)\vf_0^n(z-x)(\hat{\be},\la^{-s}\al)\om^{ns}
\gt^{k}_\al\otimes \gt^{n-k}_{-\al}\otimes \gh_{\,\,\be}^{-n}- \eq
$$
\sum\limits_{k,\,n,s=0}^{l-1}\,\,\sum\limits_{\alpha\in\,
R,\be\in\,\Pi}\frac{|\al|^2|\be|^2}{4\sqrt{l}}
\vf_\al^k(z-w)\vf_0^n(w-x)(\hat{\be},-\la^{-s}\al)\om^{ns}
\gt^{k}_\al\otimes \gt^{n-k}_{-\al}\otimes \gh_{\,\,\be}^{-n}+
$$
$$
\sum\limits_{k,\,n=0}^{l-1}\,\,\sum\limits_{\alpha,\be=-\la^r\al\in\,R}
\frac{|\al|^2|\be|^2}{4\sqrt{l}}p_{-\al}\vf_\al^k(z-x)\vf_\be^{n-k}(w-x)\om^{nr}
\gt^{k}_\al\otimes \gt^{n}_{\,\be}\otimes \gh_{-\al}^{\,-n-k}
$$
first lines contain a sum
Let us analyze the first line.
We have the following sum over $\be$:
$\sum\limits_{\be\in\Pi}\frac{|\be|^2}{2}
(\hat{\be},\la^{-s}\al)\om^{ns}\gh_{\be}^{-n}$. For arbitrary root
$\ga$: $\sum\limits_{\be\in\Pi}\frac{|\be|^2}{2}(\hat{\be},\ga)
\gh_\be=\frac{|\ga|^2}{2}\gh_\ga$. Now, using the property
$\gh^k_{\lambda^s(\al)}=\om^{-ks}\gh^k_\al$ and $|\la\al|=|\al|$
we have: $\sum\limits_{\be\in\Pi}
\frac{|\be|^2}{2}(\hat{\be},\la^{-s}\al)\om^{ns}\gh_{\be}^{-n}=
\frac{|\al|^2}{2}\om^{ns}\gh^{-n}_{\lambda^{-s}(\al)}=\frac{|\al|^2}{2}\gh^{-n}_{\al}$.
Note that the result does not depend on $s$. Thus the sum over $s$
makes the common multiple $l$ and the first line equals:
$$
-l\sum\limits_{k,\,n=0}^{l-1}\,\,\sum\limits_{\alpha\in\,
R}\frac{|\al|^2}{2\sqrt{l}} \vf_\al^{k-n}(z-w)\vf_0^n(z-x)
\gt^{k}_\al\otimes \gt^{n-k}_{-\al}\otimes \bar{\gh}_{\al}^{-n}
$$
In the same way for the second line of (\ref{dfg9}) we have:
$$
l\sum\limits_{k,\,n=0}^{l-1}\,\,\sum\limits_{\alpha\in\,
R}\frac{|\al|^2}{2\sqrt{l}} \vf_\al^k(z-w)\vf_0^n(w-x)
\gt^{k}_\al\otimes \gt^{n-k}_{-\al}\otimes \bar{\gh}_{\al}^{-n}
$$
In the third line of (\ref{dfg9}) we collect all terms with tensor
structure of type $\gt^{k}_\al\otimes \gt^{n-k}_{-\al}\otimes
\gh_{\al}^{-n}$. In this line $\be=-\lambda^{r}(\al)$ for some
$r$. Then $\om^{nr} \gt^{k}_\al\otimes \gt^{n}_{-\la^r\al}\otimes
\gh_{-\al}^{-n}=-\gt^{k}_\al\otimes \gt^{n-k}_{-\al}\otimes
\gh_{\al}^{-n}$ From the definition it follows that
$p_{-\al}=p_\al=\frac{l}{l_\al}$, where $l_\al$ describes the
minimal orbit, i.e. $l_\al$ is a minimal nonzero number with the
property $\la^{l_\al}\al=\al$. The number of desired terms equals
$l_\al$.

Finally we have: \beq{dfg10}
\sqrt{l}\sum\limits_{k,\,n=0}^{l-1}\,\,\sum\limits_{\alpha\in\,
R}-\frac{|\al|^2}{2} \vf_\al^{k-n}(z-w)\vf_0^n(z-x)
\gt^{k}_\al\otimes \gt^{n-k}_{-\al}\otimes \bar{\gh}_{\al}^{-n}+
\eq
$$
\frac{|\al|^2}{2}\vf_\al^k(z-w)\vf_0^n(w-x) \gt^{k}_\al\otimes
\gt^{n-k}_{-\al}\otimes \bar{\gh}_{\al}^{-n}-
$$
$$
\frac{|\al|^2}{2}\vf_\al^k(z-x)\vf_{-\al}^{n-k}(w-x)
\gt^{k}_\al\otimes \gt^{n-k}_{-\al}\otimes \bar{\gh}_{\al}^{-n}
$$
For $n\neq 0$ the sum vanishes due to Fay identity:
$$
\vf_\al^{k-n}(z-w)\vf_0^n(z-x)-\vf_\al^k(z-w)\vf_0^n(w-x)+\vf_\al^k(z-x)\vf_{-\al}^{n-k}(w-x)=0
$$
For $n=0$ the common multiple equals:
$$
-\vf_\al^k(z-w)E_1(z-x)+\vf_\al^k(z-w)E_1(w-x)-\vf_\al^k(z-x)\vf^k_{-\al}(w-x)=
$$
$$
\vf^\al_k(z-w)(E_1(z-w+\langle u-\eta\tau,\al\rangle)-E_1(\langle
u-\eta\tau,\al\rangle))= \p_1\vf_\al^k(z-w)
$$
The later is exactly compensated by "dynamical" part of YB:
$$
-\sqrt{l}\sum\limits_{k=0}^{l-1}\,\,\sum\limits_{\alpha\in\,
R}\frac{|\al|^2}{2}\gt^{k}_\al\otimes \gt^{-k}_{-\al}\otimes
\bar{\gh}_{\al}^{0}\,\p_1\vf_\al^k(z-w)
$$
$\Box$

\section{Appendix A. Simple Lie groups. Facts and notations,
  \cite{Bou,OV}}
\setcounter{equation}{0}
\def\theequation{A.\arabic{equation}}

\emph{\textbf{Roots and weights.}}\\

Let $V$ be a finite-dimensional vector space over $\mR$, $\dim\,V=n$ and $V^*$ is its
dual and
  $\lan~,~\ran$ is a pairing between $V$ and  $V^*$.
A finite system of vectors $R=\{\al\}$ in $V^*$ is called a root system, if\\
1. $R$ generates $V^*$;\\
2. For any $\al\in R$ there exists a coroot $\al^\vee\in V$ such that
$\lan\al,\al^\vee\ran=2$ and the reflection in $V^*$
\beq{ref}
s_\al\,:~x\mapsto x-\lan x,\al\ran\al^\vee
\eq
preserving $R$;\\
3. $\lan\be,\al^\vee\ran\in\mZ$ for any $\be\in R$;\\
4. For $\al\in R$ $\,n\al\in R$ iff $n=\pm 1$.

The dual  system $R^\vee=\{\al^\vee\}$ is the root system in $V$.
If $V$ and $V^*$ are identified by a scalar product $(~,~)$, then
$\al^\vee=\frac{2\al}{(\al,\al)}$.
The group of automorphisms of $V^*$ generated by reflections (\ref{ref}) is \emph{the Weyl group} $W(R)$. The groups $W(R)$ and $W(R^\vee)$ are isomorphic and $W(R^\vee)$ acts on $V$ as
$$
s_\al\,:~x\mapsto x-\lan x,\al^\vee\ran\al\,,~~~x\in V^*\,.
$$

A basis $\Pi=(\al_1,\ldots,\al_l)$ of \emph{simple roots} in $R$ is defined in
such a way  that any $\al\in R$ is decomposed as
\beq{ale1}
\al=\sum_{j=1}^nf^\al_j\al_j\,, ~~f^\al_j\in\mZ\,,
\eq
and all $f^\al_j$ are positive (in this case $\al$ is a positive root), or negative
($\al$ is a negative root). In other words, the root system is an union of positive and negative roots $R=R^+\cup R^-$.
The sum
\beq{ale}
f_\al=\sum_{\al_j\in\Pi}f_j
\eq
is called \emph{the level of $\al$}.

The matrix of order $n$
\beq{CMa}
a_{jk}=\lan \al_j,\al_k^\vee\ran\,,~~\al_j\in\Pi\,,~\al_k^\vee\in\Pi^\vee
\eq
is called \emph{the Cartan matrix}. The Dynkin diagram is constructed by means of $a_{jk}$.

Let $S^W$ be an algebra of polynomials on $V$ invariant with respect to $W$-action.
There exists a basis in $S^W$ of $l$ homogeneous polynomials of degrees
$d_1=2, d_2,\ldots,d_l$. The degrees are uniquely defined by the root system $R$.
The number of roots can be read off from the degrees
\beq{nro}
\sharp\,R=2\sum_{i=1}^l(d_i-1)\,.
\eq

 The simple roots generate the root lattice in $V^*$
 $$
 Q=\sum_{j=1}^nn_j\al_j\,,~~(n_j\in\mZ\,,\,\al_j\in\Pi)\,.
 $$

There exists a unique  \emph{a maximal root} in $-\al_0\in R^+$
\beq{mro}
-\al_0 =\sum_{\al_j\in\Pi}n_j\al_j\,.
\eq
Its level is equal to $h-1$, where
\beq{co11}
h=1+\sum_{\al_j\in\Pi}n_j
\eq
 is \emph{the Coxeter number}. The minimal root $\al_0$ is defined as a
minimal element in $-R^+=R^-$. \\
The extended system $\Pi^{ext}=\Pi\cup(-\al_0)$ generates the affine Cartan matrix $a_{jk}^{(1)}$ of order $n+1$ and extended Dynkin diagram.

Let $X$ be an union of hyperplane $\lan x,\al\ran=0$, $\al\in R$, $x\in V$. The connected
components of $V\setminus X$ are called the Weyl chambers. One of them
\beq{wca}
C^+=\{x\in V\,|\,\lan x,\al\ran>0\,,~\al\in R^+\}
\eq
 is the positive Weyl chamber. The Weyl group acts simply-transitively  on the set of the  Weyl chambers.
 The simple coroots $\Pi^\vee=(\al^\vee_1,\ldots,\al^\vee_l)$ form a basis in $V$ and
generate the coroot lattice
\beq{cjrl}
Q^\vee=\sum_{j=1}^nn_j\al^\vee_j\subset V\,,~~~n_j\in\mZ\,.
\eq

The \emph{fundamental weights} $\varpi_j\in V^*\,,$ $\,(j=1,\ldots,n)$
 are defined by condition
$\lan\varpi_j,\al^\vee_k\ran=\de_{jk}$. In this way \emph{the weight lattice}
$P=\sum_{j=1}^lm_j\varpi_j\subset V^*$ is dual to the coroot lattice (\ref{cjrl}).\\
The simple roots are related to fundamental weights by the Cartan matrix
\beq{raw}
\al_k=a_{kj}\varpi_j\,.
\eq

Similarly, \emph{the fundamental coweights} are defined as
\beq{bor}
\lan\al_k,\varpi_j^\vee\ran=\de_{kj}\,.
\eq
\emph{The coweight lattice}
\beq{cwl}
 P^\vee=\sum_{j=1}^lm_j\varpi^\vee_j\,,~~m_j\in\mZ\,,~~~
\lan\varpi^\vee_j,\al_k\ran=\de_{jk}
\eq
 is dual to the root lattice $Q$.

The half-sum of positive roots
$\rho=\oh\sum_{\al\in R^+}\al$ is equal to sum of fundamental weights
$\rho=\oh\sum_{j=1}^n\varpi_j$.
We define the dual vector in $V$
\beq{arho}
\rho^\vee=\oh\sum_{\al\in R^{\vee +}}\al^\vee=\sum_{j=1}^n\varpi^\vee_j\,.
\eq
Then from (\ref{ale1}) and (\ref{ale}) the level of $\al$ is equal
\beq{arho1}
f_\al=\lan \rho^\vee,\al\ran\,.
\eq

\bigskip
\noindent
\emph{\textbf{Affine Weyl group.}}\\

\emph{The affine Weyl group} $W_a$ is a semidirect product $Q^\vee\rtimes W$ of
 the Weyl group $W$ and the group $Q^\vee$. It acts on $V^*$ as
\beq{sh}
x\to x-\lan\al,x\ran\al^\vee+k\be^\vee\,,~~~\al^\vee,\,\be^\vee\in R^\vee~~k\in\mZ\,.
\eq
This transformation is an affine reflection and the hyperplanes
$\{\lan\al,x\ran\in\mZ\}$ are invariant with respect to this action.
Connected components of the set $V\setminus\{\lan\al,x\ran\in\mZ\}$
are called \emph{the Weyl alcoves}. Their closure are fundamental
domains of the $W_a$-action.
Let us choose an alcove belonging to $C^+$ (\ref{wca})
\beq{gran}
C_{alc}=\{\,x\in V\,|\,\lan\al,x\ran> 0\,,~\al\in\Pi\,,~(\al_0,x)>- 1\,\}\,.
\eq
It has nodes
\beq{na}
C_{alc}=\{0,\varpi^\vee_1/n_1,\ldots,\varpi^\vee_n/n_j\}\,.
\eq
Here $n_j$ are the coefficients of expansion of the maximal root (\ref{mro}).

 Consider a semidirect product
 \beq{q25}
 W'_a=P^\vee\rtimes W\,.
 \eq
  In particular,
 the shift operator
 \beq{shi}
 x\to x+\ga\,,~~\ga\in P^\vee
 \eq
 is an element from $ W'_a$. It follows from this construction that
 the factor group
 \beq{fwe}
  W'_a /W_a\sim P^\vee/Q^\vee\sim\clZ(\bG)\,.
 \eq

\bigskip

\emph{\textbf{Chevalley basis in $\gg$}.}\\
Let  $\gg$ be  a simple Lie algebra over $\mC$ of rank $n$ and $\gH$ is a Cartan subalgebra. Let $\gH=V+iV$, where $V$ is the vector space defined above with the
root system $R$.
The algebra $\gg$ has the root decomposition
\beq{CD}
\gg=\gH+\gL\,,~~\gL=\sum_{\be\in R}\gR_\be\,,  ~~\dim_\mC\,\gR_\be=1\,.
\eq
The Chevalley basis in $\gg$ is generated by
\beq{CBA}
\{E_{\be_j}\in\gR_{\be_j}\,,~\be_j\in R\,,~~H_{\al_k}\in\gH\,,~\al_k\in\Pi\}\,,
\eq
where $H_{\al_k}$ are defined by the commutation relations
$$
[E_{\al_k},E_{-\al_k}]=H_{\al_k}\,,~~[H_{\al_k},E_{\pm\al_j}]= a_{kj}E_{\pm\al_k}\,,~~\al_k\,,\al_j\in\Pi\,.
$$
\beq{cbcr}
[H_{\al_j},E_{\al_k}]=a_{kj}E_{\al_k}\,,~~~[E_\al,,E_{\be}]=C_{\al,\be}E_{\al+\be}\,,
\eq
where $C_{\al,\be}$ are structure constants of $\gg$.

If $(~,~)$ is a scalar product in $\gH$ then $H_\al$ can be identified with coroots as
$$
H_\al=\al^\vee=\frac{2\al}{(\al,\al)}\,.
$$
Therefore,
\beq{kilh}
(H_\al,H_\be)=\frac{4(\al,\be)}{(\al,\al)(\be,\be)}=\frac{2}{(\al,\al)}a_{\al,\be}\,.
\eq
The Killing form in the subspace $\gL$ is expressed in terms of $(\al,\al)$
\beq{are6}
(E_\al,E_\be)=\de_{\al,-\be}\frac{2}{(\al,\al)}\,.
\eq

The  structure constants $C_{\alpha,\beta}$
 possess  the obvious properties.  Then
 \beqn{scp}
 \begin{array}{l}
C_{\alpha,\beta}=-C_{\beta,\alpha}\\
\\
C_{\lambda\alpha,\beta }=C_{\alpha,\lambda^{-1}\beta}\,,~~~~\la\in W\,,
\\
\\
C_{\alpha+\beta,-\alpha}=\frac{|\beta|^2}{|\alpha+\beta|^2}\,C_{-\alpha,-\beta}
\end{array}
 \eqn
The first property is obvious from definition (\ref{cbcr}), the
second one reflects the fact that $\lambda$ is automorphism of the
algebra and the third is the consequence of the invariance of the
Killing form. Indeed, using the Killing form,  we can define
the structure constants as the product:
$$
C_{\alpha,\beta}= \frac{|\alpha+\beta|^2}{2}\, \Big(\,
E_{-\alpha,-\beta}, [E_{\alpha}, E_{\beta}]\, \Big)\,.
$$
From invariance of the Killing form we get:
$$
C_{\alpha+\beta,-\alpha}=\frac{|\beta|^2}{2}\,\Big( E_{-\beta},
[E_{\alpha+\beta}, E_{-\alpha}] \Big)=\frac{|\beta|^2}{2}\,\Big(
E_{\alpha+\beta},[ E_{-\alpha},E_{-\beta}]
\Big)=\frac{|\beta|^2}{|\alpha+\beta|^2}\,C_{-\alpha,-\beta}
$$

\bigskip

Consider the ring of invariant polynomials on $\gg$
 It has $n=rank\gg$   generators, which we can take
to be homogeneous of degrees $d_1\,$, $\dots,d_r=h$.
Let $B$ be a Borel subgroup of $G$. It is generated
by Cartan subgroup of $G$ and by negative root subspaces $\exp\,(\sum_{\al\in R^-}E_\al)$. The coset space $Fl=G/B$ is called\emph{ the flag variety}.
It has dimension (see (\ref{nro}))
\beq{fld}
\dim\,Fl=\sum_{j=1}^l(d_j-1)\,.
\eq
  The coadjoint orbits
\beq{co}
\clO=\{Ad^*_gS_0\,|\,g\in G\,,~S_0~{\rm is~a fixed~element~of~}\gg^*\}\,.
\eq
is a generalization of a cotangent bundle to the flag varieties,
\footnote{It is a cotangent bundle if $S_0$ is a Jordan element. If $S_0$ is
semisimple,then $\clO$ is the  torsor over $Fl$.} and for generic orbits
\beq{dio}
\dim\,\clO=2\sum_{j=1}^l(d_j-1)\,.
\eq

Consider a Cartan subgroup $\clH\subset G$. Let $\clN(\clH)$ be a normalizer of $\clH$.
Then
\beq{wen}
W(R)\sim\clN(\clH)/\clH\,.
\eq

\bigskip

\noindent
\emph{\textbf{Centers of simple groups}}.\\
Let  $\bar G$ be an universal covering of $G$. The group $\bar G$ is simply-connected
and in all cases apart $G_2$, $F_4$ and $E_8$ has a non-trivial center $\clZ(\bar G)$.

\begin{center}

\begin{tabular}{|c|c|c| }
  \hline
   $\bG$ &Lie $(\bar{G})$ & $\clZ(\bar{G})$ \\
 \hline
SL$(n,\mC)$ &  $A_{n-1}$ & $\mu_n$  \\
Spin$_{2n+1}(\mC)$&  $B_n$ & $\mu_2$  \\
Sp$_n(\mC)$&  $C_n$ & $\mu_2$   \\
Spin$_{4n}(\mC) $&  $D_{2n}$& $\mu_2\oplus\mu_2$     \\
Spin$_{4n+2}(\mC) $&  $D_{2n+1}$ & $\mu_4$   \\
$E_6(\mC)$ &  $E_6$ & $\mu_3$   \\
$E_7(\mC)$ &  $E_7$ & $\mu_2$   \\
  \hline
\end{tabular}
\\
\vspace{3mm}
\textbf{Table 7}\\
Centers of universal covering groups
\\
($\mu_N=\mZ/N\mZ$)
\end{center}
\vspace{5mm}

The factor-group $P^\vee/Q^\vee$ is isomorphic to the center $\clZ(\bG)$ of simply-connected
group $\bG$. It is a cyclic group except $\gg=D_{4l}$. The order of $\clZ(\bG)$
is defined in terms of the Cartan matrix
\beq{ord}
ord \,(\clZ(\bG))=\det\,(a_{kj})\,.
\eq
The adjoint group  $G^{ad}$ is the factor group
\beq{adg}
G^{ad}=\bar G/\clZ(\bar G)\,.
\eq

In the cases $A_{n-1}$ ($n$ is non-prime), and $D_n$ the center $\clZ(\bar G)$ has
non-trivial subgroups $\clZ_l\sim\mu_l=\mZ/l\mZ$. Then there exists the factor groups
\beq{fgl}
G_l=\bG/\clZ_l\,,~~~G_p=G_l/\clZ(G_l)\,,
\eq
where $\clZ(G_l)$ is the center of $G_l$ and $\clZ(G_l)\sim\mu_p=\clZ(\bar G)/\clZ_l$.

Consider in detail the group $\bar G=Spin_{4n}(\mC)$. It has a non-trivial center
\beq{cd2}
 \clZ(Spin_{4n})=(\mu^L_2\times\mu^R_2)\,,~~\mu_2=\mZ/2\mZ\,,
\eq
where three subgroups can be described in terms of their generators as
$$
\mu^L_2=\{(1,1)\,,(-1,1)\}\,,~~\mu^R_2=\{(1,1)\,,(1,-1)\}\,,~~\mu^{diag}_2=\{(1,1)\,,(-1,-1)\}\,.
$$
 Therefore there are three intermediate subgroups between $\bar G=Spin_{4n}(\mC)$
 and $G^{ad}$
\beq{hsog}
\begin{array}{ccccc}
   &   & Spin_{4n} &  \\
   & \swarrow & \downarrow & \searrow &  \\
  Spin_{4n}^{R}= Spin_{4n}/\G^L &   & SO(4n)= Spin_{4n}/\G^{diag}&  &Spin_{4n}^{L}=Spin_{4n}/\G^R\\
   & \searrow & \downarrow & \swarrow &  \\
   &  & G^{ad}=Spin_{4n}/(\mu^L_2\times\mu^R_2) &  &
\end{array}
\eq

\bigskip

\noindent
\emph{\textbf{Characters and cocharacters.}}\\
Let $\clH$ be a Cartan subgroup $\clH\subset G$.
 Define the group of characters
\footnote{
The holomorphic maps of $\clH$ to $\mC^*$ such that $\chi(xy)=\chi(x)\chi(y)$ for $x,y\in\clT$.}
\beq{cha}
\G( G)=\{\chi\,:\,\clH\to\mC^*\}\,.
\eq
This group can be identified with a lattice group in $\gH^*$ as follows.
Let $\bfx=(x_1,z_2,\ldots,x_n)$ be an element of $\gH$, and
$\exp\,2\pi i\bfx\in\clH$. Define $\ga\in V^*$ such that
$\chi_\ga=\exp 2\pi i \lan\ga,\bfx\ran\in\G( G)$. Then
\beq{char}
\G(\bar G)=P\,,~~\G(G^{ad})=Q\,,
\eq
and $\G(G^{ad})\subseteq\G(G_l)\subseteq\G(\bar G)$.
The fundamental weights $\varpi_k\,$ $(k=1,\ldots,n)\,$
(simple roots $\al_k$) form a basis in $\G(\bar G)$
($\G(G^{ad})$). Let $\clZ(\bar G))$ be a cyclic group and $p$ is a divisor of $ord\,(\clZ(\bar G))$
such that $l=ord\,(\clZ(\bar G))/p$.
Then the lattice $\G(G_l)$ is defined as
\beq{cg}
\G(G_l)=Q+\varpi\mZ\,,~~ p\varpi\in Q\,.
\eq

Define the dual groups   of cocharacters  $t(G)=\G^*(G)$
as holomorphic maps
\beq{coch}
t(G)=\{\mC^*\to\clH\}\,.
\eq
In another way
\beq{tb1}
t(G)=\{\bfx\in \gH\,|\,\chi(e^{2\pi i\bfx})=1\}\,.
\eq
A generic element of $t(G)$ takes the form
\beq{cocharc1}
z^\ga=\exp\,2\pi i \ga\ln z\in\clH_G\,,~~\ga\in \G(G)\,,~~~z\in\mC^*\,.
\eq
In particular, the groups $t(\bar G)$ and $t(G_{ad})$
 are identified with the coroot and the coweight lattices
\beq{tb}
t(\bar G)=Q^\vee\,,~~t(G_{ad})=P^\vee\,,
\eq
and $t(\bar G)\subseteq t(G_l)\subseteq t(G^{ad})$.
It follows from (\ref{cg}) that
\beq{coch1}
t(G_l)=Q^\vee+\varpi^\vee\mZ\,,~~l\varpi^\vee\in Q^\vee\,.
\eq
The center $\clZ(G)$ of $G$ is isomorphic to the quotient
\beq{cG}
\clZ(G)\sim P^\vee/t(G)\,,
\eq
while $\pi_1(G)\sim t(G)/Q^\vee$.
In particular,
\beq{center}
\clZ(\bar G)=P^\vee/t(\bar G)\sim P^\vee/Q^\vee\,.
\eq
Similarly, the fundamental group of $G^{ad}$ is
$\pi_1(G^{ad})\sim t(G^{ad})/Q^\vee\sim P^\vee/Q^\vee$.

The triple $(R,\,t(G),\,\G(G))$ is called \emph{the root data}.
\emph{A Langlands dual} to $G$ group $^LG$ is defined by
the root data $(R^\vee,\,t(^LG),\,\G(^LG))$, where
\beq{ldg}
t(^LG)\sim\G(G)\,,~~~\G(^LG)\sim t(G)\,.
\eq
In  particular, in the simply-laced cases  $^L\bG=G^{ad}$.

\begin{center}
\vspace{3mm}

\begin{tabular}{|c|c|c|}
  \hline
  Root system & $G$ & $^LG$ \\
  \hline
  $A_n$, $N=n+1=pl$ & $G_l=\SLN/\mu_l$ & $G_p=\SLN/\mu_p$ \\
  $B_n$ & Spin$(2n+1)$ & Sp$(n)/\mu_2$ \\
  $C_n$ &  Sp$(n)$ & SO$(2n+1)$ \\
  $D_{n}$, $n=2l+1$ & Spin$(4l+2)$ & SO$(4l+2)/\mu_2$ \\
   &  SO$(4l+2)$ & SO$(4l+2)$ \\
  $D_{n}$, $n=2l$ & Spin$(4l)$ & SO$(4l)/\mu_2$  \\
   & SO$(4l)$ &  SO$(4l)$ \\
  $l=2m$ & Spin$^L(8m)$ & Spin$^L(8m)$ \\
  & Spin$^R(8m)$ & Spin$^R(8m)$ \\
  $l=2m+1$ & Spin$^L(8m+2)$ & Spin$^R(8m+2)$ \\
   $E_6$ & $E_6$ & $E_6/\mu_3$ \\
  $E_7$ & $E_7$ & $E_7/\mu_2$ \\
  \hline
\end{tabular}
\\
\vspace{3mm}
\textbf{Table 8}\\
Duality in simple groups.
\end{center}


\section{Appendix B. Elliptic Functions,
\cite{Mum,We} }

\setcounter{equation}{0}
\def\theequation{B.\arabic{equation}}

  The basic function is the theta-function.
\beq{theta}
\vth(z|\tau)=q^{\frac
{1}{8}}\sum_{n\in {\bf Z}}(-1)^ne^{\pi i(n(n+1)\tau+2nz)}\,.
\eq
It is a holomorphic function on $\mC$ with simple poles at the lattice
$\tau\mZ+\mZ$ and the quasi-periodicities
\beq{qpt}
\vth(z+1)=-\vth(z)\,,~~~\vth(z+\tau)=-q^{-\oh}e^{-2\pi iz}\vth(z)\,,
\eq
Define the relation of the theta-functions
\beq{phiz}
\phi(u,z)=\frac{\vth(u+z)\vth'(0)}{\vth(u)\vth(z)}\,.
\eq
It follows from (\ref{theta}) and (\ref{qpt}) that it
is a meromorphic function of $z\in\mC$ with simple poles at the lattice
$\tau\mZ+\mZ$ and
\beq{resp}
Res\,\phi(u,z)|_{z\in(\tau\mZ+\mZ)}=1\,,
\eq
 and the quasi-periodicities
\beq{qpp}
\phi(u,z+1)=\phi(u,z)\,,~~~\phi(u,z+\tau)=e^{-2\pi iu}\phi(u,z)\,.
\eq
Since $\phi(u,z)=\phi(z,u)$
\beq{qpp1}
\phi(u+1,z)=\phi(u,z)\,,~~~\phi(u+\tau,z)=e^{-2\pi iz}\phi(u,z)\,.
\eq
We also need two Fay identities for $\phi(z,w)$, the first one:
\beqn{fay1}
\phi(u_{1},z_{1})\phi(u_{2},z_{2})-\phi(u_{1}+u_{2},z_{1})\phi(u_{2},z_{2}-z_{1})
-\phi(u_{1}+u_{2},z_{2})\phi(u_{1},z_{1}-z_{2})=0
\eqn and its
degenerate form:
\beqn{fay2}
\phi(u_{1},z)\phi(u_{2},z)-\phi(u_{1}+u_{2},z)
(E_{1}(u_{1})+E_{1}(u_{2})) -\partial_{z}\phi(u_{1}+u_{2},z)=0\,,
\eqn
where $E_{1}(z)$ is the first Eisenstein function
\beq{e1f}
E_{1}(z)=\p_z\log\vartheta(z)\,.
\eq
The second  Eisenstein function is
\beq{e2f}
E_{2}(z)=\p^2_z\log\vartheta(z)=-\p_zE_{1}(z)\,.
\eq
They are related to  the Weierstrass functions as follows
\beq{zeta}
\ze(z|\tau)=E_{1}(z|\tau)+2\eta_1(\tau)z\,,
\eq
and
\beq{wef}
\wp(z|\tau)=E_{2}(z)-2\eta_1(\tau)\,.
\eq
Here
$$
\eta_1(\tau)=\frac{24}{2\pi i}\frac{\eta'(\tau)}{\eta(\tau)}\,,~~~
\eta(\tau)=q^{\frac{1}{24}}\prod_{n>0}(1-q^n)\,,
$$
and  $\eta(\tau)$ is the Dedekind function.

 $E_1(z)$ is quasi-periodic
\beq{qpz}
E_1(z+1|\tau)=E_1(z|\tau)\,,~~~E_1(z+\tau|\tau)=E_1(z|\tau)-2\pi i\,,
\eq
and has simple poles at at the lattice
$\tau\mZ+\mZ$
\beq{rze}
Res\,\ze(z|\tau)|_{z\in(\tau\mZ+\mZ)}=1\,.
\eq
$E_2(z)$ is double-periodic with second order poles
at the lattice.
It is related to $\phi(u,z)$ as
\beq{wpphi}
\phi(u,z)\phi(-u,z)=E_2(z)-E_2(u)\,.
\eq
 $E_2(z)$ and its derivatives $\p_z^kE_2(z)$ form a basis
in a space of double periodic function on $\Si_\tau=\mC/(\tau\mZ+\mZ)$.

 The most important object for
construction of Lax operators and $r$-matrices is the function
defined as follows:
$$
\varphi^{k}_{\alpha}(z)=e^{2\,\pi\,i\,<\kappa,\alpha>\,z}\phi\left(
<u+\kappa\,\tau\,,\, \alpha>+\frac{k}{N},\,z \right)\,.
$$
Here $u$ and $\kappa$ are vectors defined in Proposition 3.1, $\alpha$ is a root of the corresponding Lie algebra.
Note, that to safe space we omit the $u$-dependence of the
function in its definition.

An important property of function $\varphi^{k}_{\alpha}(z)$ is its
$\lambda$-invariance, more precisely, this function satisfies:
\beqn{erb} \varphi^{k}_{\lambda\alpha}(z)=
\varphi^{k}_{\alpha}(z)\ \ \
 \varphi^{k}_{\alpha+\lambda\beta}(z)=
\varphi^{k}_{\alpha+\beta}(z) \,.
\eqn
Indeed:
 \beqn{fipr} \nonumber
\varphi^{k}_{\lambda\alpha}(z)=e^{2\,\pi\,i\,<\kappa,\lambda\alpha>\,z}\phi\left(
<u+\kappa\,\tau\,,\, \lambda\alpha>+\frac{k}{N},\,z \right)=\\
\nonumber e^{2\,\pi\,i\,<\lambda^{-1}\kappa,\alpha>\,z}\phi\left(
<\lambda^{-1}(u+\kappa\,\tau)\,,\, \alpha>+\frac{k}{N},\,z
\right)=\\
\nonumber
=e^{2\,\pi\,i\,<\kappa+\varpi^{\vee},\alpha>\,z}\phi\left(
<u+\kappa\tau+\varpi^{\vee}\,\tau\,,\, \alpha>+\frac{k}{N},\,z
\right)\,,
 \eqn
 where we use the invariance of vector $u$: $\lambda u = u$. By
 definition of the
 Cartan element $\kappa$, it transforms under $\lambda$-action as
 $\lambda^{-1} \kappa=\kappa+ \varpi^{\vee}$ where $\varpi^{\vee}\in
 Q^{\vee}$, therefore $<\kappa,\alpha>=n$ - is an
 integer number and we have:
\beqn{fipr2} \nonumber
e^{2\,\pi\,i\,(<\kappa,\alpha>+n)\,z}\phi\left(
<u+\varpi^{\vee}\,\tau\,,\, \alpha>+n\tau+\frac{k}{N},\,z
\right)=\\
\nonumber =e^{2\,\pi\,i\,<\kappa,\alpha>\,z}\phi\left(
<u+\varpi^{\vee}\,\tau\,,\, \alpha>+\frac{k}{N},\,z
\right)=\varphi^{k}_{\alpha}(z)\,.
 \eqn
 Here, in the last step we have used
 $\phi(w+\tau,z)=\exp(-2\pi i z)\phi(w,z)$. The proof of the
 second identity in (\ref{erb}) is equivalent.


\small{

\end{document}